%% file: main.tex
\RequirePackage{rotating}
\documentclass[acmsmall]{acmart}
\usepackage{algorithm}
\usepackage{algcompatible}
\usepackage{colortbl}
\usepackage{enumerate}
\usepackage{lscape}
\usepackage{multirow}
\usepackage{wrapfig}
\usepackage{float}
\floatstyle{plaintop}
\restylefloat{table}

\newcommand\NoThen{\renewcommand\algorithmicthen{}}
\DeclareMathOperator*{\argmin}{arg\,min}
\newlength\myindent
\AtBeginDocument{%
  }

\setcopyright{acmcopyright}
\copyrightyear{2023}
\acmYear{2023}
\acmDOI{XXXXXXX.XXXXXXX}

\setcopyright{rightsretained}
\acmDOI{10.1145/3632905}
\acmYear{2024}
\copyrightyear{2024}
\acmSubmissionID{popl24main-p312-p}
\acmJournal{PACMPL}
\acmVolume{8}
\acmNumber{POPL}
\acmArticle{63}
\acmMonth{1}
\received{2023-07-11}
\received[accepted]{2023-11-07}

\begin{document}

\title{Inference of Probabilistic Programs with Moment-Matching Gaussian Mixtures}

\author{Francesca Randone}
\orcid{}
\affiliation{%
  \institution{IMT School for Advanced Studies Lucca}
  \city{}
  \country{Italy}
}
\email{francesca.randone@imtlucca.it}

\author{Luca Bortolussi}
\orcid{}
\affiliation{%
  \institution{University of Trieste}
  \city{}
  \country{Italy}
}
\email{lbortolussi@units.it}

\author{Emilio Incerto}
\orcid{}
\affiliation{%
  \institution{IMT School for Advanced Studies Lucca}
  \city{}
  \country{Italy}
}
\email{emilio.incerto@imtlucca.it}

\author{Mirco Tribastone}
\orcid{}
\affiliation{%
  \institution{IMT School for Advanced Studies Lucca}
  \city{}
  \country{Italy}
}
\email{mirco.tribastone@imtlucca.it}

%\renewcommand{\shortauthors}{Randone et al.}

\begin{abstract}
  Computing the posterior distribution of a probabilistic program is a hard task for which no one-fit-for-all solution exists. We propose Gaussian Semantics, which approximates the exact probabilistic semantics of a bounded program by means of Gaussian mixtures. It is parametrized by a map that associates each program location with the moment order to be matched in the approximation. We provide two main contributions. The first is a universal approximation theorem stating that, under mild conditions, Gaussian Semantics can approximate the exact semantics arbitrarily closely. The second is an approximation that matches up to second-order moments analytically in face of the generally difficult problem of matching moments of Gaussian mixtures with arbitrary moment order.  
  We test our second-order Gaussian approximation (SOGA) on a number of case studies from the literature. We show that it can provide accurate estimates in models not supported by other approximation methods or when exact symbolic techniques fail because of complex expressions or non-simplified integrals. On two notable classes of problems, namely collaborative filtering and programs involving mixtures of continuous and discrete distributions, we show that SOGA significantly outperforms alternative techniques in terms of accuracy and computational time. 
\end{abstract}






%\newcommand{\f}[1]{{\color{blue} FR: #1}}
%\newcommand{\m}[1]{{\color{red} MT: #1}}
%\newcommand{\e}[1]{{\color{red} EI: #1}}

\maketitle

\section{Introduction}
\label{sec:intro}
Probabilistic programming languages are programming languages augmented with primitives expressing probabilistic behaviours ~\citep{gordon2014probabilistic}. Examples are random assignments (``program variable $x$ is distributed according to the probability distribution $D$''), probabilistic choices (``do $P_1$ with probability $p$ else $P_2$) or conditioning (``variable $x$ is distributed according to $D$, under the constraint that it can only take positive values''). This has enabled a variety of applications such as the analysis of randomized algorithms, machine learning and biology~\citep{gordon2014probabilistic}. 

Given a probabilistic program, there are different equivalent ways in which its semantics can be defined~\citep{kozen1983probabilistic}. Following Kozen's Semantics 2 \citep{kozen1979semantics}, in this paper we see a program as a transformer: given an initial joint distribution over the program variables, each instruction in the program transforms that joint distribution into a possibly different one, for example,  due to the presence of probabilistic assignments or conditional statements. In this framework, we are interested in the \emph{inference problem}: given a program $P$ and an initial distribution $D$ over program variables, what is the distribution over program variables after executing $P$? Borrowing from Bayesian inference, we will sometimes refer to the initial distribution $D$ as the \emph{prior distribution} over program variables and to the distribution obtained after executing $P$ as the \emph{posterior distribution}. Then, the inference problem boils down to computing the posterior.

Over the years, many approaches have tackled this problem: numerical methods based on Monte Carlo Markov chain (MCMC) sampling \citep{hastings1970monte,nori2014r2,goodman2008church,mansinghka2014venture,pfeffer2001ibal,chaganty2013efficiently}, variational inference (VI) \citep{bingham2019pyro,jordan1999introduction,kucukelbir2015automatic}, symbolic execution \citep{gehr2016psi,narayanan2016probabilistic,saad2021sppl}, volume computation \citep{holtzen2020scaling,filieri2013reliability,huang2021aqua}, and approaches based on moment-based invariants~\citep{barthe2016synthesizing,chakarov2014expectation,katoen2010linear,bartocci2020mora,moosbrugger2022moment}. %However, no single solution can still solve the general inference problem, and each is tailored to some specific program properties.

%As far as the author know, there are current no methods to compute the semantics of unbounded loops in presence of continuous distributions, therefore  we restrict our attention to bounded programs, similarly to what is done in \citet{gehr2016psi}, \citet{huang2021aqua}, \citet{carpenter2017stan}, \citet{albarghouthi2017fairsquare} and \citet{nori2014r2}. Even for this class of programs inference can be challenging.

%\begin{algorithm}[t]
%\caption{\emph{Altermu} from \citep{huang2021aqua}} \label{alg:altermu}
%\begin{algorithmic}[1]
%\REQUIRE Observations $(o_i)_{i=1,\hdots,40}$ 
%\ENSURE  Posterior means~$\mathbb{E}[m_j|(o_i)_{i=1,\hdots,40}], 1 \leq j \leq 3$
%\STATE $m_1 = gauss(0,5)$ 
%\STATE $m_2 = gauss(0,5)$
%\STATE $m_3 = gauss(0,5)$
%\FOR{$i$ \textbf{in} $range(40)$}
%    \STATE $y_i = 3 m_1 m_2$
%    \STATE $y_i = y_i - m_3 + gauss(0,1)$ 
%\ENDFOR
%\FOR{$i$ \textbf{in} $range(40)$}
%    \STATE\textbf{observe}($y_i == o_i$)
%\ENDFOR
%\end{algorithmic}
%\end{algorithm}

\subsection{Motivating Example}\label{sec:tracking}

\begin{wrapfigure}{L}{0.33\textwidth}
    \begin{minipage}{0.33\textwidth}
        %\begin{figure}%[H]
        %\caption{\emph{Tracking\_n} adapted from \citep{wu2018discrete}} \label{alg:tracking}
            \begin{algorithmic}[1]
            \STATE $x = 2, y = -1$
            \FOR{$i$ \textbf{in} $range(n)$}
                \STATE $x = x + gauss(0,1)$
                \STATE $y = y + gauss(0,1)$ 
            \ENDFOR
            \STATE $dist = x^2 + y^2$
            \IF{$dist > 10$}
                \STATE $out = 1$
            \ELSE 
                \STATE $out = 0$
            \ENDIF
            \IF{$out = 1$}
                \STATE $obs\_dist = 10$
            \ELSE
                \STATE $obs\_dist = gauss(dist,1)$
            \ENDIF
            \STATE $\textbf{observe}(out == 1)$
            \STATE $\textbf{return } obs\_dist$
            \end{algorithmic}
        %\end{figure}
    \end{minipage}
\end{wrapfigure}
%\hfill
%\begin{minipage}{0.55\textwidth}
As a motivating example, let us consider the \emph{Tracking\_n} model reported in left inset and adapted from \citet{wu2018discrete}. It describes a Gaussian process evolving on a bi-dimensional space for $n$ steps and starting from coordinates (2, -1) (lines 1-5). A radar is positioned in (0,0), and can sense the process if it is at a squared distance ($dist$, line 6) of less than 10 units from the radar. Therefore the process can be either out of scope ($out=1$, line 8) or in scope ($out=0$, line 10). When the process is out of scope, the radar returns an observed distance of $10$ (line 13) and a noisy measurement of the true distance else (line 15). Therefore, the distribution over $obs\_dist$ is a mixture of $\delta_{10}$, i.e., a Dirac delta centered in 10, and a Gaussian with mean $dist$. However, if we observe that the process is out of scope (line 17), the posterior over $obs\_dist$ is just $\delta_{10}$ because any continuous distribution puts zero mass on a single point. Therefore, the exact posterior over $obs\_dist$ is a distribution placing probability 1 on 10. While this program may seem quite simple, performing inference may be challenging. 

Using PSI~\citep{gehr2016psi}, an exact symbolic solution returns a formula for the posterior mean of $obs\_dist$ in less than a second, which, however, contains several non-simplified integrals. This is because, in line~6, computing $dist$ requires computing the probability density function (pdf) of the product of two continuous distributions, and this requires symbolic integration.  Attempting to integrate it numerically  using Mathematica [\citeauthor{Mathematica}] did not terminate after 30 minutes on common machine. One can resort to approximate approaches; however, many methods, such as AQUA's quantization~\citep{huang2021aqua} and STAN's MCMC sampling~\citep{carpenter2017stan} and Pyro's VI~\citep{bingham2019pyro} do not support discrete posteriors, therefore this particular program cannot be encoded in their syntax. BLOG is a probabilistic programming language relying on probabilistic relational model representation and likelihood weighting sampling~\citep{milch2004blog}, that has been extended by~\citet{wu2018discrete} for mixtures of continuous and discrete distribution such as the one in our example. It computes the exact posterior in 0.516\,s for $n = 1$ and about 5\,s for $n=100$. A similar behavior is exhibited by applying Pyro's variable elimination \citep{obermeyer2019tensor}, which computes the exact posterior in 0.192\,s for $n = 1$ and about 9\,s for $n=100$ (see Section~\ref{sec:mix}).

\subsection{Proposed Approach}

The difficulty in performing inference on the previous program stems from various factors: PSI's exact engine returns non-simplified integrals, requiring computationally expensive numerical integration. STAN's MCMC, Pyro's VI and AQUA's quantization cannot be applied in this case, but in general, can incur long computational times and out-of-memory errors (see Section \ref{sec:eval}). BLOG's and Pyro's ad hoc sampling perform best, but increasing the number of steps hinders scalability. 

To complement all these techniques, we present a new \emph{approximate analytical} method that does not require integration or sampling and that relies on a compact representation of the joint distribution using moment-matching Gaussian mixtures (GMs). Our choice of representation is based on some desirable properties of GMs, and in particular the following three: i) they can encode both continuous and discrete distributions (using degenerate GMs); ii) their moments can be computed exactly and efficiently; iii) they are universal approximators, so we can always increase the number of components in our representation to get a better approximation. These considerations lead to the definition of a family of approximating semantics called \emph{Gaussian Semantics}.  

\begin{figure}
    \centering
    \includegraphics[width=0.90\textwidth]{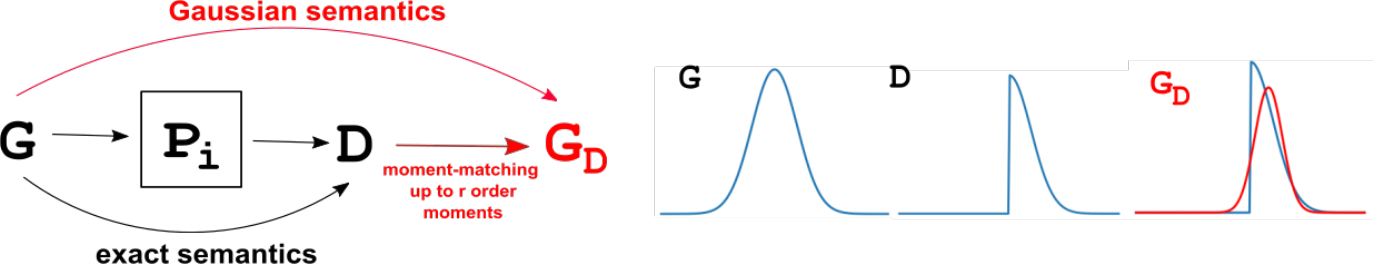}
    \caption{Left part: general approximation scheme used in Gaussian Semantics. In program location $P_i$ the exact semantics of the program transforms a GM $G$ into a non-GM distribution $D$. In the same program location, Gaussian Semantics transform $G$ into another GM $G_D$, that approximates $D$ using moment-matching. Right part: concrete example. The Gaussian distribution $G$ is transformed by the exact semantics into a truncated Gaussian $D$ and by the second-order Gaussian Semantics into the red Gaussian distribution $G_D$.}
    \label{fig:fig1}
\end{figure}

More in detail, we define Gaussian Semantics so that it is closed with respect to the class of (degenerate) GMs, meaning that, at every program location, the Gaussian Semantics of a program transforms a GM into a GM. In particular, we proceed as in the general approximation scheme proposed by~\citet{boyen1998tractable}: given a GM $G$, the exact semantics of a program location would transform it in a different distribution $D$, which is not necessarily a GM. However, we approximate $D$ with a new GM $G_D$ and define the Gaussian Semantics as semantics that transform $G$ into $G_D$ at that program location. This process is represented in Figure \ref{fig:fig1}. Performing this at every program location approximates the whole program semantics. In particular, we choose to approximate $D$ with $G_D$ using moment-matching, meaning that $G_D$ is a GM having the same moments of $D$ up to a certain order $r$. This is convenient for two reasons: first, it avoids computing the full pdf of $D$, as only its first $r$ moments are needed to find $G_D$; second, since $D$ is obtained as a transformation of a GM, it can be expressed as a linear combination of transformed Gaussians, and its moments can be computed analytically using the results summarized in Table~\ref{tab:tab1}. 

\begin{table}[]
    \centering
    \resizebox{0.9\textwidth}{!}{
    \begin{tabular}{ccc}
    \toprule 
    \textbf{Operation} & \textbf{Theoretical Result} & \textbf{Computes moments for:} \\
    \midrule 
    \multirow{2}{3.5cm}{Sum of Gaussians} & Closed w.r.t linear transformations & \multirow{2}{*}{$c_1 \mathcal{N}(\mu_1, \Sigma_1) + c_2 \mathcal{N}(\mu_1, \Sigma_2)$} \\
    & \cite{billingsley2008probability} &  \\
    \midrule
    \multirow{2}{3.5cm}{Conditioning Gaussians to $x_i==c$} & Closed w.r.t. conditioning & \multirow{2}{*}{$\mathcal{N}(\mu, \Sigma \,|\, x_i == c)$}\\
    & \cite{bishop2006pattern}  & \\
    \midrule
    \multirow{2}{3.5cm}{Conditioning Gaussians to $x \in [a,b]$} & Iterative formulas & \multirow{2}{*}{$\mathcal{N}(\mu, \Sigma \,|\, x \in [a,b])$}\\
    & \cite{kan2017moments}  & \\
    \midrule
    \multirow{2}{3.5cm}{Product of Gaussian} & Isserlis' Theorem & \multirow{2}{*}{$\mathcal{N}(\mu_1, \Sigma_1)\mathcal{N}(\mu_2, \Sigma_2)$}\\
    & \cite{wick1950evaluation}  & \\
    \bottomrule
    \end{tabular}}
    \caption{Summary of the theoretical results used to compute the moments of transformed Gaussian Mixtures. $\mathcal{N}(\mu, \Sigma)$ denotes a Gaussian distribution with mean $\mu$ and covariance matrix $\Sigma$, $c$ is any real constant and $a,b$ are vectors in $\mathbb{R}^d$ defining the hyper-rectangle $[a,b] = \{x \in \mathbb{R}^d : a_i \le x_i \le b_i\}$.}
    \label{tab:tab1}
\end{table}

To sum up, in a Gaussian Semantics each program location is associated with an integer $r$, and the semantics acts on a GM $G$ performing two steps: first, it computes the first $r$ order moments of the transformed distribution $D$, using the results in Table \ref{tab:tab1}; then, it finds a new GM $G_D$ having same moments as $D$ up to order $r$. %Actually, if $D$ is a mixture, we do not just moment-match $D$ with a GM, but each of its components \m{really needed to highlight this here?}. 
More than one moment-matching GM $G_D$ can exist, therefore, we give a heuristic to determine a unique $G_D$ for any $r$.  In particular, we base our heuristics, called \emph{max entropy matching}, on the maximum entropy principle \cite{kullback1951information}. 

Our first technical contribution is theoretical: we provide a universal approximation result stating that, under mild conditions, when the order of moments matched at each program location grows, the family of Gaussian Semantics converges to the exact probabilistic semantics. %To prove this, we apply measure theory to the moment problem~(\citep{schmudgen2017moment}), which can be seen as a generalization of the celebrated method of moments~(\citep{pearson1894contributions}). 
While our result exploits the well-known universal approximation power of GMs \citep{lo1972finite}, it is a non-trivial consequence of it. The density of GMs guarantees the existence of a GM arbitrarily close to a target distribution; however, for a probabilistic program the target distribution is generally not known. Here we give a \emph{constructive method} to build the approximating GM.

Besides the definition of Gaussian Semantics, we look at how they can be practically computed. Unfortunately, it turns out that while the formulas in Table \ref{tab:tab1} allow us to compute moments up to any order, finding a moment-matching GM is a hard task. In fact, finding a moment-matching GM for moment orders higher than two requires the solution of a constrained system of polynomial equations, for which no analytical solution is known \citep{lasserre2009moments}. Despite this, when only the first two orders of moments are matched, our matching boils down to using a single Gaussian distribution with a given mean and covariance, and no system of equations needs to be solved. 

We call this particular instance \emph{Second Order Gaussian Approximation} (SOGA) and present an algorithm that implements it. In our motivating example, at line 6, to approximate the distribution of $dist$ after the assignment $dist = x^2 + y^2$, SOGA proceeds as follows. It first computes the means and covariance matrices of $x^2$ and $y^2$ using Isserlis' theorem \cite{wick1950evaluation} (observe that $x$ and $y$ are Gaussian, but $x^2$ and $y^2$ are not). Then, it approximates the distributions of $x^2$ and $y^2$ with two Gaussians having the computed means and covariance matrices. Finally, it exploits the closedness of Gaussians with respect to sum to approximate the distribution of $dist$ with the sum of the Gaussians approximating $x^2$ and $y^2$. Therefore, while in the exact semantics, after line 6, $dist$ does not have a GM distribution, in SOGA it does. This significantly simplifies the subsequent computations. Indeed, when entering the if statement at line 7, $dist$ is conditioned to $dist > 10$. Performing conditioning in the exact semantics requires computing the integral of the pdf of $dist$ over the set of vectors satisfying $dist > 10$. Instead, in SOGA $dist$ is Gaussianly distributed, therefore we can compute the moments of the conditioned distribution using the formulas from \citet{kan2017moments}, and then approximate the conditioned distribution with a Gaussian having given mean and covariance matrix. Overall, for Algorithm~1 SOGA computes the output, which in this case is exact, in 0.042\,s for $n=1$ and in 0.192\,s for $n=100$, performing significantly better than BLOG and Pyro.

In general, the posterior computed by SOGA is a GM whose number of components grows exponentially in the number of conditional statements. To help cope with this, we introduce a pruning strategy that keeps the number of components in the GMs below a user-specified threshold by merging components with minimal cost. Using a prototype implementation, we compare SOGA on a corpus of benchmarks against state-of-the-art tools representative of different inference methods: MCMC sampling (STAN), symbolic execution (PSI), quantization (AQUA), VI (Pyro). %Our results show that SOGA performs inference on various models in competitive runtimes and with high accuracy. \m{questa frase la toglierei perché è il concetto è ripetuto nel paragrafo successivo?} 
Even when it is not the best-performing method, it still provides the flexibility to model both continuous and discrete posteriors, unlike STAN, Pyro and AQUA, which only support the former. Additionally, it enables reaching numerical solutions in reasonable runtimes when PSI returns non-simplified integrals that demand computationally prohibitive times for numerical integration. When applied to the analyzed benchmarks, pruning significantly reduced the computational time without incurring noticeable approximation errors.

Importantly, we highlight that SOGA is particularly useful for performing inference on two classes of programs: those involving mixtures of continuous and discrete distributions and collaborative filtering models. Most state-of-the-art approaches do not support the first class, even though it is known that this kind of distribution arises in various application domains \cite{gao2017estimating,kharchenko2014bayesian,pierson2015zifa}. Thanks to its GM representation, SOGA can easily encode these distributions. When tested on benchmarks introduced specifically for this problem, SOGA is able to perform inference faster than dedicated methods such as \citet{wu2018discrete}, while identifying the exact posterior. Collaborative filtering models are an established framework to model recommendation systems and have been extensively investigated in the machine learning community~\cite{koren2021advances}. 
%\m{rimuovere questa frase?} They involve many latent parameters that are generally non-identifiable and therefore, difficult to infer. 
SOGA can deal with a large number of variables without incurring large computational times or out-of-memory errors, as happens with alternative methods. 

\paragraph{Contributions.} In summary, the contribution of this paper is threefold:
\begin{enumerate}[i)]
\item From the theoretical point of view, we define a family of approximating semantics called Gaussian Semantics and prove that they approximate the exact semantics of a bounded probabilistic program arbitrarily well.
\item From the practical perspective, we present an implementation of a particular instance of Gaussian Semantics, called SOGA, and evaluate it against other state-of-the-art implementations of alternative techniques (PSI~\cite{gehr2016psi}, STAN~\cite{carpenter2017stan}, AQUA~\cite{huang2021aqua}, Pyro~\cite{bingham2019pyro}) on a set of benchmarks taken from the literature. 
While not always the best performer, SOGA can handle models with discrete posteriors while STAN, AQUA and Pyro only support continuous ones. On the other hand, SOGA can provide accurate and computationally tractable approximations when symbolic analysis by PSI may fail due to complex formulas or non-simplified integrals that cause high computational costs for their numerical integration. 
\item We focus on two classes of models taken from the machine learning literature --- collaborative filtering and inference involving mixtures of continuous and discrete distributions --- where SOGA clearly outperforms the other methods, complementing the current state-of-the-art.
\end{enumerate}
 
\paragraph{Paper Structure} Notation and background notions are presented in Section \ref{sec:background}. Control-flow syntax and exact semantics are introduced in Section \ref{sec:source}. Gaussian Semantics is introduced in Section~\ref{sec:gaussian}, while the universal approximation theorem is presented in Section~\ref{sec:convergence}. We present SOGA in Section~\ref{sec:gaussapp} and evaluate it in Section~\ref{sec:eval}. We cover further related works in Section~\ref{sec:related}, while conclusions and future works are drawn in Section~\ref{sec:conclusion}.

\section{Background}
\label{sec:background}

We now introduce some of the notation and the concepts that will be used in the rest of the paper (we refer the reader to the Supplementary Material for additional background material).

\paragraph{Notation.} Given a Boolean value $B$, $\neg B$ denotes its negation. For a vector $x \in \mathbb{R}^d$, $x\setminus x_i$ denotes a vector in $\mathbb{R}^{d-1}$ obtained from $x$ by suppressing the $i$-th component; $x[x_i=E(x)]$ denotes a vector $x \in \mathbb{R}^d$ in which the $i$-th component is replaced with the expression $E(x)$; $\| x\|$ denotes the 2-norm; $\mathit{diag}(d_1, \hdots, d_s)$ denotes the $\mathbb{R}^{s \times s}$ diagonal matrix having $d_1, \hdots, d_s$ as diagonal elements.

\paragraph{Probability Distributions.} We always deal with distributions over $\mathbb{R}^d$ and use $x \sim D$ to denote that the stochastic vector $x$ is distributed according to distribution $D$. We always assume that a distribution $D$ can be specified by its probability density function (pdf) $f_D:\mathbb{R}^d \to \mathbb{R}_{\ge0}$. For $x \sim D$ and a set $A \subseteq \mathbb{R}^d$, the probability of $A$ under $D$, denoted by $P_D(A)$, can be expressed as the Lebesgue integral $P_D(A) = \int_A f_D(x)dx$. Sometimes, we will find it more convenient to refer to the probability measure induced by $D$ on the measurable space $(\mathbb{R}^d, \mathcal{B}(\mathbb{R}^d))$, where $\mathcal{B}(\mathbb{R}^d)$ is the Borel $\sigma$-algebra on $\mathbb{R}^d$. By probability measure, we mean a function $m : \mathcal{B}(\mathbb{R}^d) \to [0,1]$ that satisfies the following two properties: i) $m(\emptyset) = 0$ and $m(\mathbb{R}^d)=1$; ii) $m(\cup_{i \in \mathbb{N}} A_n) = \sum_{i \in \mathbb{N}} m(A_n)$ for any countable collection of disjoint sets $A_n \subset \mathbb{R}^d$. For a distribution $D$, the associated measure $m_D(A)$ is given by $m_D(A) = P_D(A) = \int_A f_D(x) \, dx$ for every $A \in \mathcal{B} (\mathbb{R}^d)$. Moreover, due to the presence of conditional branches and observe statements in a probabilistic program, we consider distributions conditioned to subsets of $\mathbb{R}^d$. Letting $\mathbb{I}_A$ be the characteristic function of a set $A \subset \mathbb{R}^d$ such that $P_D(A) > 0$, $D|A$ will denote the \emph{distribution of $D$ truncated (or conditioned) to $A$}, whose pdf is given by
$ f_{D|A} = \frac{1}{P_D(A)} f_D\mathbb{I}_{A}$. Observe that $f_{D|A}$ is obtained by setting $f_D$ to $0$ outside $A$, and then, by dividing it by $P_D(A)$, so that the induced measure is still a probability measure. Given a $d$-dimensional random vector $x \sim D$ and a subvector $x' = (x_{i_1}, \hdots, x_{i_s})$ with $i_1, \hdots, i_s \in \{1, \hdots, d\}$, we denote by $\mathit{Marg}_{x'}(D)$ the marginal distribution of $D$, obtained integrating out the components not in $x'$, i.e. $x' \sim \mathit{Marg}_{x'}(D) = \int_{\mathbb{R}^{d-s}} f_D(x) d(x\setminus x')$.

%For our main theorem we need to introduce a topology on the space of distributions, namely the \emph{weak topology}, induced by weak convergence. 
%\begin{definition}[Weak Convergence] 
%For a sequence of random vectors $X_n \sim D_n$ with cdfs $F_{D_n}$ we say that $D_n$ {\it converge weakly to} $D$, with $F$, if for every continuity point $x$ of $F$ (i.e. points for which $\lim_{y \to x} F(y) = F(x)$) it holds:
%$$ \lim_n F_{D_n}(x) = F_D(x).$$
%We denote weak convergence with $D_n \xrightarrow{n \to \infty} D$.
%\end{definition} 
%Equivalently we say that the corresponding measure converges weakly, denoted by $m_{D_n} \xrightarrow{n \to \infty} m_D$.
%Interestingly, the space of distributions with the weak topology is metrizable, i.e. we can define a metric such that weak convergence is equivalent to convergence in the metric. This metric is the Levy-Prokhorov distance that for two measures $m, m'$ on $(\mathbb{R}^d, \mathcal{B}(\mathbb{R}^d))$ is defined as:
%\begin{align}  \nonumber
%d_{LP}(m,m') = &\inf\{\epsilon > 0 \, | \, m(A) \le m'(A^{\epsilon}) + \epsilon \text{ and } m'(A) \le m'(A^{\epsilon}) + \epsilon, \, \\
%& \hspace{7cm} \forall \, A \in \mathcal{B}(\mathbb{R}^d) \} \label{eq:lp}
% \end{align}
%where for $A \subseteq \mathbb{R}^d$, $A^{\epsilon} = \{ x \in \mathbb{R}^d \, | \, \exists \, y \in A \text{ s.t. } \| x - y \| \le \epsilon\}$.

\paragraph{Gaussian Distributions and Mixtures.}
%and have pdf
%$$ f_{\mathcal{N}(\mu, \Sigma)} = \frac{1}{\sqrt{(2\pi)^ddet(\Sigma)}} \exp{\left(-\frac{1}{2}(x-\mu)^T\Sigma^{-1}(x-\mu)\right)}.$$

\begin{figure}
    \centering
    \includegraphics[width=\textwidth]{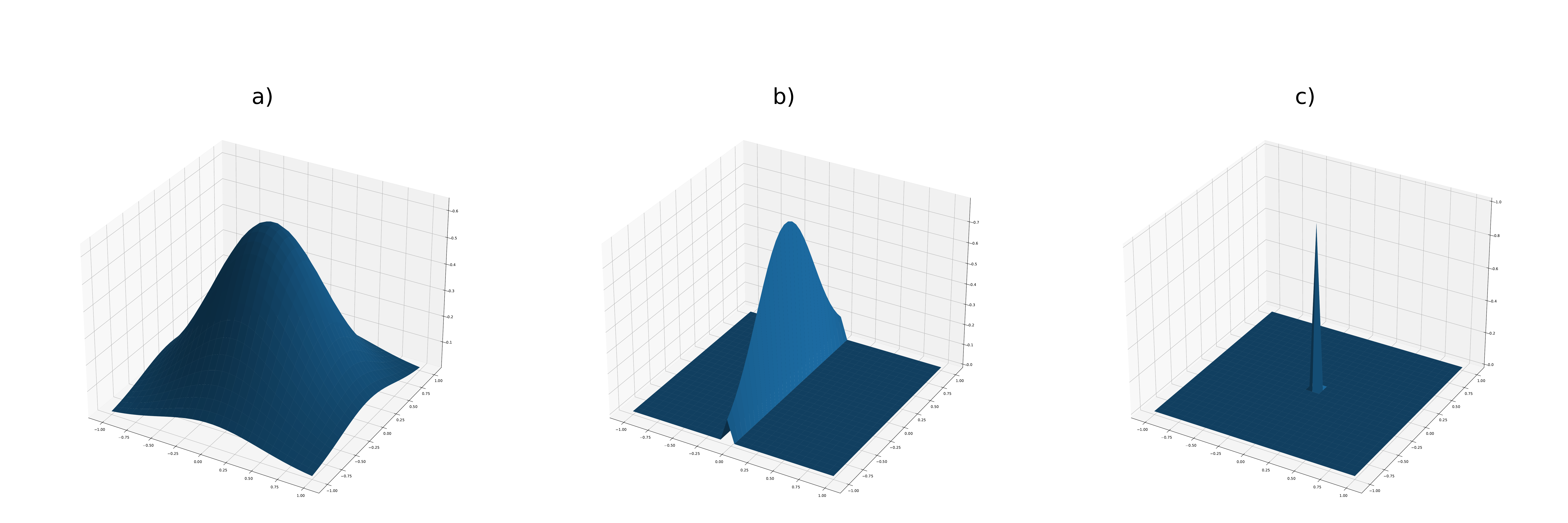}
    \caption{Plot of: a) a non-degenerate 2-dimensional Gaussian; b) a degenerate Gaussian whose covariance matrix has rank 1; c) a degenerate Gaussian with null covariance matrix (Dirac delta). Considering mixtures of possibly degenerate Gaussians, allow us to capture (mixtures of) both continuous and discrete distributions.}
    \label{fig:gauss}
\end{figure}
%While this pdf is only well definite if $\Sigma$ is non-singular
Gaussian distributions with mean $\mu$ and covariance matrix $\Sigma$ are denoted by $\mathcal{N}(\mu, \Sigma)$. We assume that $\Sigma$ can be singular, which corresponds to Gaussian distributions having support in a subspace of $\mathbb{R}^d$, as shown in Figure \ref{fig:gauss}. In particular, when $\Sigma$ is the null matrix we consider the associated random variable to be a Dirac delta centered in $\mu$, referred to as $\delta_\mu$. Gaussian distributions enjoy many useful properties; some that we will use are listed in Table \ref{tab:tab1} while we refer  to~\cite[Chapter 2.3]{bishop2006pattern} for a detailed treatment.
%In this case it is non-trivial to express the density of such distributions, and for a detailed treatment we refer the reader to Appendix \ref{app:comp}. 

We refer to \emph{mixtures} as the scalar products of two vectors $(p_1, \ldots, p_C)$ and $(D_1, \ldots, D_C)$ such that $\sum_{i=1}^C p_i = 1$, $0 < p_i \le 1 $,  and $D_i$ is the distribution of the $i$-th \emph{component}, with $i=1, \ldots,C$. The numbers $p_i$ are called \emph{mixing coefficients}. We will denote a mixture as $M = p_1D_1+ \ldots +p_CD_C$, thus indicating that $M$ has pdf $f_M(x) = p_1f_{D_1}(x) + \ldots + p_Cf_{D_C}(x)$. When $C=1$, we recover the case of a single distribution.
A special case is given by Gaussian Mixtures (GMs) in which $D_i = \mathcal{N}(\mu_i, \Sigma_i)$ for $i=1,\ldots,C$, with mean vectors and covariance matrices $\mu_i, \Sigma_i$. We assume $(\mu_i, \Sigma_i) \neq (\mu_j, \Sigma_j)$ for $i \neq j.$ The set of GMs is dense in the set of probability distributions with respect to the weak topology~\citep{lo1972finite}, meaning that for any probability distribution one can always find a GM that approximates it arbitrarily closely with respect to a particular metric, the Levy-Prokhorov distance. Since we consider Dirac deltas as particular Gaussian distributions, discrete distributions over a finite set of values are included in the set of GMs.

\paragraph{Distributions Determined by Their Moments.}
Let $x \sim D$. For $r = (r_1, \ldots, r_d)$, with $r_i \in \mathbb{N}, \, \forall \, i$ define
$$ \mathbb{E}[x^{r}] = \int_{\mathbb{R}^d} x_1^{r_1}\ldots x_d^{r_d} f_D(x) dx.$$
Letting $r$ vary over all vectors in $\mathbb{N}^d$ such that $r_1 + \ldots + r_d = r$ we obtain the set of all $r$-th order central moments of $D$. Observe that, for any $D$, $\mathbb{E}[x^{0}] = 1$.
Since the construction of our semantics relies on the Method of Moments, we need to assure that this converges to the correct distribution. This is true only if no other distribution has all moments equal to those of the target one \citep{billingsley2008probability}. We say that in this case, the target distribution is \emph{determined by its moments}, formalized next. 
\begin{definition} A distribution $D$ is determined by its moments, if for any other distribution $D'$ such that for all $r_1, \ldots, r_d \ge 0$ 
$$ \int_{\mathbb{R}^d} x_1^{r_1}\ldots x_d^{r_d} f_D(x)dx = \int_{\mathbb{R}^d} x_1^{r_1}\ldots x_d^{r_d} f_{D'}(x)dx$$
it holds that $m_D=m_{D'}.$
\end{definition}

\section{Syntax and Exact Probabilistic Semantics}\label{sec:source}

\subsection{Syntax}\label{sec:syn}
Following \citet{kozen1979semantics}, we will consider probabilistic programs as transformers over distributions $D$ defined over a vector of variables taking values in $\mathbb{R}^d$.
Similarly to~\citet{chaudhuri2011smoothing}, we represent programs in a control flow-graph (cfg) syntax~\citep{cousot1977abstract}. We use as explanatory example the simple program in Algorithm $\ref{alg:running}$.

A program is a directed graph $P=(V,E)$ where $V$ is set of nodes and $E$ is the set of edges. Specifically, we consider directed acyclic graphs (DAGs) of bounded depth. Each node belongs to one of five types in $\Gamma = \{ \textit{entry, state, test, observe, exit} \}.$
We denote the fact that a node $v \in V$ is of a given type $\gamma \in \Gamma$ with $v \colon \gamma$. The nodes satisfy the following properties.
\begin{itemize}
    \item A node $v\colon entry$ has no incoming edge and one outgoing edge. 
    \item A node $v \colon state$ has any number of incoming edges and one outgoing edge. A function $cond$ is defined on the set of state nodes, such that $cond: \{ v \in V \, | \, v \colon state \} \to \{true, false, none\}$ and $cond(v) = none$ if and only if the parent of $v$ is not a test node.
    \item A node $v \colon test$ has one incoming edge and two outgoing edges toward state nodes $v_1, v_2$ such that $cond(v_1) = true$ and $cond(v_2) = \textit{false}$.
    \item A node $v \colon observe$ has one incoming edge and one outgoing edge.
    \item A node $v \colon exit$ has any number of incoming edges and no output edge. 
\end{itemize}
Moreover, for each program, there is exactly one $v \in V$ such that $v \colon entry$ and one $v \in V$ such that $v \colon exit$, and they correspond to the root and the only leaf of the DAG representing the program, respectively. The control-flow syntax for Algorithm \ref{alg:running} is represented in Figure \ref{fig:cfg}. %\m{trasponiamo Fig 3 sfruttando la larghezza della pagina e con carattere più grande? se rimane tempo lo farei con tikz} \f{non ho capito cosa intend con trasponiamo}

\begin{minipage}{0.45\textwidth}
\begin{algorithm}[H]
\caption{Example} \label{alg:running}
\begin{algorithmic}[1]
\STATE \textbf{entry} \hfill\COMMENT{$v_0:entry$}
\STATE $x_1 = gauss(0,1)$ \hfill\COMMENT{$v_1:state$}
\NoThen \IF{$x_1 > 0$ \hfill\COMMENT{$v_2:test$}}
    \STATE$ x_2 = 2x_1 + 1 + gauss(0,0.1)$ \hfill\COMMENT{$v_3:state$}
\ELSE
    \STATE $x_2 = -2x_1 + 1 + gauss(0,0.1)$ \hfill \COMMENT{$v_4: state$}
\ENDIF
\STATE $\textbf{skip}$ \hfill\COMMENT{$v_5:state$}
\STATE $\textbf{observe}(x_2 < 3) $ \hfill\COMMENT{$v_6:observe$}
\STATE \textbf{exit} \hfill\COMMENT{$v_7:exit$}

\hfill
\end{algorithmic}
\end{algorithm}
\end{minipage}
\hfill
\begin{minipage}{0.54\textwidth}
    \centering
    \includegraphics[angle=270,width=0.95\linewidth]{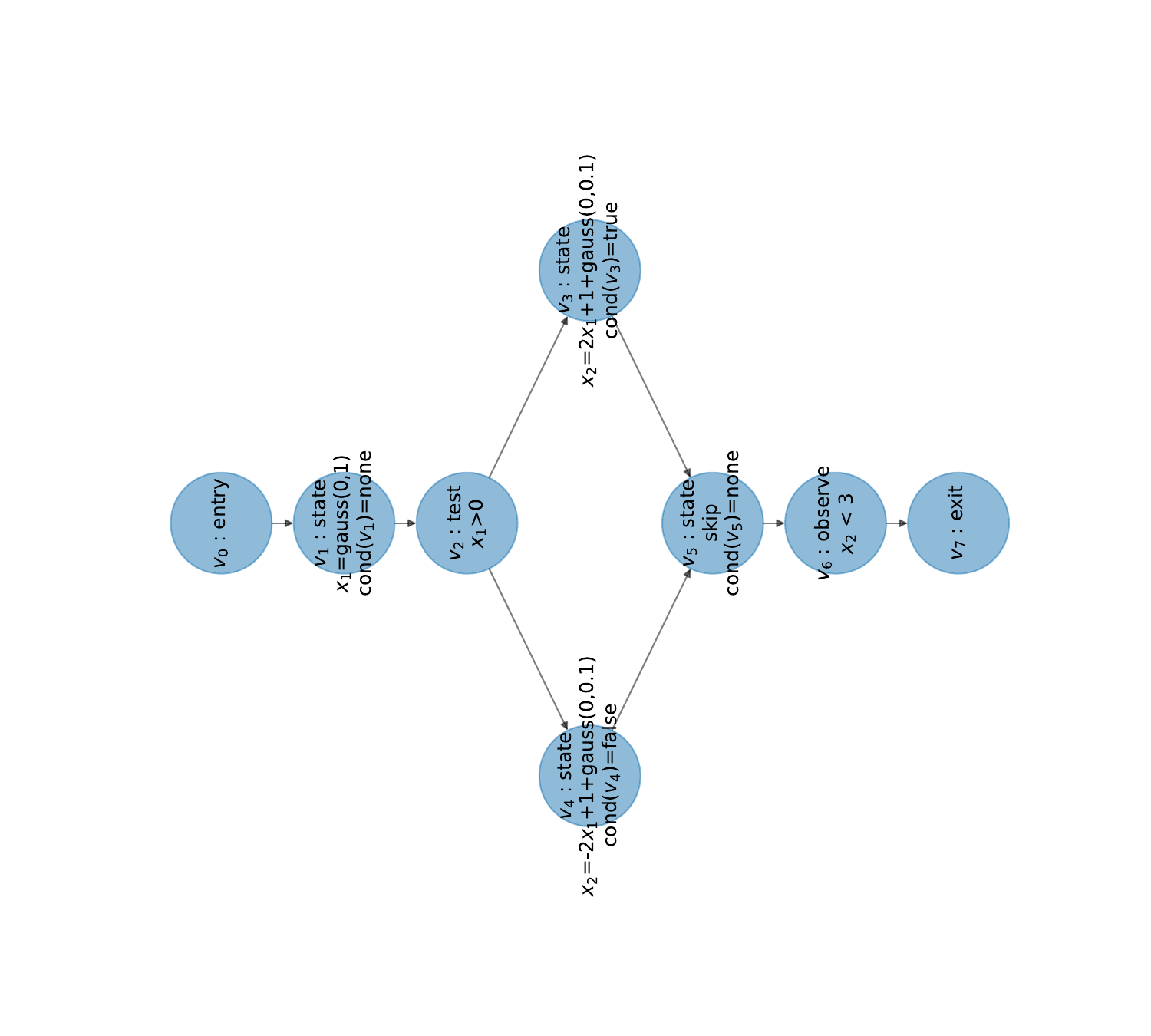}
    \captionof{figure}{Cfg representation of Algorithm \ref{alg:running}}
    \label{fig:cfg}
\end{minipage}

Variables are defined as
$$ \mathtt{z := x \mid g \qquad g:= gm([\pi_1, \hdots, \pi_s], [\mu_1, \hdots, \mu_s], [\sigma_1, \hdots, \sigma_s]) \mid gauss(\mu, \sigma)}$$ 
where $\mathtt{x}$ is an \emph{output variable}, i.e., a variable on which to compute the posterior distribution, and $\mathtt{g}$ denotes 
a fresh \emph{read-only variable} distributed according to univariate GMs with mixing coefficients $\pi_i$, means $\mu_i$ and variances $\sigma_i^2$, $i = 1, \ldots, s$. For the sake of brevity, in our examples, we will also use read-only variables denoted by $\mathtt{gauss(\mu,\sigma)}$ which is syntactic sugar for $\mathtt{gm([1], [\mu], [\sigma])}$.
We use read-only variables to  
perform random assignments, as it is done in lines 2, 4, and 6 of Algorithm \ref{alg:running} and to encode Boolean conditions depending on arbitrary distributions. 

The vector of output variables is denoted by $x = (x_1, \hdots, x_d)$. The vector $z$ augmented with read-only variables is denoted by $z = (x_1, \ldots, x_d, g_1, \ldots, g_{n-d})$.
We denote the distribution of the augmented vector with $D_z$. We assume read-only variables are dropped after the assignment is performed or the condition is evaluated, marginalizing them out. For example, in line 4 of Algorithm \ref{alg:running}, the assignment $x_2 = 2x_1 + 1 + gauss(0,0.1)$ is performed by augmenting the vector $x = (x_1,x_2)$ to $z = (x_1, x_2, g)$, with $g$ being an independent standard Gaussian, and assigning $x_2$ with $2x_1 + 1 + g$. After the new posterior on $z$ is computed, $g$ is marginalized out, returning to the vector $x = (x_1, x_2)$. 

State nodes are labelled by either $\mathtt{skip}$ or assignment instructions of the type $\mathtt{x_i = E(z)}$, where $\mathtt{E(z)}$ is an expression of the following form:
\begin{equation} \label{eq:expr} \mathtt{E(z) := c_1\cdot z_{1} + \ldots + c_n\cdot z_{n} + c \mid z_{i_1}\cdot z_{i_2} }
\end{equation}
where $\mathtt{c, c_1, \ldots, c_n}$ are scalar.

Test and observe nodes are labelled by linear Boolean conditions (LBCs) of the following form:
\begin{equation} \label{eq:LBC}
\mathtt{B(z) := true  \mid false \mid c_1\cdot z_1 + \ldots + c_n\cdot z_n  \, \bowtie \, c \mid z_i  \, \square \, c} 
\end{equation}
 where $\mathtt{c, c_1, \ldots, c_n}$ are scalar constants, $\mathtt{\bowtie} \in \{<, \le, \ge, >\}$ and $\mathtt{\square} \in \{ ==, {!\!=} \}$. We associate an LBC with the set, defined on the space of augmented variables,
\begin{equation} \label{eq:LBCset}
\mathtt{\llbracket B(z) \rrbracket} = \{ z \in \mathbb{R}^n \textit{ s.t. } \mathtt{B}(z) \textit{ holds } \}. 
\end{equation}
where $\mathtt{\llbracket true \rrbracket} = \mathbb{R}^n$ and $\mathtt{\llbracket false \rrbracket} = \emptyset$. Observe that an expression or LBC can have at most $d$ output variables but any finite number of read-only variables. 

\subsection{Supported Programs} 
\label{sec:supported}

Our syntax rules out general distributions depending on non-constant parameters, unbounded loops, and non-polynomial functions. We briefly comment on the limitations of this approach, how they can be mitigated, and when they are shared by other techniques. %{\color{red} \m{possiamo anche rimuovere la frase:} Further comments are provided in Section \ref{sec:conclusion}.}

\paragraph{Probabilistic Assignments.}
Probabilistic assignments are performed by assigning univariate read-only variables to output variables. This is not a limitation since dependence between variables can be encoded using multiple assignments. For what concerns the restriction on GM distributions, instead, we exploit the already discussed density of GMs in the space of distribution \citep{lo1972finite}, and assign a GM arbitrarily close to the target distribution. In this sense, we will assume that we are approximating non-GM distributions with a GM whenever we refer to non-GM distributions. 

Finally, probabilistic assignments will involve only distributions depending on constant parameters. %While not completely general, this approach has also been followed in \citet{holtzen2020scaling}, \citet{albarghouthi2017fairsquare}, \citet{bartocci2020mora}, \citet{moosbrugger2022moment} and \citet{sankaranarayanan2013static}, since it easies the analysis. 
This restriction is more difficult to overcome and is shared with other tools based on moment-based techniques, such as \citet{bartocci2020mora} and \citet{moosbrugger2022moment}. This is because it is not always possible to derive how the moments change if one or more parameters of a distribution are probabilistic. As in \citet{moosbrugger2022moment}, this limitation can be mitigated by performing suitable reparametrizations (see Supplementary Material).

\paragraph{Iterations.}
We restrict our attention to loops bounded by deterministic constants (as in our illustrating example in Algorithm~1), similarly to \citet{gehr2016psi}, \citet{huang2021aqua}, \citet{holtzen2020scaling}, \citet{albarghouthi2017fairsquare} and \citet{nori2014r2}.
If guarantees on almost sure termination can be given, the true distribution of the loop could be approximated by a bounded unrolled loop with a sufficiently large number of iterations~\citep{kozen1979semantics}. 

\paragraph{Polynomial Programs.}
Differently from \citet{gehr2016psi,huang2021aqua,carpenter2017stan}, we consider programs involving only the arithmetic operations $+,-,^*$, and \textasciicircum. This assumption is common to other approaches relying on moment-based techniques such as \citet{bartocci2020mora} and \citet{moosbrugger2022moment}, due to the fact that non-polynomial functions (such as the logarithm) may generate distributions that are not determined by their parameters. We remark that from expressions such as~\eqref{eq:expr} and~\eqref{eq:LBC}, general polynomial expressions and Boolean conditions can be obtained, respectively, by chaining state nodes and nesting conditional statements. %For example, a conditional statement depending on $c + x_1 \cdot x_2 > 0$ can be encoded assigning $y = x_1 \cdot x_2$ and then $y = y + c$. Then, $y>0$ is an LBC of the form \eqref{eq:LBC}. Conjunctions and disjunctions of LBCs can be encoded by suitably nesting test nodes. 
A probabilistic choice, i.e., $x$ is assigned $e_1$ with probability $p$ or $e_2$ with probability $1-p$, is encoded using the LBC $y < q$ where $y$ is a standard Gaussian and $q$ is the Gaussian $p$-quantile.

\subsection{Exact Probabilistic Semantics}
\label{sec:sem}
The ``exact'' semantics follows Kozen's \emph{Semantics 2}~\citep{kozen1979semantics}. 
Since we are using the control-flow syntax of~\citet{cousot1977abstract}, we are close to the \emph{collecting semantics} in~\citet{chaudhuri2011smoothing}: we combine the semantics of the nodes to define the semantics of the paths, then define the semantics of the program as a sum over the semantics of the paths.  This semantics is particularly convenient for our method because it gives the posterior distribution as a mixture, similarly to \citet{zhou2020divide}. %In~\citep{chaudhuri2011smoothing} this was done by collecting the guarded linear expressions encountered during the execution of a path. Here, we collect pairs $(p, D)$, in which $p$ is the probability of executing a given path and $D$ is the output distribution of the path.

Given a program $P=(V,E)$, we define a path $\pi$ as an ordered sequence of nodes $\pi = v_0 \cdots v_n$ with $v_0 \colon entry$, $v_n \colon exit$ and $(v_{i-1}, v_i) \in E, \, \forall \, i=1, \ldots n$. The successor of node $v_i$ in path $\pi$ is denoted as $s_{\pi}(v_{i}) = v_{i+1}$. The set of all paths of $P$ is denoted by $\Pi^P$.  We define the semantics of a path $\pi$, denoted by $\llbracket \pi \rrbracket$, as a pair $(p, D)$, where $p \ge 0$ and $D$ is a probability distribution on $\mathbb{R}^d$. 
The semantics $\llbracket \pi \rrbracket$ composes the semantics of the nodes along path $\pi$, i.e.,
$\llbracket \pi \rrbracket = \llbracket v_n \rrbracket_\pi \circ \ldots \circ \llbracket v_0 \rrbracket_\pi. $
The semantics of each node is defined as follows. 
\begin{itemize}
    \item The \textit{entry} node outputs the pair $(1,\delta_{0})$, $\delta_0$ being a Dirac delta centered on the zero vector: 
    $$\text{ if } v \colon entry \text{ then } \llbracket v \rrbracket_\pi = (1, \delta_0).$$
    
    \item A \textit{state} node $v$ takes as input a pair $(p,D)$ and returns a pair $(p, D')$ depending on its label. If it is labelled by $\mathtt{skip}$, it returns $(p,D)$. If it is labelled by $\mathtt{x_i = E(z)}$, it returns $(p,D')$ with $D'$ the distribution of the vector $x[x_i=E(z)]$: 
    $$ \text{ if } v \colon state \text{ then } \llbracket v \rrbracket_\pi (p, D) = \begin{cases} (p, D) & \text{ if } v \text{ is labelled by } \mathtt{skip} \\
    (p, D') & \text{ if } v \text{ is labelled by } \mathtt{x_i = E(z)}. \end{cases} $$
    
    \item A \textit{test} node $v$ labelled by $\mathtt{B(z)}$ takes as input a pair $(p,D)$ returns $(p', D')$ depending on the value of $cond(s_{\pi}(v))$. In particular, first, the augmented vector $z$ and its distribution $D_z$ are considered. If $cond(s_{\pi}(v)) = \mathit{true}$, then the node computes the probability of the Boolean condition evaluating to true, i.e. $P_{D_z}(\mathtt{\llbracket B(z)\rrbracket})$. Then, it conditions the current distribution to such event, i.e. $D_z \,|\, \mathtt{\llbracket B(z)\rrbracket}$. The result is the output pair $( p \cdot P_{D_z}(\mathtt{\llbracket B(z)\rrbracket}), \mathit{Marg}_x(D_z \,|\, \mathtt{\llbracket B(z)\rrbracket}))$. Similarly, if $cond(s_{\pi}(v))=false$ the output is $(p \cdot P_{D_z}(\mathtt{\llbracket \neg B(z)\rrbracket}), \mathit{Marg}_x(D_z \,|\, \mathtt{\llbracket \neg B(z)\rrbracket}))$. To overcome conditioning with respect to zero-probability events we assume that whenever $P_{D_z}(\mathtt{\llbracket B(z)\rrbracket}) = 0$ (resp. $P_{D_z}(\mathtt{\llbracket \neg B(z)\rrbracket}) = 0$) the output pair is $(0, D)$: %Observe that this does affect the output distribution of the program.
    $$ \text{ if } v\colon test \text{ then } \llbracket v \rrbracket_\pi(p,D) = \begin{cases} ( p \cdot P_{D_z}(\mathtt{\llbracket B(z)\rrbracket}), \mathit{Marg}_x(D_z \,|\, \mathtt{\llbracket B(z)\rrbracket})) \\ 
    \hspace{2.5cm} cond(s_{\pi}(v)) = \mathit{true} \mathop{\land} P_{D_z}(\mathtt{\llbracket B(z)\rrbracket}) \neq 0 \\ (p \cdot P_{D_z}(\mathtt{\llbracket \neg B(z)\rrbracket}), \mathit{Marg}_x(D_z \,|\, \mathtt{\llbracket \neg B(z)\rrbracket})) \\
    \hspace{2.4cm} cond(s_{\pi}(v)) = \mathit{false} \mathop{\land} P_{D_z}(\mathtt{\llbracket \neg B(z)\rrbracket}) \neq 0 \\ (0, D) \hspace{1.6cm} \textit{else}. \end{cases}$$

    \item 
    For an \textit{observe} node $v$ labelled by $\mathtt{B(z)}$ we condition the current distribution to $\mathtt{\llbracket B(z)\rrbracket}$. Observe that if $\mathtt{B(z)}$ only contains read-only variables, conditioning does not affect the distribution of the output variables $x$. If $B(z) = \mathtt{ x_i == c}$ the node returns a distribution $D'$ having pdf $\frac{1}{I}f_D(x,x_i=c)\delta_c(x_i)$ with $I = \int_{\mathbb{R}^{d-1}} f_D(x,x_i=c) d(x\setminus x_i)$. In all other cases conditioning is treated as usual:
    $$ \text{ if } v\colon observe \text{ then } \llbracket v \rrbracket_\pi(p,D) = \begin{cases} ( p \cdot I, D') \hspace{1.15cm} B(z) = \mathtt{x_i == c}\\
    (p \cdot P_{D_z}(\llbracket \mathtt{B(z)} \rrbracket), D \,|\, \mathtt{\llbracket B(z)\rrbracket}) \\
    \hspace{2.5cm} B(z) \neq \mathtt{x_i == c} \mathop{\land} P_{D_z}(\mathtt{\llbracket B(z)\rrbracket}) > 0\\
    (0, D) \hspace{1.6cm} \textit{ else} \end{cases}$$
    \item The \textit{exit} node $v$ takes as input $(p,D)$ and outputs the same pair $(p, D)$:
     $$ \text{ if } v \colon exit \text{ then } \llbracket v \rrbracket_\pi(p, D) = (p, D).$$
\end{itemize}

The semantics of the program $P$ is then defined as:
\begin{equation} \label{eq:sem} \llbracket P \rrbracket= \sum_{\substack{(p,D) = \llbracket \pi \rrbracket \\ \pi \in \Pi^P}} \frac{p}{\displaystyle{\sum_{\substack{(p',D') = \llbracket \pi \rrbracket\\ \pi \in \Pi^P}} p'}}D. \end{equation}

\begin{example}
    For the program in Algorithm \ref{alg:running} we have only two paths, $\pi_T = v_0v_1v_2v_3v_5v_6v_7$ and $\pi_F = v_0v_1v_2v_4v_5v_6v_7$ corresponding to evaluations of the conditional statement as true or false, respectively. To compute the semantics of $\pi_T$ we start from $v_0$, that outputs $(1, \delta_0)$. This pair is taken as input by $v_1$, which is a state node assigning $gauss(1,0)$ to $x_1$, so its output is $(1, \mathcal{N}(0,\Sigma_1))$ with $\Sigma_1 = \mathit{diag}(0,1)$ (corresponding to the distribution in Figure 2b). This pair is taken as input by the test node $v_2$. Since we are considering $\pi_T$, for which $s_{\pi_T}(v_2) = v_3$, and $cond(v_3) = true$, the semantics of $v_2$ in this path conditions $\mathcal{N}(0, \Sigma_1)$ to $x_1>0$. Therefore in this path the output of $v_2$ is $(0.5, \mathcal{N}(0, \Sigma_1) | \llbracket x_1 > 0\rrbracket)$, where $0.5 = P_{\mathcal{N}(0, \Sigma_1)}(\llbracket x_1 > 0\rrbracket)$. This pair is taken as input by $v_3$, which updates the distribution of $x_2$ and therefore outputs a new pair $(0.5, D_3)$. We can proceed until we compute the output of $v_7$, which gives the final pair $\llbracket \pi_T \rrbracket = (p_T, D_T)$. In the same way, we compute $\llbracket \pi_F \rrbracket = (p_F, D_F)$, and finally, the semantics of the whole program as the mixture $(p_TD_T + p_FD_F)/({p_T + p_F})$.
\end{example}

\section{Gaussian Semantics}
\label{sec:gaussian}

Gaussian Semantics is a  family of semantics closed with respect to GMs. Each node takes as input and returns a GM over the program variables. This is done by composing the exact semantics of a node with an operator $T^{GM}_r$ acting on the output distribution of $\llbracket v \rrbracket_\pi$. In particular, $T^{GM}_r$ transforms any distribution $D$ into a GM $G$, having the same moments of $D$ up to order $r$. Therefore, we call $T^{GM}_r$ the \emph{moment-matching operator}. Formally, we use a map $R:V \to \mathbb{N}_0$ to associate each node $v \in V$ with the highest order of moments that will be matched at $v$. The semantics of a node is then:
\begin{equation} \label{eq:gsem} 
\llbracket v \rrbracket^{R}_\pi = \begin{cases} \llbracket v \rrbracket_\pi & \text{ if } v\colon entry, exit \\
 \left(\mathbb{I}, T^{GM}_{R(v)}\right) \circ \llbracket v \rrbracket_\pi & \text{ if } v \colon state, test, observe \end{cases}
\end{equation}
where $\mathbb{I}$ is the identity acting on the first element of the pair $(p,D)$, and $T^{GM}_{R(v)}$ is the moment-matching operator. The Gaussian Semantics of paths and programs are defined similarly to the exact semantics as:
$$ \llbracket \pi \rrbracket^{R} = \llbracket v_n \rrbracket^{R}_\pi \circ \ldots \circ \llbracket v_0 \rrbracket^{R}_\pi \qquad \text{and} \qquad \llbracket P \rrbracket^{R}(D_0) = \sum_{\substack{(p,D) = \llbracket \pi \rrbracket^R \\ \pi \in \Pi^P}}  \frac{p}{\displaystyle{\sum_{\substack{(p',D') = \llbracket \pi \rrbracket^R\\ \pi \in \Pi^P}} p'}}D. $$

\begin{example}
We have seen in Algorithm \ref{alg:running} that the exact semantics is not closed with respect to GMs. For example, node $v_2$ takes as input a Gaussian but returns a truncated Gaussian, which is not a GM. If, instead, we consider Gaussian Semantics with $R(v_2)=3$ the output of $v_2$ will be a GM matching the first three order moments of $\mathcal{N}(0, \Sigma_1) \, | \, \llbracket x > 0 \rrbracket$. Two steps are required to compute $\llbracket v_2 \rrbracket^R_\pi$. In the first step, we compute the first $R(v_2) = 3$ order moments of the output distribution of $\llbracket v_2 \rrbracket_\pi$. Therefore, we compute the first order moments $\mathbb{E}[x_1] = 0.7979$ and $\mathbb{E}[x_2] = 0$, the second order moments $\mathbb{E}[x_1^2] = 1, \mathbb{E}[x_1x_2] = \mathbb{E}[x_2^2] = 0$ and the third order moments $\mathbb{E}[x_1^3] = 1.5958, \mathbb{E}[x_1^2x_2] = \mathbb{E}[x_1x_2^2] = \mathbb{E}[x_2^3] = 0$. Observe that, thanks to the results in Table~\ref{tab:tab1}, this is significantly easier than computing the whole pdf of the output distribution. The second step involves finding a GM having the computed moments. This is generally more complex and is performed by the operator $T^{GM}_r$. In the rest of the section, we will assume that the computation of the moments is done using the aforementioned formulas, and we focus on the definition of the operator $T^{GM}_r$ and the derivation of its properties.
\end{example}

\paragraph{Moment-Matching Operator}  In general, the operator $T^{GM}_r(D)$ acts on distributions $D = \sum_{i=1}^C p_iD_i$ that are mixtures (possibly of a single component). The moments of $D$ are computed as a linear combination of the moments of its components: if $x \sim D = \sum_{i=1}^C p_iD_i$ and $x_i \sim D_i$, then $\mathbb{E}[x^n] = \sum_{i=1}^C p_i\mathbb{E}[x_i^n]$. Therefore, when computing the moments of $D$, we first compute the moments of every component $D_i$. Then, it makes sense to define $T^{GM}_r(D)$ so that when it acts on a mixture $D$ ($C>1$), it recursively acts on each component of the mixture, moment-matching each of them. %This is more accurate \m{possiamo provarlo? (non dico di farlo ora, magari semplicemente non lo diciamo?} than just moment-matching the whole mixture in the sense that more information about the initial distribution is preserved by $T^{GM}_r(D)$. As noticed, doing this does not require extra computation \m{wrt what?}. 
When, instead, $T^{GM}_r$ acts on a non-mixture distribution $D$ ($C=1$), it returns a GM having moments up to order $r$ equal to those of $D$. This second action is encoded by a second operator $\textbf{match}_r$. \begin{equation} \label{eq:mmo}
T^{GM}_r(D) = \begin{cases}
p_1T^{GM}_r(D_1) + \ldots +  p_CT^{GM}_r(D_C) & \text{if~} D = p_1D_1 + \ldots + p_CD_C \text{~and~} C > 1 \\
\textbf{match}_r(D) & \text{otherwise.}
\end{cases}
\end{equation}

We require that $\textbf{match}_r(D)$ satisfies the following two conditions:
\begin{itemize}
    \item[R1)] for any distribution $D$, $\textbf{match}_r(D)$ is a GM;
    \item[R2)] $\textbf{match}_r(D)$ has central moments up to order $r$ equal to those of $D$.
\end{itemize}
The existence of the operator $\textbf{match}_r$ is guaranteed by the following result, derived from~\citet[Theorem 17.2]{schmudgen2017moment}, stating that that, for any finite sequence of moments, there exists a moment-matching discrete distribution putting positive mass on a number of points smaller than or equal to the number of matched moments. Since discrete distributions are GMs, the Proposition holds. For detailed proof, see the Supplementary Material.

\begin{proposition}
\label{prop:exist}
For any $r \in \mathbb{N}_0$, there exists an operator $\textbf{match}_r$ satisfying R1 and R2.
\end{proposition}

\begin{example}
In our example, we want to match a total of $10$ moments, a zeroth-order moment (which is always 1), two first-order, three second-order, and four third-order moments. Theorem 17.2 in \citet{schmudgen2017moment} ensures that there exists at least one discrete distribution (therefore a GM) with $C \le 10$ components that has exactly the given moments. %\m{se serve spazio questo esempio si può rimuovere}
\end{example}

Proposition \ref{prop:exist} ensures the existence of at least one GM matching the moments of $D$ up to order $r$. In general, letting $GM_r(D)$ denote the set of all finite GMs matching the moments of $D$ up to order $r$, we may have that $|GM_r(D)| > 1$. For $\textbf{match}_r$ to be well-defined, we need to uniquely identify a moment-matching GM in $GM_r(D)$. This can be done using different heuristics: we propose one based on the principle of maximum entropy, which we call the max entropy matching (MEM).

\paragraph{Max Entropy Matching}

MEM can be summed up as follows: if $|GM_r(D)| > 1$ we choose the GM $G$ having the least number of components (in order to minimize the number of parameters to be fit) and minimizing a certain cost function. Any remaining tie is resolved by comparing the vectors of parameters $P$ that identify the GMs with respect to lexicographic ordering (we give an ordering on the parameters of GMs in the Supplementary Material). We select our cost function as the sum of the opposite of the differential entropy plus a penalty term, where the differential entropy for a distribution $D$ with pdf $f_D$ is defined as~\citep{cover1999elements}
\begin{equation} \label{eq:entropy}
    H(D) = -\int_{\mathbb{R}^d} f_D(x) \log(f_D(x)) \, dx.
\end{equation}
Intuitively, the \emph{principle of maximum entropy} asserts that the distribution maximizing entropy is the one that minimizes the number of assumptions on the distribution~\citep{cover1999elements}. Therefore, maximizing $H(D)$ we are choosing the most general moment-matching distribution. We add to $H(D)$ a penalty term to avoid uncontrolled growth of the parameter values.

Then, the procedure to compute $\textbf{match}_r(D)$ is the following.
\begin{itemize}
    \item[1)] Find $ c^* = \min\left\{ c : \exists \, G=\sum_{s=1}^c p_s \mathcal{N}(\mu_s, \Sigma_s) \in GM_r(D)\right\}.$
    \item[2)] Find the set $\mathcal{P}$ such that $P = (p_1, \hdots, p_{c^*}, \mu_1, \ldots, \mu_{c^*}, \Sigma_1, \ldots, \Sigma_{c^*}) \in \mathcal{P}$ if and only if the GM with parameters $P$ matches the moments of $D$ up to the $r$-th order.
    \item[3)] Find the set $\mathcal{P}^*$ given by: 
    \begin{equation} \label{eq:normcost} 
    \mathcal{P}^* = \argmin_{\mathcal{P}} \left\{ - H\left(\sum_{i=1}^{c^*} p_i\mathcal{N}(\mu_i, \Sigma_i) \right) + \sum_{i=1}^{c^*} \left( \| \mu_i \|^2_2 + \|\Sigma_i\|^2_2 \right)\right\}. \end{equation}
    \item[4)] If $|\mathcal{P}^*| > 1$ choose $P^* \in \mathcal{P}^*$ maximum with respect to lexicographic ordering.
\end{itemize}

The following proposition guarantees that MEM leaves us with a well-defined operator $\textbf{match}_r$. It is again proved using Theorem 17.2 from~\citet{schmudgen2017moment}, and by noticing that $\mathcal{P}$ is a compact set, therefore Eq.~\eqref{eq:normcost} is well-defined. Again, we defer a detailed proof to the Supplementary Material while we explain how MEM works using an example.

\begin{proposition} \label{prop:welldef}
For any $D$ and $r$ the max entropy matching uniquely identifies $\textbf{match}_r(D)$.
\end{proposition}

We remark that the choice of MEM is arbitrary, as other cost functions could be introduced. However it has various benefits. \emph{(i)} To guarantee that Proposition~\ref{prop:welldef} holds, one needs a bounded cost function. \emph{(ii)} Using entropy leads to a parallelism with VI: SOGA itself can be seen as a form of VI since it involves the minimization of the reverse differential entropy~\citep{kullback1951information}. However, correspondence with VI is lost when higher-order moments are considered, because the minimizer of the reverse differential entropy is not analytically expressible for GMs. \emph{(iii)} In the spirit of minimizing the number of assumptions made on the approximating distribution, the approach looks more pleasing mathematically.

\begin{example}
    While \citet{schmudgen2017moment} ensures that we can find a moment-matching GM with 10 components, it is easy to check that $c^* = 2$ is the minimum number of components required to match three order moments. In fact, $c^* > 1$, since for a single Gaussian, given the mean and the covariance matrix, all the other moments are fixed (so, we can match the first two order moments but not the third). For $c = 2$ instead we can consider the GM $G$ with pdf $p\mathcal{N}(m, S) + (1-p)\mathcal{N}(m', S')$ such that $m, m', S, S'$ satisfy the following system:
    \begin{align*}
    & pm_i + (1-p)m_i' = \mathbb{E}[x_i] & \text{ for } i=1,2\\
    & p(m_i^2 + S_{i,i}) + (1-p)(m_i'^2 + S_{i,i}') = \mathbb{E}[x_i^2] & \text{ for } i=1,2 \\
    & p(m_1m_2 + S_{1,2}) + (1-p)(m_1'm_2' + S_{1,2}') = \mathbb{E}[x_1x_2] & \\
    & p(m_i^3 + 3m_iS_{i,i}) + (1-p)(m_i'^3 + S_{i,i}') = \mathbb{E}[x_i^3] & \text{ for } i = 1,2 \\
    %& p(m_iS_{j,j} + 2m_jS_{i,j} + m_im_j^2) + (1-p)(m_i'S_{j,j}' + 2m_j'S_{j,j}' + m_i'm_j'^2) = \mathbb{E}[x_ix_j^2] & \text{for } i,j = 1,2, i \neq j \\
    & p(m_1S_{2,2} + 2m_2S_{1,2} + m_1m_2^2) + (1-p)(m_1'S_{2,2}' + 2m_2'S_{1,2}' + m_1'm_2'^2) = \mathbb{E}[x_1x_2^2] & \\
    & p(2m_1S_{1,2} + m_2S_{1,1} + m_1^2m_2) + (1-p)(2m_1'S_{1,2}' + m_2'S_{1,1}' + m_1'^2m_2') = \mathbb{E}[x_1^2x_2] & \\
    & 0 < p < 1, \quad S_{1,1}S_{2,2} - S_{1,2}^2 \ge 0, \quad S_{1,1}'S_{2,2}' - S_{1,2}'^2 \ge 0 & 
    \end{align*}

In the system we equate the moments of $G$ (l.h.s) with those of $x\sim D$ (r.h.s, computed in Example 4.1). Moreover, we look for solutions such that $0 < p < 1$ and $S, S'$ are positive semidefinite (last line). Since the system is polynomial, using SMT solvers over reals we can check it is satisfiable; therefore, $c^* = 2$. Now we should determine the set $\mathcal{P}$ of all solutions and find those that minimize the cost function. Since finding all solutions is generally impossible, we directly proceed to optimize our cost function numerically, constraining the variables to satisfy the previous system. We find the approximate solution $p=0.572$, $m=(0.471,0)$, $m'=(1.236, 0)$ and $S=\mathit{diag}(0.066,0)$, $S'=\mathit{diag}(0.426,0)$. The approximating GM is shown in the green line of Figure \ref{fig:approx}, while the blue line shows the true non-Gaussian distribution.
\end{example}

The example shows that the difficult step in computing a Gaussian Semantics of arbitrary order is 2). Finding the parameters of a moment-matching GM requires the solution of a system of polynomial equations, like the one in the example. This problem is notoriously hard to solve, as no analytical solution exists \citep{lasserre2009moments}. Performing numerical optimization can solve the problem approximately, but is in general numerically unstable and requires relatively long computational times (in our example, using scipy \citep{2020SciPy-NMeth}, it took around 7\,s to match a single Gaussian!). While we leave open the problem of solving 2) efficiently in the general case, the following lemma gives two important properties of $\textbf{match}_r$, which will be used to derive our second-order approximation. The proof is quite trivial and reported in the Supplementary Material.

\begin{lemma}
\label{lem:order2}
The following two properties hold:
\begin{itemize}
    \item[i)] when $r=2$, $\textbf{match}_2(D)$ is a single Gaussian distribution with mean and covariance matrix equal to those of $D$;
    \item[ii)] if $D$ is Gaussian, for any $r \ge 2$ $\textbf{match}_r(D) = D.$
\end{itemize}
\end{lemma}

%One may wonder why in our semantics we have not directly applied $\textbf{match}_r$ to the mixtures, preferring to apply it to each component separately. The answer lies in the fact that, by choosing the second procedure, 

We conclude the section with a consequence of Lemma \ref{lem:order2}. It follows from ii) that $T^{GM}_r$ has the desirable property of leaving GMs unaltered, i.e. if $M$ is a  GM $T^{GM}_r(M) = M$ for all $r \ge 2$. %On the contrary, by Lemma \ref{lem:order2}, if $M$ has $C>1$ components $\textbf{match}_2(M)$ would consist in a single Gaussian distribution with same mean and covariance matrix as $M$. In general both $\textbf{match}_r(M)$ and $T^{GM}_r(M)$ preserve the first $r$-th order moments of $M$, but in addition $T^{GM}_r(M)$ preserve also the moments of the components. Observe that no additional computations are needed to compute $T^{GM}_r(M)$ with respect to $\textbf{match}_r(M)$, as the central moments of $M$ are given by the combinations of the central moments of its components.
As a consequence, Gaussian Semantics coincides with the exact semantics for programs only involving GMs, and in particular, for programs involving only discrete distributions (mixtures of deltas).

\begin{proposition} \label{prop:exact1}
Let $P=(V,E)$ be such that every read-only variable in the program is a finite discrete distribution. Then, for any $R$, $\llbracket P \rrbracket^R = \llbracket P \rrbracket$. 
\end{proposition}

\proof
Since truncations, linear combination and products of discrete distributions are discrete distributions, only discrete distributions are generated in the execution of $\llbracket P \rrbracket$. By Lemma \ref{lem:order2} applying the moment-matching operator $T^{GM}_r$ to them leaves them unaltered, so conclusion follows.
\endproof

\begin{example}
    Consider a second Gaussian semantics that maps $v_2$ to $R(v_2) = 2$. In this case, we want to match only the first two order moments, namely $\mathbb{E}[x_1], \mathbb{E}[x_2], \mathbb{E}[x_1^2], \mathbb{E}[x_2^2], \mathbb{E}[x_1x_2]$. As noticed before, in this case $c^*=1$, since we can take the Gaussian $\mathcal{N}(\mu, \Sigma)$ with $\mu = (0.7979, 0)$ and $\Sigma = diag(1,0)$ and it will have required moments. Observe that we do not need to solve any system or optimization problem. 

\begin{figure}
\centering
    \includegraphics[width=0.9\linewidth]{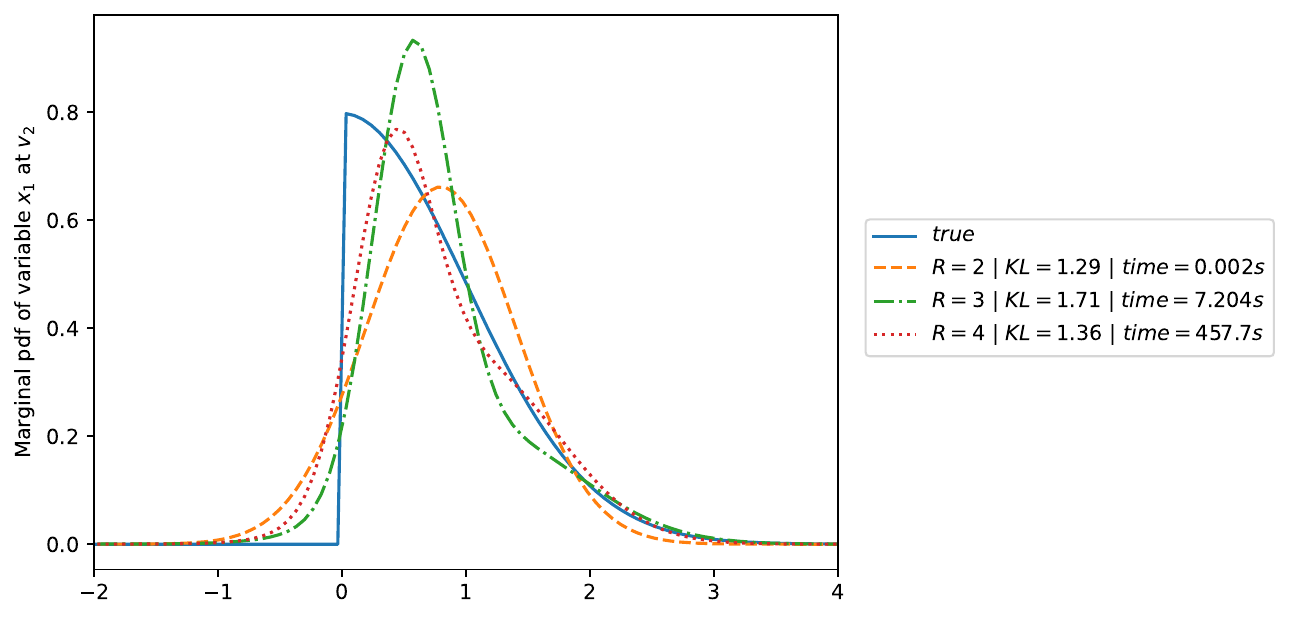}
    \captionof{figure}{Marginal pdf of $x_1$ at node $v_2$ in Algorithm~\ref{alg:running} given by exact semantics ($true$) and the Gaussian Semantics with $R=2, 3$ and $4$. In the legend we report the KL divergence with the true distribution and the time need to compute the approximating GM. }
    \label{fig:approx}
\end{figure}

%\begin{tikzpicture}
%    \node[shape=circle,draw=black] (A) at (0,0) {$v_0 \colon entry$};
%    \node[shape=circle,draw=black] (B) at (3,0) {$v_1 \colon state$ \\ $x_1 = gauss(0,1)$ \\ $cond(v_1) = none$};
%    \node[shape=circle,draw=black] (C) at (6,0) {$v_2 \colon test$ \\ $ x_1 > 0$};
%    \node[shape=circle,draw=black] (D) at (9,1) {$v_3: state$ \\ $x_2 = 2x_1 + 1 + gauss(0,0.1)$ \\ $cond(v_3) = true$};
%    \node[shape=circle,draw=black] (E) at (9,-1) {$v_4: state$ \\ $x_2 = -2x_1 + 1 + gauss(0,0.1)$ \\ $cond(v_4) = false$};
%    \node[shape=circle,draw=black] (F) at (12,0) {$v_5: state$ \\ $skip$ \\ $cond(v_5) = none$} ;
%    \node[shape=circle,draw=black] (G) at (15,0) {$v_6: observe$ \\ $x_2 < 3$} ;
%    \node[shape=circle,draw=black] (H) at (18,0) {$v_7: exit$} ;
%
%    \path [->](A) edge node[left] {} (B);
%    \path [->](B) edge node[left] {} (C);
%    \path [->](C) edge node[left] {} (D);
%    \path [->](C) edge node[left] {} (E);
%    \path [->](E) edge node[left] {} (F);
%    \path [->](D) edge node[left] {} (F);
%    \path [->](F) edge node[left] {} (G);
%    \path [->](G) edge node[left] {} (H);  
%\end{tikzpicture}

    In general, for a fixed number of moments matched, we expect Gaussian Semantics to approximate reasonably well the matched moments but not necessarily the whole distribution (see Section~\ref{sec:map} for further discussion).
    Indeed, let us compare the exact posterior distribution with the ones obtained distribution when $R = 2,3,4$,  respectively, using Kullback-Leibler (KL) divergence \citep{kullback1951information}. The KL divergence between two distributions $P$ and $Q$ is a standard way to evaluate the error committed in approximating $P$ with $Q$. In our case, we take as $P$ the truncated Gaussian and as $Q$ the GMs obtained matching different order moments. The respective values are 2.29 (R=2), 1.71 (R=3) and 1.36 (R=4),  so indeed higher-order Gaussian semantics improve the approximation. Figure~\ref{fig:approx} compares the true marginal pdf of $x_1$ at node $v_2$ (blue solid line) with the second- (orange dashed), third- (green dash-dotted) and fourth (red dotted) approximations. The advantages of fitting only a small number of moments are mainly computational. Indeed, it can be seen from the legend that as the number of moments matched grows, the increased KL accuracy comes with an increased computational cost.
\end{example}

%{\color{red}
%\paragraph{Remark}\m{Still needed after refactoring?}
%    In the following section we will give a convergence result that holds asymptotically as the number of the moments matched grows to infinity. However, it will not provide any information on a fixed finite number of moments matched. . This will be true in particular for SOGA, a particular instantiation of Gaussian Semantics, in which only the first two orders of moments are fitted. 

\section{Universal Approximation Theorem}
\label{sec:convergence}

Our main convergence result states that, for well-behaved programs, it is possible to find a map $R$ so that the output distribution yielded by the semantics $\llbracket \cdot \rrbracket^{R}$ is arbitrarily close to that of $\llbracket \cdot \rrbracket$ in the Levy-Prokhorov metric \citep{ethier2009markov}. By ``well-behaved'' we mean that the distributions in the exact semantics  are determined by their moments and that they can measure continuously sets in the form \eqref{eq:LBCset}. To formalize the latter requirement we introduce $m$-continuity sets. Since both this definition and that of Levy-Prokhorov metric are borrowed from measure theory we refer the reader to the Supplementary Material for a more detailed background on these concepts.

\begin{definition}Given a measure $m$, a set $A \subseteq \mathbb{R}^d$ is called an $m$-continuity set if $m(\partial A) = 0$ where, $\partial A$ is the boundary of set $A$, defined as the closure of the set $A$ minus its interior. 
\end{definition} 

Then, we can state our main theorem.
\begin{theorem} \label{thm:conv}
Assume that $P=(V,E)$ is a program such that for each $v \in V$ and each path $\pi \in \Pi^P$ the output distribution $D$ of $\llbracket v \rrbracket_\pi$ satisfies the following:
\begin{itemize}
    \item[H1)] $D$ is determined by its moments;
    \item[H2)] if $D$ is the input distribution for a test or observe node $v'$, then the set defined by the LBC labelling $v'$ is an $m_{D_z}$-continuity set.
\end{itemize}

Then there exists a sequence of maps $(R^k: V \to \mathbb{N}_0)_{k \in \mathbb{N}}$ such that:
\begin{equation} \label{eq:conv}
\llbracket P \rrbracket^{R^k} \xrightarrow{k \to \infty}  \llbracket P \rrbracket. \end{equation}
where the convergence is intended in the weak topology, or equivalently, in the Levy-Prokhorov metric.
\end{theorem}

\subsection{Satisfaction of the Hypotheses}
Before giving an outline of the proof, we briefly comment on the hypotheses. 

First, observe that H1 and H2 are sufficient but not necessary. In particular, if the hypotheses of Proposition \ref{prop:exact1} are satisfied convergence holds trivially, even when H1 or H2 are violated.

Hypothesis H1 is common to other works considering moment-based approximation, such as in \citet{bartocci2020mora} and \citet{moosbrugger2022moment} and is needed to guarantee that the method of moments converges to the true distribution~\citep{billingsley2013convergence}. For a given program, it is possible to perform static analysis to check whether the arising distributions are determined by their moments, exploiting known results on moment determinacy~(see, for example, the moment-generating function characterization in \citep{billingsley2013convergence}). Notably, to apply these results is not necessary to compute the exact pdf of the arising distributions, but it is sufficient to keep track of their type. In fact, for some classes of distributions, moment-determinacy is established: this is true for finite discrete distributions, Gaussians, uniforms, Poissons, exponentials, truncations and mixtures thereof~\citep{billingsley2013convergence}. The case studies analyzed in this paper feature such distributions. On the contrary, log-normal distributions are not determined by their moments. However, as long as moments are computable, Gaussian Semantics can still be applied: in this case no formal guarantee of convergence towards the true distribution is given, but the method still provides an analytical approximation for the moments. 

Hypothesis H2 guarantees that when distributions are conditioned to sets in the form \eqref{eq:LBCset}, weak convergence is preserved. This requirement can be falsified if $D$ has degenerate components that place positive probability mass on the boundary of the set defined by the LBC. This could happen, for instance, if a component is a Dirac measure centered on any point of $\partial \mathtt{\llbracket B(x) \rrbracket}$.  
For example, consider line 12 of $\mathit{Tracking}$\_$n$ in Section~\ref{sec:tracking}, where $out$ can be 1 or 0 with probability $> 0$. This falsifies H2. However, such cases can be statically detected and the program can be transformed into one that uses the equivalent condition as $out > 0.5$, so that H2 holds. More in general, continuity corrections such as those performed in \citet{laurel2020continualization} can be adopted.

\begin{example}
    Algorithm \ref{alg:running} satisfies H1 since the joint at each location is either a mixture of Gaussians or truncated Gaussians, for which moment-determinacy is known. Moreover, the two LBCs checked in the program are $x_1 > 0$ (line 3) and $x_2 < 3$ (line 9). Before checking $x_1 > 0$, $x_1$ has non-degenerate Gaussian distribution, and therefore the set $x_1 = 0$ (border of $\llbracket x_1 > 0 \rrbracket$) has measure 0. Similarly, before line 9, $x_2$ is Gaussian-distributed with $\sigma > 0$, therefore $x_2 = 3$ (border of $\llbracket x_2 < 3 \rrbracket$) has again probability 0. We conclude that also H2 is verified.

    For $\mathit{Tracking}$\_$n$ in Section~\ref{sec:tracking}, each marginal is obtained by Gaussians, performing sums, squares, or conditioning. Here, to check moment-determinacy, we use a result in \citet{billingsley2008probability}, which states that if the moment-generating function (mgf) of a distribution is defined in an interval of $0$, then the distribution is determined by its moments. Using symbolic integration, we can compute the mgf of the product of two Gaussians and verify that it is defined in an interval of 0. Therefore, the distribution is determined by its moments and H1 holds. For H2, we have already shown how to correct the condition in line 12 so that H2 is verified. For the if statement in line 7, observe that before entering it, the marginal w.r.t. to $dist$ is the distribution of $x^2 + y^2$, which is continuous, and therefore the point $10$ has measure 0 with respect to it. Therefore also H2 holds.
\end{example}

\subsection{Outline of the Proof}

The proof of Theorem \ref{thm:conv} is rather technical and involves a number of results from measure theory. Thus, here we provide a sketch, and refer the reader to Supplementary Material for a detailed proof.

First of all, recall that the semantics of a program is defined as a (finite) mixture of the semantics of the paths in the program. Therefore, \eqref{eq:conv} holds if it holds for every path $\pi \in \Pi^P$. Moreover, it can be shown that we can consider only the paths such that $\llbracket \pi \rrbracket = (p, D)$ with $p \neq 0$, since paths for which $p = 0$ contribute to the semantics of the program nor in the exact, neither in the Gaussian Semantics. So, the proof amounts to showing that for every path $\pi \in \Pi^P$ such that $\llbracket \pi \rrbracket = (p, D)$ with $p \neq 0$ we can choose a sequence of maps $R^k$ such that $\llbracket \pi \rrbracket^{R^k} \to \llbracket \pi \rrbracket.$

We can build the sequence of maps $R^k$ by specifying $R^k(v)$ for each $v \in \pi$. In particular, for $v_0$ we can choose any value of $R^k$. For $v_1$, if it is not the exit node, we can assume that $\llbracket v_1 \rrbracket$ transforms the pair $(1, \delta_0)$ into $(p_1, D_1)$ with $p_1 \neq 0$. Then, by definition of Gaussian Semantics, $\llbracket v_1 \rrbracket^{R^k}$ transforms $(1, \delta_0)$ into the pair $(p_1, T^{GM}_{R^k(v_1)}(D_1))$. To ensure convergence in this case, we use Theorem \ref{thm:bill} of \citet{billingsley2008probability}, which we state below.

\begin{theorem}[\citet{billingsley2008probability}] \label{thm:bill}
Suppose $X \sim D$, $X_n \sim D_n$, and $D$ is a distribution determined by its moments, while $D_n$ has have moments of all orders. 
If for all $r > 0$
$$\lim_n \mathbb{E}[X^r_n] = \mathbb{E}[X]$$
then $D_n \to D$ in the Levy-Prokhorov metric. 
\end{theorem}

Then, if $R^k(v_1)$ is an increasing sequence in $k$ (for example $R^k(v_1) = k$) we can use Theorem 30.2 of \citet{billingsley2013convergence} and H1 to say that $\llbracket v_1 \rrbracket^{R^k}(1, \delta_0) \to \llbracket v_1 \rrbracket(1, \delta_0)$.

For $v_2$ and for any node $v_i$ after that, we assume that, in the exact semantics, the node takes as input a pair $(p, D)$ such that $p \neq 0$, while in the Gaussian Semantics associated with $R^k$, it takes as input $(p_k, D_k)$ such that $p_k \to p$ in $\mathbb{R}$ and $D_k \to D$ in the Levy-Prokhorov metric. Then, we need to prove that we can choose $R^k(v_i)$ so that $\llbracket v_i \rrbracket^{R^k}(p_k, D_k) \to \llbracket v_i \rrbracket(p,D).$

We do this in three steps. First, we prove that the exact semantics preserves the convergence, i.e. $\llbracket v_i \rrbracket(p_k, D_k) = (p_k', D_k') \to (p', D') = \llbracket v_i \rrbracket(p, D)$ for each node type. Then, we use again Theorem \ref{thm:bill} and H1 to say that $T^{GM}_r(D_k') \to D'$ as $r \to \infty$. Finally, since $\llbracket v \rrbracket^{R^k}(p_k, D_k)$ is obtained from $(p_k', D_k')$ applying the operator $(\mathbb{I}, T^{GM}_{R^k(v_i)})$, we can use a diagonal argument to ensure that we can fix $R^k(v_i)$ for each $k$ so that $\llbracket v_i \rrbracket^{R^k} = (p_k', T^{GM}_{R^k(v_i)}(D_k')) \to (p', D')$.

The most industrious step is the first one, i.e. proving that $(p_k', D_k') = \llbracket v_i \rrbracket(p_k, D_k) \to (p', D') = \llbracket v_i \rrbracket(p, D)$ for each type of node. To do this, we use the Mapping Theorem \citep{billingsley2008probability}.
\begin{theorem}[Mapping Theorem] \label{thm:mapping}
    Suppose $h: \mathbb{R}^d \to \mathbb{R}^d$ is measurable and that the set $H$ of its discontinuities is such that $m_D(H) = 0$. If $D_n \to D$ in the Levy-Prokhorov metric, then $h(D_n) \to h(D)$.
\end{theorem}
In particular, when $v_i : state$, an assignment is of the type $x_j = E(x)$ with $E(x)$ in the form \eqref{eq:expr} (here we do not consider read-only variables for simplicity, but they can easily taken into account). Since $E(x)$ is continuous, convergence follows from Theorem \ref{thm:mapping} by taking 
$$ [h(x)]_k = \begin{cases} x_k & \text{if } k \neq j \\ E(x) &\text{if } k = j. \end{cases} $$

For $v_i : test$ and $v_i : observe$ we use again the Mapping Theorem, in two different ways. First, we use it with $h = \mathbb{I}_{\llbracket B(x) \rrbracket}$, where $B(x)$ is the LBC labelling the node, to show that $p_k' \to p'$. Then, we fix a point $x^*$ such that $m_{D_k}(x^*) = 0$ for all $k$ and we choose $h$ to be:
$$ [h(x')]_k = \begin{cases} x' & \text{if } x' \in \llbracket B(x) \rrbracket \\ x^* &\text{else}  \end{cases} $$
which proves that $D_k' = D_k \, | \, \llbracket B(x) \rrbracket \to D' = D \, |  \, \llbracket B(x) \rrbracket.$
In both cases, H2 is fundamental in guaranteeing that the Mapping Theorem still holds.

\section{Second Order Gaussian Approximation}
\label{sec:gaussapp}

As discussed in Examples 4.5 and 4.8, while implementing an arbitrary order Gaussian Semantics may be difficult, it is straightforward to compute the Gaussian Semantics associated with $R = 2$ (i.e., such that at each node of the control-flow graph matches the first two order moments (mean and covariance matrix). We propose \emph{Second Order Gaussian Approximation (SOGA)}, an algorithm that implements this particular case. 

A prototype implementation can be found at \url{https://zenodo.org/records/10026970}. 

\subsection{Overview}

In our implementation, SOGA accepts programs in a Python-like syntax, then compiled into a formal control-flow graph. SOGA recursively visits the nodes of the control-flow graph in a breadth-first fashion to compute the semantics of all paths. Furthermore, each node has two attributes, $p$ and $dist$: $p$ is a non-negative scalar proportional to the probability of reaching that node, while $dist$ stores the output distribution (in the form of a GM) computed by the semantics of that node. 

    \begin{algorithm}[t]
            \caption{$\mathit{SOGA}(\mathit{node})$}\label{alg:SOGA}
            \begin{algorithmic}[1]
            \STATE $input\_list = []$
            \FOR{$par$ \textbf{in} $node.parents$}
                \STATE $input\_list.append((par.p,par.dist))$
            \ENDFOR
             \STATE $input\_p, input\_dist =\text{merge\_dist}(input\_list)$
            \STATE $node.p, node.dist =\text{node\_semantics} (node, input\_p, input\_dist)$
       \FOR{$child$ in $node.children$}
        \STATE SOGA($child$)
    \ENDFOR
            \end{algorithmic}
    \end{algorithm}
The procedure applied for each node is summarized in Algorithm \ref{alg:SOGA}, where we assume that each node stores its parents and children in suitable attributes.
When entering a new node, SOGA retrieves the pairs $(p, dist)$ computed by the parents of the current node and merges them in a single pair $(input\_p, input\_dist)$ using the function \textit{merge\_dist} (line 1-5). Then, it computes how the node semantics transforms the latter pair and stores the result in the attributes $node.p, node.dist$ (line 6). Finally, it calls itself recursively on the children node (lines 7-9). When the $exit$ node is reached the algorithm ends, leaving the posterior distribution  stored in its attribute $node.dist$.

The core of the algorithm is the function node\_semantics, that, for each node type, transforms the pair $input\_p, input\_dist$ into a new pair $p, dist$. When $v \colon entry, exit$ node\_semantics leaves the pair unaltered; when $v \colon state$ or $v \colon test, observe$ the functions \textit{apply\_rule($input\_dist, expr$)} and \textit{approx\_trunc($input\_dist, trunc$)} are invoked, respectively. We detail the functions below.

\paragraph{Function~\textit{apply\_rule}.} It implements the semantics of a \textit{state} node. In particular, it takes as input the current mixture $input\_dist$ and an expression $expr$ of type \eqref{eq:expr}. It returns a new distribution $dist$ obtained applying $expr$ and $T^{GM}_2$ to $input\_dist$. To compute the moments of the transformed distribution $dist$, and therefore its second-order approximation, it uses the results in Table \ref{tab:tab1}: when $expr$ is a linear transformation, it applies the formulas for the sum of multivariate Gaussians~\citep{billingsley2013convergence}. When $expr$ involves products, it applies Isserlis' theorem~\citep{wick1950evaluation}.

\paragraph*{Function approx\_trunc} It implements the semantics of a \textit{test} or an \textit{observe} node. It takes as input the current mixture $input\_dist$ and a set $trunc$ defined by an LBC of type \eqref{eq:LBC}. It returns the probability mass $p$, given by the probability that $input\_dist$ satisfies $trunc$, and a new mixture distribution $dist$, representing the GM approximating $input\_dist$ conditioned to $trunc$. 
%In particular, the function first computes the vector of augmented variables $z$ and the associated distribution $D_z$ to take into account random terms appearing in the LBC. Since $D_z$ is the product distribution of $input\_dist$ and a finite number of independent GMs, it is a GM itself. Therefore, when conditioned to an LBC, $D_z$ can be rewritten as a mixture of Gaussians truncated to $\llbracket \mathtt{B(z)}\rrbracket$, for which the probability mass and first two order moments of each component can be computed exactly. 
Again, it applies the results in Table \ref{tab:tab1} to compute $dist$: in particular, when the LBC expresses inequality constraints the formulas in~\citet{kan2017moments} are used; when instead the LBC has the form $\mathtt{x_i == c}$ it uses the formulas from \citet{bishop2006pattern}.

\paragraph{Function merge\_dist} Merging is performed whenever a node is accessed prior to applying its semantics: \textit{merge\_dist} collects all the output pairs $(p, D)$ computed at the parent nodes, and merges them together in a single GM. Given the set of $s$ parents' pairs  $(p_1, D_1), \ldots ,(p_s, D_s)$, the function returns probability mass $input\_p = p_1 + \ldots + p_s$ and a new GM $input\_dist = \frac{1}{input\_p}(p_1D_1 + \ldots + p_sD_s)$. For an \textit{exit} node, the output of this function is the output distribution of the program.

\subsection{Distributivity of Transfer Functions}
SOGA explores the control-flow graph in a breadth-first fashion, performing merges when required. On the other end, the exact and the Gaussian semantics are defined as a sum over all execution paths, leading to an apparent discrepancy. To ensure that SOGA indeed computes the Gaussian Semantics associated with the map $R(v) = 2$ we show in Proposition \ref{prop:distributivity} that the transfer function of the exact semantics is distributive with respect to the merge operation. 

To do this, for a set of pairs $(p_i, D_i)$ we define the merge operator
$$ merge((p_1, D_1), \hdots, (p_s, D_s)) = \left(\sum_{i=1}^s p_i, \sum_{i=1}^s \frac{p_i}{\sum_{j=1}^s p_j}D_i \right) = (P, D).$$
We show that computing the semantics of a node after performing a merge gives the same output distribution as computing the semantics of each pair and then merging the results. This distributivity transfers straightforwardly to Gaussian Semantics, since the latter is computed by composing the exact semantics with the operator $T^{GM}_r$, which is distributive with respect to merging by Definition~\ref{eq:mmo}. This, in turn, justifies computing the semantics exploring the control-flow graph in a breadth-first fashion as SOGA does.

\begin{proposition} \label{prop:distributivity}
    Let $(p_i, D_i)$ be pairs with $p_i \ge 0$ and $D_i$ a distribution for $i=1, \hdots, s$. Let $v$ be a node of type state, test, observe or exit. Then
    \begin{equation} \label{eq:comm}
        merge(\llbracket v \rrbracket(p_1, D_1), \hdots, \llbracket v \rrbracket(p_s, D_s)) = \llbracket v \rrbracket(merge((p_1, D_1), \hdots, (p_s, D_s))) 
    \end{equation}
\end{proposition}

\proof 
Let $(\tilde{p}_i, \tilde{D}_i) = \llbracket v \rrbracket(p_i, D_i)$ for $i =1, \hdots, s$. Then the L.H.S. of Equation \ref{eq:comm} becomes 
$$merge(\llbracket v \rrbracket(p_1, D_1), \hdots, \llbracket v \rrbracket(p_s, D_s)) = \left(\sum_{i=1}^s \tilde{p}_i, \sum_{i=1}^s \frac{\tilde{p}_i}{\sum_{j=1}^s \tilde{p}_j}\tilde{D}_i\right) = (\tilde{P}, \tilde{D}).$$
For the R.H.S. let:
$$ \llbracket v \rrbracket(merge((p_1, D_1), \hdots, (p_s, D_s))) = \llbracket v \rrbracket(P, D) = (\hat{P}, \hat{D}).$$

Let us show for each type of node that $\tilde{P} = \hat{P}$ and $\tilde{D} = \hat{D}$. We observe that for $v \colon state$, with $v$ labeled by $\mathtt{skip}$, and for $v \colon exit$ conclusion follows trivially. We examine the remaining cases separately.

\begin{itemize}
    \item Let $v \colon state$ and suppose $v$ is labelled by $\mathtt{x_k := E(z)}$ then $\tilde{p}_i = p_i$ and $\tilde{D}_i$ is the distribution of $x[x_k = E(z)]$ where $x \sim D_i$. Then $\tilde{P} = P$ and $\tilde{D} = \sum_{i=1}^s \frac{p_i}{P}\tilde{D}_i$. On the other hand $\hat{P} = P = \tilde{P}$ and $\tilde{D}$ is the distribution of $y[y_k = E(z)]$ where $y \sim D = \sum_{i=1}^k \frac{p_i}{P}D_i$. Therefore $\hat{D} = \sum_{i=1}^k \frac{p_i}{P}\tilde{D}_i = \tilde{D}$.

    \item Let $v \colon test$. To ease the notation, let us assume $cond(s_\pi(v)) =true$ and $\mathtt{B(z)}=\mathtt{B(x)}$, but the argument works analogously in the other cases.  In this case $\tilde{p}_i = p_i \cdot P_{D_i}(\llbracket \mathtt{B(x)} \rrbracket)$ and $\tilde{D}_i = D_i \mid \llbracket \mathtt{B(x)} \rrbracket$. Therefore $\tilde{P} = \sum_{i=1}^s p_i \cdot P_{D_i}(\llbracket \mathtt{B(x)} \rrbracket)$ and $\tilde{D} = \sum_{i=1}^s \frac{p_i \cdot P_{D_i}(\llbracket \mathtt{B(x)} \rrbracket)}{\tilde{P}} (D_i \mid \llbracket \mathtt{B(x)} \rrbracket)$. On the other hand $\hat{P} = P \cdot P_{D}(\llbracket \mathtt{B(x)} \rrbracket) = P \cdot \sum_{i=1}^s \frac{p_i}{P}P_{D_i}(\llbracket \mathtt{B(x)} \rrbracket) = \sum_{i=1}^s p_i P_{D_i}(\llbracket \mathtt{B(x)} \rrbracket) = \tilde{P}.$ Moreover, $\hat{D} = D \mid \llbracket \mathtt{B(x)} \rrbracket$. Therefore, $\hat{D}$ has density 
    $$\frac{f_D(x)\mathbb{I}_{\llbracket \mathtt{B(x)} \rrbracket}}{P_D(\llbracket \mathtt{B(x)} \rrbracket)} = \sum_{i=1}^s \frac{p_iP_{D_i}(\llbracket \mathtt{B(x)} \rrbracket)}{P_D(\llbracket \mathtt{B(x)} \rrbracket)} \frac{f_{D_i}(x)\mathbb{I}_{\llbracket\mathtt{ B(x}) \rrbracket}}{P_{D_i}(\llbracket \mathtt{B(x)} \rrbracket)}$$
    which is the same density as the one of $\sum_{i=1}^s \frac{p_i P_{D_i}(\llbracket \mathtt{B(x)} \rrbracket)}{\tilde{P}} (D_i \mid \llbracket \mathtt{B(x)} \rrbracket) = \tilde{D}$. Observe that we have assumed that for at least one $i$ $\tilde{p}_i \neq 0$. However, if that is not the case $P_{D}(\llbracket \mathtt{B(x)} \rrbracket) = 0$ and the conclusion still holds.
    
    \item  Let $v \colon observe$. If $\mathtt{B(x)}$ has a probability greater than zero conclusion follows as in the previous case. If $\mathtt{B(x)}$ has the form $\mathtt{x_k == c}$ we can use the same argument, but we need to replace $P_{D_i}(\llbracket B(x) \rrbracket)$ with the normalization constant $I_i = \int_{\mathbb{R}^{d-1}} f_{D_i}(x, x_i=c) d(x\setminus x_i)$.

\end{itemize}
\endproof

\subsection{SOGAprune}

To improve the scalability of SOGA we propose a second version of the algorithm, called SOGAprune, in which the user can introduce at script level the instruction $\textbf{prune}(K)$, $K$ being an integer number. When the script is compiled in a cfg, the prune instruction is compiled in a new node of type $prune$. When accessed, the function node\_semantics invokes the 
function \textit{prune\_dist($input\_dist, K$)}.

\paragraph{Function prune\_dist} It prunes the current distribution $input\_dist$ to keep the number of its components below a user-specified bound $K$. 
The pruning is performed similarly to~\citet{chaudhuri2010smooth}. In particular, for each pair of components $i,j$ in the input distribution \textit{input\_dist}, having mixing coefficients $\pi_i, \pi_j$, means $\mu_i, \mu_j$ and covariance matrices $\Sigma_i, \Sigma_j$, we compute the mean $\mu' = \frac{\pi_i \mu_i + \pi_j \mu_j}{\pi_i + \pi_j}$ and the cost $cost(i,j) = \pi_i \| \mu' - \mu_i \| + \pi_j \| \mu' - \mu_j \|.$ 

After computing the cost for all pairs $(i,j)$ such that $i \neq j$ and $i,j < K$, the pair $(i, j)$ with minimal cost is substituted with a single component having mean $\mu'$ and covariance matrix $\Sigma'$ with
$$\Sigma' = \frac{\pi_i}{\pi_i + \pi_j}\left(\Sigma_i + \mu_i^T\mu_i\right) + \frac{\pi_j}{\pi_i + \pi_j}\left(\Sigma_j + \mu_j^T\mu_j\right) - \mu'^T\mu'.$$ 
Observe that $\mu'$ and $\Sigma'$ are exactly the mean ad the covariance matrix of the mixture $\frac{\pi_i}{\pi_i + \pi_j}\mathcal{N}(\mu_i, \Sigma_i) + \frac{\pi_j}{\pi_i + \pi_j}\mathcal{N}(\mu_j, \Sigma_j)$. This produces the best possible approximation of the two components \citep{chaudhuri2010smooth}. The procedure is iterated until the number of components is less than $K$ (observe that after the first two components have been merged into a new one, we need to recompute the cost only for the pairs in which the new component appears). 

A summary of how the semantics of each node is implemented is reported in Table \ref{tab:SOGA}, while detailed algorithms for SOGA implementation can be found in the Supplementary Material.

\begin{table}
\centering
 \resizebox{\textwidth}{!}{
 \begin{tabular}{rccc}
 \toprule
     \emph{Type} & \emph{Function} &  \emph{Input}   &  \emph{Computing}  \\
      \midrule
     \multirow{2}{1cm}{$state$} & \multirow{2}{1.75cm}{\textit{apply\_rule}} & \multirow{2}{1.5cm}{$input\_dist$, $expr$} & First two order moments of the components of \\
     & & & the distribution obtained applying $expr$ to $input\_dist$ \\
     \midrule
     \multirow{3}{1cm}{$text$, $observe$} & \multirow{3}{1.75cm}{\textit{approx\_trunc}} & \multirow{3}{1.5cm}{$input\_dist$, $trunc$}  & Probability mass  (or normalization constant) and   \\
     & & & first two order moments of the components of the  \\
     & & & distribution obtained truncating $input\_dist$ to $trunc$  \\
     \midrule
     \multirow{2}{1cm}{$prune$} & \multirow{2}{1.75cm}{\textit{prune\_dist}} & \multirow{2}{1.5cm}{$input\_dist$, $K$} & Distribution $input\_dist$ iteratively pruned \\
     & & & until the number of components is $\le K$ \\
     \bottomrule
     \end{tabular}
     }
     \caption{Function implementing node semantics in SOGA and SOGAprune. The input arguments $input\_p, input\_dist$ are retrieved by the parent nodes' attributes $p, dist$. The input arguments $expr$, $trunc$ and $K$ are stored in node attributes when the cfg is compiled from the program script.}
     \label{tab:SOGA}
\end{table}

\subsection{Computational Cost}

We first compute the computational cost without pruning, then we discuss how pruning affects it. 

\paragraph{Cost without pruning.} Let $|V|$ denote the total number of nodes, $|T|$ the number of \textit{test} nodes, $|TO|$ the number of \textit{test} and \textit{observe} nodes and $|S|$ the number of \textit{state} nodes. W.l.o.g. we assume for simplicity that all read-only variables are pushed to an initial distribution $D_0$ over $\mathbb{R}^n$; thus the output of the entry node is $(1, D_0)$ and all assignments only use output variables. By doing this we compute an upper bound on the true computational cost since the dimensions corresponding to read-only variables are dropped after marginalization. Letting $C_0$ denote the number of components of $D_0$, the output distribution will have at most $C \le C_{max} = 2^{|T|}C_0$ components. 

We consider the cost to access a node and perform elementary operations, such as assignments and products, constant. Expressions $expr$ and $trunc$ are assumed to be stored in suitable data structures accessible in constant time, so storage and reading of them are also considered elementary operations. Overall, elementary operations contribute to the total computational cost with a term $O(|V|)$, which is however dominated by the computational cost of executing \textit{approx\_trunc}, \textit{apply\_rule} and \textit{merge\_dist}. We examine their cost separately.

The function \textit{approx\_trunc} is invoked once when an \textit{observe} node is accessed and twice when a \textit{test} node is accessed, for the true and the false branch respectively. When $\mathtt{B(z)}$ is in the form $\mathtt{c_1\cdot z_1 + \hdots + c_n\cdot z_n \bowtie c}$ a singular value decomposition is performed to change coordinates, so that in the new set of coordinates the truncation set is a hyper-rectangle (cost $O(n^3)$, \citep{gu1995divide}). Then, a new mixing coefficient has to be computed for each component to convert the truncated GM into a mixture of truncated Gaussians (cost $O(n)$). Finally, for each truncated Gaussian, the first two order moments are computed using the formulas in \citet{kan2017moments} (cost $O(n^4)$, for a detailed account see Supplementary Material). When $\mathtt{B(z)}$ is in the form $\mathtt{x_i == c}$, to apply the formulas in \citet{bishop2006pattern}, matrix multiplication must be performed, amounting to cost $O(n^3)$ \citep{skiena2008algorithm}. Overall, we have a cost of $O\left(|TO|C_{max}n^4\right)$.

The function \textit{apply\_rule} is invoked every time a \textit{state} node is accessed. Since affine transformations require matrix multiplication (cost $O(n^3)$), the total cost is
$O\left(|S|C_{max}n^3\right)$.

Finally, the function \textit{merge\_dist} is invoked whenever a node is accessed and performs a scalar product. It contributes for a cost
$O\left(|V|C_{max}\right).$

The total cost of SOGA is therefore 
\begin{equation} O\left(|TO|C_{max}n^4\right) + O\left(|S|C_{max}n^3\right) + O\left(|V|C_{max}\right) \leq O(|V|2^{|T|}C_0n^4), \label{eq:compcost}
\end{equation}
that is, linear in the number of nodes $|V|$ and in the initial number of components $C_0$, polynomial in the dimensionality of the augmented input space $n$ and exponential in the number of \textit{test} nodes $|T|$, i.e., linear in the number of paths.

\paragraph{Effect of pruning.} Let us now consider the effect of introducing $\textbf{prune}(K)$ instructions. Let $|P|$ be the number of \textit{pruning} nodes and $|T|_{bet}$ be the maximum number of subsequent \textit{test} nodes without \textit{pruning} instructions between them. Then $|T|_{bet} \le |T|$ and $|T|_{bet} = |T|$ if no pruning instructions have been introduced in the program. Then, the maximum number of components a mixture can have before pruning occurs is $C_{max} = K2^{|T|_{bet}}$ (assuming w.l.o.g. $|C_0| < K$).

The function \textit{prune\_dist} is invoked at most $|P|$ times. When invoked, it first computes the cost for all possible pairs of components, which is at most $C_{max}(C_{max} - 1)$. The computation of the cost function for each pair has cost $O(n)$, while the computation of the covariance matrix (cost $O(n^2)$) is performed for a single pair. At its first iteration, the computational cost of \textit{prune\_dist} is, therefore, $O(C_{max}^2n)$. After this, new costs are computed for at most $C_{max}-K$ times, but each time only for $C < C_{max}$ pairs of components. The whole cost of the function is therefore $O(C_{max}^2n)$.

Substituting in \eqref{eq:compcost} one gets that the computational cost with pruning is bounded by:
\begin{equation} O(|V|K2^{|T|_{bet}}n^4) + O(K^22^{2|T|_{bet}}n) \le O(|V|K^22^{2|T|_{bet}}n^4). \label{eq:compcost2}
\end{equation}

Comparing \eqref{eq:compcost} with \eqref{eq:compcost2} one can conclude that pruning is only effective in reducing the computational cost when the overhead introduced by pruning ($K^22^{2|T|_{bet}}$) is less demanding than dealing with the full space of paths ($C_02^{|T|}$). To keep the overhead contained one could use small values of $K$ while keeping also $|T|_{bet}$ small (e.g. by introducing many \textit{pruning} instructions). However, this introduces an additional level of approximation which can hinder the accuracy of SOGA.

\section{Numerical Evaluation}
\label{sec:eval}

We split the numerical evaluation into four parts. In Section 7.1 we compare SOGA with four baseline tools representative of different inference methods for estimating the posterior mean: STAN for MCMC~\citep{carpenter2017stan}, PSI for exact symbolic analysis~\citep{gehr2016psi}, AQUA for quantization of posterior distributions~\citep{huang2021aqua} and Pyro for VI~\citep{bingham2019pyro}. %We show that SOGA has comparable performances with respect to other state-of-the-art approaches, outperforming them in some cases.
%In the second and third subsections, we propose two applications in which SOGA clearly outperforms other approaches: 
%\e{In the second and third subsections, we comprehensively evaluate SOGA's performance on two specific applications that have been extensively studied in the literature, owing to their significant practical impact. In these evaluations, it becomes evident that SOGA consistently outperforms other approaches, providing compelling evidence of its relevance and usefulness.} 
In Section 7.2 we compare SOGA against Pyro in performing Maximum A Posteriori (MAP) estimation \citep{gelman2013bayesian}, to test how well our method is able to capture the posterior distribution, in addition to its moments. Finally, in Sections 7.3 and 7.4, we evaluate SOGA's performance on two applications that have been extensively studied in the literature, owing to their significant practical impact. The first application is inference on models involving mixtures of continuous and discrete distributions, as in \citet{kharchenko2014bayesian,pierson2015zifa,gao2017estimating}; the second application is Bayesian inference on collaborative filtering. \cite{zhao2013interactive}. %These models are usually not supported by state-of-the-art tools but can be easily encoded in SOGA using the GM representation. The second application is Bayesian Inference on polynomial models. In this case, we consider a Collaborative Filtering model \cite{minka2001family} and show that even when the number of latent variables in the models increases to >40, SOGA can perform inference more efficiently and accurately than its competitors.

\subsection{Posterior Mean Estimation} 
We start by comparing SOGA with STAN for MCMC~\citep{carpenter2017stan}, PSI for exact symbolic analysis~\citep{gehr2016psi}, and AQUA for quantization of posterior distributions~\citep{huang2021aqua} and Pyro for VI~\citep{bingham2019pyro}. 
We consider the case studies from these tool's reference papers~\citep{carpenter2017stan,gehr2016psi,huang2021aqua}, excluding those which could not be encoded in our syntax. This choice is intended to stress SOGA in the analysis of programs that were not designed to enhance its properties. Overall out of 31 total models, 13 were left out: 9 because of non-parametrizable distributions depending on variable parameters, and 4 because of the presence of non-polynomial functions (taken from: STAN - 1, PSI - 3, AQUA - 8, Pyro - 1). The remaining 18 models can be found in \citet{carpenter2017stan} (\textsl{Bernoulli}), \citet{gehr2016psi} (\textsl{BayesPointMachine}, \textsl{Burglar}, \textsl{ClickGraph}, \textsl{ClinicalTrial}, \textsl{CoinBias}, \textsl{DigitRecognition}, \textsl{Grass}, \textsl{MurderMistery}, \textsl{NoisyOr}, \textsl{SurveyUnbias}, \textsl{TrueSkills}, \textsl{TwoCoins}) and \citet{huang2021aqua} (\textsl{Altermu}, \textsl{Altermu2}, \textsl{NormalMixtures}, \textsl{RadarQuery}, \textsl{TimeSeries}).

The considered programs are listed in Table~\ref{tab:results}. 
Pruning was applied after every test and observe nodes (repeating it only once if they occur subsequently) for programs whose computation time was greater than $1$\,s and at least ten times larger than the worst performing tool. We set $K= 0.015 C_{max}$ except for \textsl{NormalMixtures}; there, since $C_{max}$ exceeded the tens of thousands, we set $K = 30$. With this strategy, the pruning algorithm was invoked only in 4 out of the 18 considered programs. 

The experiments were performed on a laptop equipped with a 2.8\,GHz Intel i7 quad-core processor and 16\,GB RAM, CmdStan v2.30.1 and Wolfram Mathematica 13.1~(\citeauthor{Mathematica}), setting a time-out threshold at 600\,s. 

\begin{sidewaystable}
    \centering
    \caption{ Results using STAN, PSI, AQUA and SOGA. \emph{`---'}: discrete posterior not supported; \emph{`mem'}: out of memory error; \emph{`err'}: tool returns error state. For SOGA, $C$: final number of components; $d$: dimensionality of the output vector.}
     \begin{tabular}{c|c|cc|cc|cc|cc|cccc}
     \toprule
        \emph{Model} & \emph{Dist.} & \multicolumn{2}{c|}{\emph{STAN}}  & \multicolumn{2}{c|}{\emph{AQUA}} & \multicolumn{2}{c|}{\emph{Pyro (VI)}} & \multicolumn{2}{c|}{\emph{PSI}} & \multicolumn{4}{c}{\emph{SOGA}} \\
    \midrule
     & &
     \ \ \emph{time}   &  \emph{value} \ \ & 
     \ \ \emph{time}   &  \emph{value} \ \ & 
     \ \ \emph{time}   &  \emph{value} \ \ &
     \ \ \emph{time}   &  \emph{value} \ \ & 
     \ \ \emph{time}   &  \emph{value} \ \ & \emph{$C$} \ \ & \emph{$d$}\\
    \midrule
    \textsl{Bernoulli} & B,U & \cellcolor{lightgray} 0.17 & \cellcolor{lightgray} 0.250 & 0.84 & 0.247 & 4.51 & 0.250 & 0.38 & 0.250 & 1.28$^*$ & 0.252 & 27 & 2\\
    \textsl{BayesPointMachine} & G$^*$ & 51.0 & 0.056 & \multicolumn{2}{c|}{mem} & 60.49 & 0.046 & \multicolumn{2}{c|}{err} & \cellcolor{lightgray} 2.20 & \cellcolor{lightgray} 0.011 & \cellcolor{lightgray} 1 & \cellcolor{lightgray} 9 \\
    \textsl{Burglar} & B & \multicolumn{2}{c|}{---} & \multicolumn{2}{c|}{---} & \multicolumn{2}{c|}{---} & 0.12 & 0.003 & \cellcolor{lightgray}0.06 & \cellcolor{lightgray}0.003 & \cellcolor{lightgray}4 & \cellcolor{lightgray}6 \\
    \textsl{ClickGraph} & B,U & 102 & 0.540 & \multicolumn{2}{c|}{mem} & 3.13 & 0.566 & \cellcolor{lightgray}1.10 & \cellcolor{lightgray} 0.614 &  208$^*$ & 0.630 & 35 & 6 \\
    \textsl{ClinicalTrial} & B,U & \multicolumn{2}{c|}{---} & \multicolumn{2}{c|}{---} & \multicolumn{2}{c|}{---} & \cellcolor{lightgray} 0.97 & \cellcolor{lightgray} 0.755 & 92.2$^*$ & 0.753 & 23 & 5\\
    \textsl{CoinBias} & B,Be & \cellcolor{lightgray}0.07 & \cellcolor{lightgray}0.420 & 0.91 & 0.383 & 0.91 & 0.419 & 0.34 & 0.417 & 0.61 & 0.415 & 64 & 2 \\
    \textsl{DigitRecognition} & D & \multicolumn{2}{c|}{---} & \multicolumn{2}{c|}{---} & \multicolumn{2}{c|}{---} & \multicolumn{2}{c|}{err} & \cellcolor{lightgray}4.46 & \cellcolor{lightgray}4.453 & \cellcolor{lightgray}10 & \cellcolor{lightgray}2 \\
    \textsl{Grass} & B & \multicolumn{2}{c|}{---} & \multicolumn{2}{c|}{---} & \multicolumn{2}{c|}{---} & \cellcolor{lightgray}0.08 & \cellcolor{lightgray}0.708 & 0.09 & 0.708 & 28 & 10 \\
    \textsl{MurderMistery} & B & \multicolumn{2}{c|}{---} & \multicolumn{2}{c|}{---} & \multicolumn{2}{c|}{---} & 0.12 & 0.016 & \cellcolor{lightgray}0.01 & \cellcolor{lightgray}0.016 & \cellcolor{lightgray}2 & \cellcolor{lightgray}2\\
    \textsl{NoisyOr} & B & \multicolumn{2}{c|}{---} & \multicolumn{2}{c|}{---} & \multicolumn{2}{c|}{---} & \cellcolor{lightgray}0.16 & \cellcolor{lightgray}0.814 & \cellcolor{lightgray}0.16 & \cellcolor{lightgray}0.814 & \cellcolor{lightgray}256 & \cellcolor{lightgray}10 \\
    \textsl{SurveyUnbias} & B,G,U & \cellcolor{lightgray}0.10 & \cellcolor{lightgray}0.800 & 1.08 & 0.567 & 2.89 & 0.770 & 18.5 & 0.800 & 1.56 & 0.799 & 128 & 4 \\
    \textsl{TrueSkills} & G$^*$ & \cellcolor{lightgray}0.04 & \cellcolor{lightgray}104.0 & \multicolumn{2}{c|}{mem} & 1.30 & 101.4 & \multicolumn{2}{c|}{to} & 0.05 & 104.7 & 1 & 6 \\
     \textsl{TwoCoins} & B & \multicolumn{2}{c|}{---} & \multicolumn{2}{c|}{---} & \multicolumn{2}{c|}{---} & 0.10 & 0.333 & \cellcolor{lightgray}0.01 & \cellcolor{lightgray}0.333 & \cellcolor{lightgray}3 & \cellcolor{lightgray}3 \\
     \textsl{Altermu} & G$^*$ & 19.0 & 0.009 & 1.32 & 0.000 & 33.1 & 0.030 & \multicolumn{2}{c|}{to} & \cellcolor{lightgray}0.16 & \cellcolor{lightgray}0.000 & \cellcolor{lightgray}1 & \cellcolor{lightgray}5 \\
     \textsl{Altermu2} & G$^*$, U & 15.0 & 0.170 & 0.79 & 0.155 & 5.50 & 0.098 & 284 & 0.155 & \cellcolor{lightgray}0.36 & \cellcolor{lightgray}0.156 & \cellcolor{lightgray}4 & \cellcolor{lightgray}3 \\
     \textsl{NormalMixtures} & G$^*$, U & \cellcolor{lightgray}0.38 & \cellcolor{lightgray}0.286 & 1.27 & 0.286 & 104.89 & 0.295 & \multicolumn{2}{c|}{to} & 50.4$^*$ & 0.298 & 30 & 4 \\
     \textsl{RadarQuery} & B,G$^*$,U & 144 & 5.000 & \cellcolor{lightgray}0.90 & \cellcolor{lightgray}6.333 & \multicolumn{2}{c|}{err} & 7.75 & 6.333 & 6.34 & 5.940 & 2016 & 8 \\
     \textsl{TimeSeries} & G$^*$,U & \cellcolor{lightgray}0.37 & \cellcolor{lightgray}-1.600 & 1.67 & -1.575 & 26.15 & -1.701 & \multicolumn{2}{c|}{to} & 3.79 & -1.590 & 19 & 4\\
    \bottomrule
    \end{tabular}
    \label{tab:results}
\end{sidewaystable}

\subsubsection{Results} Table~\ref{tab:results} collects the results where \textit{time} refers to the average runtimes (in seconds) out of 10 executions and \textit{value} refers to the computed expected value of a target variable in the model. For each model we specify the kind of distributions involved: B=Bernoulli, Be=Beta, D=Discrete, G($^*$) =Gaussian (with non-constant mean), U=Uniform. For STAN, we indicate the time needed to obtain a 5\% confidence interval whose amplitude is contained in 1\% of the mean (up to a maximum of $10^5$ samples). For PSI we report the sum of the time needed to generate the symbolic formula and that needed to integrate it when in the presence of non-simplified integrals (observe that in the original paper, only the time for symbolic computation was considered). For VI, due to high sensitivity with respect to the hyperparameters \citep{hoffman2013stochastic}, we proceed using three different learning rates (0.01, 0.005, 0.001), and we report the most accurate estimation (detailed results can be found in the Supplementary Material). The number of iterations of the stochastic gradient descent is increased from a minimum of 100 to a maximum of 10k, stopping the optimization if the difference between the estimated mean posterior and the mean posterior estimated 100 steps before is less than $1\%$ of the current estimation. For SOGA, runtimes labeled with $^*$ indicate that the pruning algorithm was invoked. Finally, we highlight the fastest method with a grey background.
For accuracy evaluation, we consider PSI's results as ground truth when available (i.e., when PSI terminates and the integration is successfully computed within the timeout threshold). We made this choice since PSI is an exact method and the only guaranteed to be exact among the evaluated tools. 

Only in one example, \emph{BayesPointMachine}, SOGA performs poorly in terms of accuracy, estimating a value of 0.011 for a parameter estimated by STAN as $0.054 \pm 2\mathrm{e-}4$. We remark, however, that this program turned out to be particularly difficult to solve for AQUA (which issued an out-of-memory error) and PSI  (which was not able to complete the symbolic computation of the posterior). On the other examples, SOGA yields very good accuracy, with a relative error below 7\% across all comparable models. 
We now discuss a detailed comparison of runtimes against each tool method. 

\paragraph{STAN} STAN does not support discrete posteriors; hence it could not analyze eight models. For the models that can be analyzed by both, SOGA outperforms STAN in terms of runtimes on \textsl{Altermu}, \textsl{Altermu2}, and \textsl{RadarQuery}. By contrast, STAN outperforms SOGA in \textsl{Bernoulli} and \textsl{NormalMixtures}. We attribute this  to the presence of non-Gaussian priors and a large number of observations, resulting in a high number of components and truncations to be computed. Both have similar performance on the remaining models. 

\paragraph{AQUA} SOGA is more flexible than AQUA in that it supports discrete posteriors. On \textsl{ClickGraph}, and \textsl{TrueSkills}  AQUA issued an out-of-memory error while SOGA could approximate the posterior mean. We ascribe this issue to the fact that AQUA uses tensors, whose dimension rapidly increases with the number of distributions. %In particular, the number of distributions used in the program is in general larger than the number of variables due, for example, to the presence of conditional branches depending on fresh variables. 
In particular, in AQUA each distribution must be stored in the tensor, while SOGA can use fresh read-only variables which are dropped once the marginal over the output variables is evaluated. Notably, SOGA outperforms AQUA also on \textsl{Altermu}, \textsl{Altermu2} and \textsl{TimeSeries} proposed in the AQUA paper~\citep{huang2021aqua}. %In addition to \textsl{Bernoulli} and \textsl{NormalMixtures} models, AQUA is more efficient than SOGA also in \textsl{RadarQuery}. 
Instead, AQUA is more efficient than SOGA in \textsl{RadarQuery}, \textsl{Bernoulli} and \textsl{NormalMixtures}, for the same reasons explained for STAN.

\paragraph{Pyro}

Being a gradient-based method, Pyro's VI offers limited support for discrete variables,\footnote{\url{https://pyro.ai/examples/enumeration.html}} so that, similarly to STAN and AQUA, we were not able to encode models with discrete posterior. In addition, we found that the encoding of \textsl{RadarQuery} incurred runtime errors. For the remaining models, VI is comparable to SOGA, when not less accurate, and taking longer runtimes. Noticeable exceptions are \textsl{BayesPointMachine}, where, as already noticed, SOGA is not able to achieve a good accuracy and \textsl{ClickGraph}, where SOGA incurs long runtimes even with pruning. On the other hand, VI exhibits a significant sensitivity with respect to the choice of the hyperparameters, which can result in non-convergence and sloppy approximations for poor choices of the parameters (results for all the tested learning rates can be found in the Supplementary Material).

\paragraph{PSI} PSI outperforms SOGA on \textsl{Bernoulli}, \textsl{ClickGraph}, and \textsl{ClinicalTrial}. However, on six models (\textsl{SurveyUnbias}, \textsl{TrueSkills}, \textsl{Altermu}, \textsl{Altermu2}, \textsl{RadarQuery}, and \textsl{TimeSeries}) PSI timed out or resulted in long runtimes. 
This behavior can be explained by the presence of distributions dependent on variable parameters (\textsl{SurveyUnbias}, \textsl{Trueskills}, \textsl{Radar}) or by the high number of observations (\textsl{Altermu}, \textsl{Altermu2}, \textsl{TimeSeries}). In \textsl{Altermu}, PSI could not compute a symbolic formula within the time-out threshold, while in \textsl{BayesPointMachine}, \textsl{TrueSkills}, and \textsl{TimeSeries} the formula contained non-simplified integrals, whose integration in Mathematica took longer than the time-out threshold. Notably, for models involving only Bernoulli distributions (\textsl{Burglar}, \textsl{Grass}, \textsl{MurderMistery}, \textsl{NoisyOr}, \textsl{TwoCoins}), for which both tools are exact, their performance is comparable.

\subsubsection{Performance of Pruning} \, \\ \noindent
\begin{minipage}{0.53\textwidth}
In the right inset  we report the runtimes, values, and number of components $C$ for SOGA without pruning applied to the four models that required the application of pruning (\textsl{Bernoulli}, 
\textsl{ClickGraph}, \textsl{ClinicalTrial}, \textsl{NormalMixtures}). All models share the occurrence of GM distributions with more than 1000 components. For \textsl{Bernoulli}, pruning allowed comparable runtimes 
\end{minipage}
\begin{minipage}{0.46\textwidth}
%\begin{table}[H]
%\centering
\vspace{5pt}
\centering
 \begin{tabular}{rccr}
 \toprule
     \emph{Model} &  \multicolumn{1}{c}{\emph{Time}}   &  \multicolumn{1}{c}{\emph{Value}} & \multicolumn{1}{c}{\emph{$C$}} \\
      \midrule
     \textsl{Bernoulli} & 11.97 & 0.252 & 1774  \\ 
     \textsl{ClickGraph} & \multicolumn{2}{c}{to} & 2304  \\
     \textsl{ClinicalTrial} & \multicolumn{2}{c}{to} & 1508  \\
     \textsl{NormalMixtures} & \multicolumn{2}{c}{to} & 97714  \\
     \bottomrule
     \end{tabular}
     %\caption{Application of pruning.}
     %\label{tab:prune}
%\end{table}
\vspace{7pt}
\end{minipage}
with respect to the best-performing tool, while base SOGA was about 9 times slower (11.97\,s). In addition, base SOGA computed an output indistiguishable from the pruned version up to the third decimal digit. For the other three cases, base SOGA was unable to compute a numerical result within the time-out threshold. For these, the number of components $C$ is the one reached before timing out. Applying pruning allowed SOGA to complete the computation within the time-out threshold while achieving excellent accuracy with respect to the ground truth.

\subsection{Maximum a Posteriori Estimation}\label{sec:map}

\begin{table}[t]
    \centering
    \begin{tabular}{cccccc}
    \toprule
    & \multicolumn{2}{c}{\emph{Pyro}} & \multicolumn{2}{c}{\emph{SOGA}} & \multirow{2}{*}{\emph{True value}} \\
    \cmidrule(l){2-3} \cmidrule(l){4-5}  
    \emph{Model} & \emph{value} & \emph{time} & \emph{value} & \emph{time} \\
    \midrule
    \textsl{Bernoulli} & \cellcolor{lightgray}0.200 & \cellcolor{lightgray}0.20 & 0.220 & 12.0 & 0.200 \\
    \textsl{Bernoulli (P)} & \cellcolor{lightgray}0.200 & \cellcolor{lightgray}0.20 & 0.290 & 1.28 & 0.200 \\
    \textsl{BayesPointMachine} & 0.000 & 7.83 & \cellcolor{lightgray}0.011 & \cellcolor{lightgray}2.20 & 0.032 $\pm$ 0.002*\\
    \textsl{ClickGraph} & 0.501 & 2.98 & \cellcolor{lightgray}0.861 & \cellcolor{lightgray}208 & 1.000\\
    \textsl{CoinBias} & \cellcolor{lightgray}0.400 & \cellcolor{lightgray}0.63 & 0.493 & 0.61 & 0.400 \\
    \textsl{SurveyUnbias} & \cellcolor{lightgray}0.964 & \cellcolor{lightgray}3.46 & 0.755 & 1.56 & 1.000\\
    \textsl{TrueSkills} & 101.6 & 0.99 & \cellcolor{lightgray}104.7 & \cellcolor{lightgray}0.05 & 104.8 $\pm$ 0.681* \\
    \textsl{Altermu} & \multicolumn{2}{c}{not converged} & \cellcolor{lightgray}0.000 & \cellcolor{lightgray}0.16 & 0.114 $\pm$ 0.092*  \\
    %Altermu2 & 0.001 & 2100 & 0.490 & 45.40 & -2.043 & 0.03 & -4.844 & 0.36 \\
    \textsl{NormalMixtures (P)} & 0.236 & 49.5 & \cellcolor{lightgray}0.276 & \cellcolor{lightgray}50.4 & 0.275 $\pm$ 0.005*\\
    \textsl{TimeSeries} & \cellcolor{lightgray}-1.564 & \cellcolor{lightgray}55.4 & -1.494 & 3.69 & -1.694 $\pm$ 0.021 \\
    \bottomrule
    \end{tabular}
    \caption{Comparison between Pyro and SOGA for MAP estimation. Models with `(P)' were pruned when SOGA was applied. True values are derived optimizing the exact posterior, or from samples (denoted with `*').
    }
    \label{tab:map}
\end{table}

Since SOGA approximates the posterior with a Gaussian mixture, it can also compute the Maximum a Posteriori (MAP) estimate by simply returning the mean of the GM component with the largest mixing coefficient. Here we compare its performance against Pyro, in which MAP estimation can be performed using a different parametrizing distribution than the one used for the mean posterior inference.\footnote{\url{https://pyro.ai/examples/mle_map.html}}
To get a baseline for the MAP value, we first generate the symbolic posterior using PSI and then optimize it numerically. For models in which PSI is not able to compute the exact posterior, we estimate the ground truth by taking 10k samples from the posterior and binning them into 50 intervals; then, MAP is the midpoint of the interval with the most samples. We tested the same models with continuous posterior reported in Table~\ref{tab:results}, except \emph{Altermu2}, since, by visual inspection, we found that it has a flat posterior. 

Results are reported in Table~\ref{tab:map}. 
Due to Pyro's sensitivity to hyperparameters observed in the previous section, we tested three different values of learning rate. Table~\ref{tab:map} only reports the closest estimation to the baseline; full results are available in the Supplementary Material, confirming the sensitivity issues. 
These experiments show that SOGA performs relatively worse than in the estimation of the posterior mean. This is expected because SOGA is designed to match means and variances, but it does not necessarily approximate the whole distribution.  However, compared to Pyro, it is still able to obtain the closest estimation for \textsl{BayesPointMachine}, \textsl{ClickGraph}, \textsl{TrueSkills}, \textsl{Altermu} and \textsl{NormalMixtures}, while it is outperformed by Pyro in \textsl{Bernoulli}, \textsl{CoinBias}, \textsl{SurveyUnbias} and \textsl{TimeSeries}. Finally, we note that analyzing \textsl{Bernoulli} with SOGAprune degrades the MAP estimation, unlike in the posterior mean.  

\subsection{Mixtures of Continuous and Discrete Distributions}\label{sec:mix}

\begin{minipage}{0.47\linewidth}
\vspace{5pt}
 Mixtures of continuous distributions and discrete probability masses appear in different domains such as in \citet{kharchenko2014bayesian,pierson2015zifa,gao2017estimating}. Languages such as STAN and AQUA do not support them. Ad hoc methods have been proposed in \citet{tolpin2016design} and \citet{nitti2016probabilistic}. More recently \citet{wu2018discrete} extended the sampling techniques used in BLOG for more accurate inference. We test SOGA on the three benchmarks proposed by \citet{wu2018discrete} and compare its runtimes against PSI, BLOG, and variable elimination (VE) as implemented in Pyro \cite{obermeyer2019tensor}. \emph{IndianGPA} and \emph{Scale}
 \end{minipage}
\begin{minipage}{0.45\linewidth}
\centering 
%\begin{table}[t]
\vspace{5pt}
    \begin{tabular}{crrrr}
    \toprule 
    & \multicolumn{4}{c}{\emph{Runtimes (s)}} \\ 
      \cmidrule(l){2-5} 
      \emph{Model} & \multicolumn{1}{c}{\emph{SOGA}} & \multicolumn{1}{c}{\emph{PSI}} & 
      \multicolumn{1}{c}{\emph{BLOG}} & 
      \multicolumn{1}{c}{\emph{VE}} \\
  \midrule
    \emph{IndianGPA} & $0.099$ & $0.180$ & $0.516$ & $0.192$ \\
    \emph{Scale} & $0.013$ & $0.120$ & $0.810$ & $0.150$ \\
    \emph{Tracking\_1} & $0.042$ & \multicolumn{1}{c}{to} & $0.803$ & $0.143$ \\
    \emph{Tracking\_5} & $0.046$ & \multicolumn{1}{c}{to} & $1.044$ & $0.394$ \\
    \emph{Tracking\_10} & $0.046$ & \multicolumn{1}{c}{to} & $1.330$ & $0.670$ \\
    \emph{Tracking\_50} & $0.110$ & \multicolumn{1}{c}{to} & $2.886$ & $4.095$ \\
    \emph{Tracking\_100} & $0.192$ & \multicolumn{1}{c}{to} & $5.054$ & $8.885$ \\  
    \emph{Tracking\_150} & $0.271$ & \multicolumn{1}{c}{to} &  $6.602$ & $13.723$ \\
    \bottomrule
    \end{tabular}
    \vspace{7pt}
 %   \caption{Runtimes (in seconds) for the models proposed \citet{wu2018discrete} comprising mixtures of discrete and continuous distributions. We do not report values, since all methods identify the exact posterior.}
%    \label{tab:mix}
%\end{table} 
\end{minipage}
   
  \noindent  are reported exactly as in the original paper, while the \emph{Tracking}\_$n$ example from Section~\ref{sec:tracking} is adapted since it was originally cast as a control problem. All examples have a Dirac delta posterior, which is computed exactly by all. However, SOGA is the fastest and the one which scales better as the number of steps $n$ increases.

\subsection{Bayesian Inference for Collaborative Filtering}
\label{sec:collabfiltering}

%\m{Ho commentato tutta una parte, non credo sia necessaria}
%SOGA performs particularly well also on a special class of Bayesian Inference problems. In Bayesian inference, one assumes that an underlying model depends on some unknown parameters and that some noisy observations sampled from the model are available. Parameters are estimated by introducing a prior distribution over them and then conditioning it to the available observations. Here the prior distribution is meant to express the initial uncertainty about the true value of the parameters. Standard choices for priors are Gaussian distributions when the parameters are not restricted to a particular interval, or Beta distributions else \citep{bishop2006pattern}. By conditioning, every observation adds information on the value of parameters. In fact, after observing a certain sample, the probability over the parameter is re-weighted so that values of the parameters for which the observation is more probable are given a higher weight and values for which the observation is impossible are weighted 0. Once the posterior is inferred, the posterior means or modes are taken as the estimated value of the parameters. 
%Here we look at how well SOGA scales on a particular class of polynomial models called collaborative filtering \cite{zhao2013interactive}. 

\begin{table}[t]
    \centering
    \begin{tabular}{ccrrrrrrrr}
    \toprule 
    & & \multicolumn{2}{c}{\emph{SOGA}} & \multicolumn{2}{c}{\emph{STAN}} & \multicolumn{2}{c}{\emph{AQUA}} & \multicolumn{2}{c}{\emph{VI}} \\
    \cmidrule(l){3-4} \cmidrule(l){5-6} \cmidrule(l){7-8} \cmidrule(l){9-10}
    $k$ & \multicolumn{1}{c}{\emph{Ground truth}} & \multicolumn{1}{c}{\emph{time}} & \multicolumn{1}{c}{\emph{value}} & \multicolumn{1}{c}{\emph{time}} & \multicolumn{1}{c}{\emph{value}} & \multicolumn{1}{c}{\emph{time}} & \multicolumn{1}{c}{\emph{value}} & \multicolumn{1}{c}{\emph{time}} & \multicolumn{1}{c}{\emph{value}} \\
    \midrule
    1 & 2 & $0.16$ & 1.86 & $0.94$ & 1.90 & 1.64 & 1.83 & 18.90 & 1.79  \\
    2 & 25 & $0.18$ & 24.28 & $4.87$ & 24.00 & \multicolumn{2}{c}{mem} &  25.10 & 23.93 \\
    3 & -5 & $0.19$ & -5.79 & $7.63$ & -5.80 & \multicolumn{2}{c}{mem} &  26.40 & -5.82 \\
    5 & -30 & $0.22$ & -31.98 & $7.22$ & -32.00 & \multicolumn{2}{c}{mem} &  23.19 & -31.47  \\
    10 & 151 & $0.30$ & 149.75 & $5.42$ & 150.00 & \multicolumn{2}{c}{mem} & 20.04 & 146.39 \\
    20 & 70 & $0.60$ & 73.76 & $14.10$ & 74.00 & \multicolumn{2}{c}{mem} & 23.18 & 69.92 \\
    \bottomrule
    \end{tabular}
    \caption{Runtimes (in seconds) and computed values for the collaborative filtering model $\mathcal{N}(cf_k, 1)$.}
    \label{tab:cf}
\end{table}

Collaborative filtering models are well-known in machine learning for applications to recommendation systems~\citep{koren2021advances}.
%A comprehensive survey on their advancement can be found in \citet{koren2021advances}. 
We target the problem of Bayesian inference on the latent factor model proposed in \citet{hofmann1999latent}, which arises after a singular value decomposition and serves as the basis for solving an optimization problem~\cite{zhao2013interactive}.
The model assumes noisy observations sampled from $\mathcal{N}(cf_k, 1)$ where $cf_k$ has the form $cf_k = a_1b_1 + \hdots + a_kb_k + c$, where $a_i, b_i$, and $c$ are unknown latent variables. As noticed in \citet{nishihara2013detecting}, performing Bayesian inference on these models is particularly difficult due to non-identifiability \cite{tsiatis1975nonidentifiability} and symmetry \cite{neal1999erroneous} of the parameters. %For example, suppose we want to infer the distribution of every latent variable. This gives poor, non-informative results: notice, 
For example, switching the distributions of $a_i$ and $b_i$ will result in the same distribution for $cf_k$, which is the only one observed. In some cases, one may still want to model each parameter separately to allow for more flexibility. In this particular case, though not solving the problem of symmetry and non-identifiability, SOGA can estimate the distribution of $cf_k$ faster than its competitors.  Results are shown in Table \ref{tab:cf} for various values of $k$. PSI results are not reported because the tool was able to produce a symbolic formula only up to $k=3$; however, even in these cases, numerical integration of the non-simplified integrals required more than 600\,s. Although STAN's estimates are accurate and close to SOGA's ones, its runtimes are longer due to the increased cost of sampling, which is exponential in the number of variables. As above, we attribute AQUA's out-of-memory error to its tensor based representation. For VI, we report results for the learning rate 0.005, which we found to be the one performing best in average, among the tested ones. A full set of experiment results can be found in the Supplementary Material. VI exhibits an accuracy comparable to SOGA's, but significantly longer runtimes. We observe, however, that thanks to vectorization, VI's runtimes do not significantly increase with $k$. Overall, the excellent runtime performance of SOGA is due the particular structure of the models, which exhibit Gaussian posteriors on variables combined in a scalar product without introducing truncations that could slow down the computations.

\section{Further Related Work}
\label{sec:related}

\paragraph{Inference.} 
%\m{Ho commentato testo già menzionato precedentemente, in caso recuperiamo i riferimenti e li citiamo prima se necessario.}
%Inference in probabilistic programs is a widely investigated problem for which no one-fit-for-all solution currently exists. MCMC and other forms of sampling \citep{hastings1970monte,nori2014r2,goodman2008church,mansinghka2014venture,pfeffer2001ibal,chaganty2013efficiently} can be used for vast classes of models, even if they can become computationally intensive in presence of conditioning because of high rejection rates. While in general sampling approaches do not support discrete posteriors and mixtures of probability densities and discrete masses, specific approaches such as likelihood-weighted sampling can be devised \cite{milch2004blog}. 
In addition to the techniques discussed earlier in this paper, volume computation can be quite efficient for discrete models~\citep{filieri2013reliability,holtzen2020scaling}; however, it cannot be applied to continuous distributions. %as shown in ,  %A way to generalize volume computation to hybrid programs has been proposed in \citep{poorva2023bit}. 
%The already mentioned AQUA \citet{huang2021aqua} uses an approximated symbolic technique in which distributions are represented through tensors and targets continuous posteriors are targeted in the same fashion as in STAN. Symbolic execution \citep{gehr2016psi,narayanan2016probabilistic,saad2021sppl} has been successfully applied to a variety of models. Despite the efficiency with which the symbolic posterior can be generated, when a numerical response is queried these methods can encounter significant time limitations due to the necessity of solving complex multi-dimensional integrals. 
All the mentioned methods use a pdf representation of the distributions. More recently, representations using generating functions have been investigated, but only for discrete distributions~\citep{chen2022does}. Finally, some approaches use moment-based invariants \citep{barthe2016synthesizing,chakarov2014expectation,katoen2010linear,bartocci2020mora,moosbrugger2022moment}. While they share the idea of computing moments up to a certain order, they differ both with respect to the supported programs and the computed information, making a direct comparison difficult.

\paragraph{Universal Approximators.} Our approach can be ascribed to the practice, common in many branches of mathematics, of studying \emph{universal approximators}, whereby one shows that a given function belonging to a certain class is shown to be approximated, arbitrarily closely, by another family of (parameterized) functions. Notable examples are polynomials~\citep{perez2006survey}, and neural networks~\citep{hornik1989multilayer,zhou2020universality}. %Proofs of universal approximation theorems may not be constructive, hence one does not know how to practically find an instance of the approximating family that suits a concrete problem. However, it is hoped that certain instances are computationally manageable as well as accurate in practice.

\paragraph{Gaussian Approximators.}
The approximation-by-Gaussian approach is also common to
%also to VI \m{perché lo citiamo solo ora?} \f{in realtà lo diciamo già in Section 4, parlando del Maximum Entropy Matching} \citep{zhang2018advances} and 
Laplace approximation \citep{tierney1986accurate}.  Laplace approximation is a mode matching strategy and is more expensive computationally than VI, as it is based on an optimization process to find the mode. Generally, however, it is inferior to VI~\citep{bishop2006pattern}.  %When VI is used to approximate more complex distributions, e.g. for Bayesian inference in neural networks, it requires solving a complex optimization problem. This is usually tackled by stochastic gradient descent, implying a large computational cost and no guarantees of convergence and accuracy. 
Another kind of Gaussian approximation is Gaussian Smoothing \citep{chaudhuri2010smooth,chaudhuri2011smoothing}, although it does target neither probabilistic programs nor the inference problem.

%\m{ho commentato questa parte, il testo mi sembrava ridondante, in caso recuperiamo citazioni perse}
%\paragraph{Applications} Models with mixtures of discrete and continuous distributions have been studied in \citet{gao2017estimating,kharchenko2014bayesian,pierson2015zifa}. \citet{tolpin2016design} and \citet{nitti2016probabilistic} discussed possible solutions to the \emph{IndianGPA} model without addressing the general problem of encoding mixed mixtures. A variation of likelihood-weighted sampling, tailored to this kind of distribution, has been proposed in \citet{wu2018discrete} and integrated into BLOG in~\citet{milch2004blog}. Collaborative filtering models are an established paradigm in machine learning to model recommendation systems. 

\section{Conclusions}\label{sec:conclusion}

Gaussian Semantics is a family of approximations parameterized by the moment order to match against a Gaussian mixture at each location of a probabilistic program. The universal approximation theorem states that such a family converges to the true semantics. Although, in principle, any program location could be treated with different moment-order matching, in practice this is a difficult problem that requires the solution of a system of nonlinear equations. While the system is guaranteed to have a solution, finding it using SMT solvers over reals or numerical methods yields poor results, due to long computational times and numerical instability. Therefore we leave open the general problem of implementing Gaussian Semantics for any order of moments. However, we provide an analytical method that matches second-order moments of the exact probabilistic semantics (SOGA). The numerical results for the case studies demonstrate high quality of the approximation and that SOGA complements state-of-the-art methods for probabilistic inference and in particular for inference on models with mixtures of discrete and continuous distributions and for Bayesian inference on collaborative filtering models. Due to the efficiency shown by SOGA, we believe that in these cases our method can effectively be used as an alternative to sampling.

As regards future work, while SOGA performed satisfactorily on all tested benchmarks, it could not be applied to some of the models from the same repositories, due to the limitations of our syntax. Extending the latter to include general distributions depending on non-constant parameters, unbounded loops and non-polynomial functions would widen its scope of applicability. A possible way to overcome the former restriction could be learning offline the approximating distributions as a function of the variable parameters, but how to do this efficiently is currently not clear, even though of great interest. For what concerns unbounded loops, we observed that for almost surely terminating programs, the loops can be unrolled for a finite number of iterations so that the error committed in the approximation is arbitrarily small. This suggests that increasing the number of unrolled iterations together with the number of moments matched should preserve our convergence theorem, even in the case of almost surely terminating unbounded programs. Similarly, one could exploit convergence results for polynomial approximations to extend the convergence result to sequences of polynomial programs that approximate programs featuring non-polynomial functions. We leave the possibility to explore these extensions of our convergence result in future work.

Finally, one might devise algorithms for higher-order moments. While an extension to \emph{exact} higher-order moment matching seems hard, a relaxed moment problem could be defined as an optimization problem~\citep{hansen2010generalized}. %Although it could be solved more  efficiently, this would require a non-trivial adaptation to the proof of our universal approximation theorem. Another direction is studying different criteria of choice for the approximating Gaussian Mixture. Although a different criterion would not affect the convergence properties, in practice it may yield a different approximation quality. Certainly, another direction is to look beyond Gaussian Mixtures as the class of  approximating distributions. 
%Finally, here we dealt with bounded programs. This is due to the fact that unbounded loops may not guarantee almost sure termination~\citep{barthe2020foundations} and introduce significant complications in the definition of the semantics \cite{zhang2022reasoning}. In addition, the proof of universal approximation exploits the finite number of paths in the control-flow graph. How to extend our framework to programs with unbounded loops would certainly be of great theoretical and practical interest. 

%From an experimental viewpoint, the encouraging numerical results call for the development of a robust compiler for Gaussian Semantics building on the prototype herein employed. Ideally, it would implement the static type analysis and program transformation strategies discussed in the paper to check the validity of the main hypothesis of convergence. 
%To cope with the dependence of the number of mixture components on the number of paths of the program, one could implement customary methods when dealing with Gaussian Mixtures that merge components when they exceed a maximum threshold~\citep{hennig2010methods}. This would increase scalability but introduce a numerical approximation in the computation of SOGA, which is otherwise computed exactly with our proposed algorithm. The trade-off between precision and accuracy will be subject of future work.

\section*{Acknowledgment}
This work was partially supported by the projects SERICS (PE00000014) and by Investment 1.5 Ecosystems of Innovation, Project Tuscany Health Ecosystem (THE, B83C22003920001) and Interconnected North-East Innovation Ecosystem (iNEST, ECS\_00000043) under the MUR National Recovery and Resilience Plan funded by the European Union - NextGenerationEU. We would like to thank Joost-Pieter Katoen for his feedback on a preliminary version of this paper and the anonymous reviewers for their valuable comments.

\bibliographystyle{ACM-Reference-Format}
\bibliography{bibfile}

\newpage
\appendix
\input{extended}

\end{document}

%% file: extended.tex
\section{Additional Background Material}

\subsection{Measurable spaces and random variables}

%Let $\Omega$ be any non-empty set. A family $\mathcal{F}$ of sets of $\Omega$ is called an algebra if satisfies the following:
%\begin{itemize}
%    \item[i)] $\Omega \in \mathcal{F}$;
%    \item[ii)] if $A \in \mathcal{F}$ then $\Omega \setminus A \in \mathcal{F}$;
%    \item[iii)] if $A, B \in \mathcal{F}$ then $A \cup B \in \mathcal{F}$.
%\end{itemize}

%In addition $\mathcal{F}$ is called a $\sigma$-algebra if it also satisfies:
%\begin{itemize}
%    \item[iv)] if $A_n \in \mathcal{F} \, \forall n \in \mathbb{N}$ then $\cup_{n=1}^{\infty} A_n \in \mathcal{F}.$
%\end{itemize}

%Given a set $\Omega$ and a $\sigma$-algebra $\mathcal{F}$ we call $(\Omega, \mathcal{F})$ a \emph{measurable space}.

%Given a family of sets $\mathcal{G}$ in $\Omega$ we call the $\sigma$-algebra generated by $\mathcal{G}$ the intersection of all $\sigma$-algebras containing $\mathcal{G}$, which is a $\sigma$-algebra itself. 

We consider the measurable space $(\mathbb{R}^d, \mathcal{B}(\mathbb{R}^d)$ where $\mathcal{B}(\mathbb{R}^d)$ is the $\sigma$-algebra of Borel, defined as the $\sigma$-algebra generated by the family of the open hyper-rectangles in $\mathbb{R}^d$, i.e. by the family of set $R = (a_1, b_1) \times \ldots \times (a_d, b_d)$. We always assume it equipped with the standard Lebesgue measure $\lambda$ \cite{billingsley2013convergence}.

Given a random variable taking values in $(\mathbb{R}^d, \mathcal{B}(\mathbb{R}^d))$ it induces a probability measure on $\mathbb{R}^d$ given by:
$$ m_X(A) = P(X^{-1}(A)) \quad \forall \, A \in \mathcal{B}(\mathbb{R}^d).$$ We say that $X$ has probability density function $f$ if $m_X$ has density $f$ with respect to $\lambda$, i.e. if
$$ m_X(A) = \int_A f(x) dx \quad \forall A \in \mathcal{B}(\mathbb{R}^d).$$

\subsection{Degenerate Gaussians}
Consider the case of a $d$-dimensional Gaussian with mean $\mu$ and singular covariance matrix $\Sigma$. In this case, if the rank of the covariance matrix is $d_0$ such that $0<d_0<d$, we can consider the following density:
$$ f_{\mathcal{N}(\mu, \Sigma)} = \frac{1}{\sqrt{(2\pi)^{d_0}det^*(\Sigma)}} \exp{\left(-\frac{1}{2}(x-\mu)^T\Sigma^{+}(x-\mu)\right)} $$
where $det^*$ is the pseudo-determinant defined as ($\mathbb{I}$ is the identity matrix)
$$ det^*(\Sigma) = \lim_{\alpha \to 0} \frac{ det(\Sigma + \alpha \mathbb{I})}{\alpha^{d-d_0}} $$
and $\Sigma^+$ is the generalized inverse (also called Moore-Penrose pseudoinverse), defined as the matrix $\Sigma^+$ satisfying the following properties
\begin{align*}
\Sigma\Sigma^+\Sigma = \Sigma, \quad & \Sigma^+\Sigma\Sigma^+ = \Sigma^+,\\
(\Sigma\Sigma^+)^T = \Sigma\Sigma^+, \quad & (\Sigma^+\Sigma)^T = \Sigma^+\Sigma.
\end{align*} 
If the rank of the covariance matrix is 0, we interpret the Gaussian as a Dirac delta distribution centered in $\mu$. For further details we refer the reader to \cite{florescu2014probability}.

\subsection{Weak Convergence}

\begin{definition}[Weak Convergence] 
For a sequence of random vectors $X_n \sim D_n$ with cdfs $F_{D_n}$ we say that $D_n$ {\it converge weakly to} $D$, with $F$, if for every continuity point $x$ of $F$ (i.e. points for which $\lim_{y \to x} F(y) = F(x)$) it holds:
$$ \lim_n F_{D_n}(x) = F_D(x).$$
We denote weak convergence with $D_n \xrightarrow{n \to \infty} D$.
\end{definition} 
Equivalently we say that the corresponding measure converges weakly, denoted by $m_{D_n} \xrightarrow{n \to \infty} m_D$.
Interestingly, the space of distributions with the weak topology is metrizable, i.e. we can define a metric such that weak convergence is equivalent to convergence in the metric. This metric is the Levy-Prokhorov distance that for two measures $m, m'$ on $(\mathbb{R}^d, \mathcal{B}(\mathbb{R}^d))$ is defined as:
\begin{align}  \nonumber
d_{LP}(m,m') = &\inf\{\epsilon > 0 \, | \, m(A) \le m'(A^{\epsilon}) + \epsilon \text{ and } m'(A) \le m'(A^{\epsilon}) + \epsilon, \, \\
& \hspace{7cm} \forall \, A \in \mathcal{B}(\mathbb{R}^d) \} \label{eq:applp}
 \end{align}
where for $A \subseteq \mathbb{R}^d$, $A^{\epsilon} = \{ x \in \mathbb{R}^d \, | \, \exists \, y \in A \text{ s.t. } \| x - y \| \le \epsilon\}$.

\subsection{\emph{m}-continuity sets}

Let $B_\epsilon(x)$ be the ball of radius $\epsilon$ centered in $x$, defined as $ B_\epsilon(x) = \{ y \in \mathbb{R}^d \, | \, \| x - y \| < \epsilon \}. $
For a set $A \subseteq \mathbb{R}^d$ we define:
\begin{itemize}
\item[-] the \emph{interior} of $A$ as the set
$$int(A) = \{ x \in A \, | \, \exists \, \epsilon > 0 \text{ s. t. } B_\epsilon(x) \subset A \};$$
\item[-] the \emph{closure} of $A$ as the set
$$ \bar{A} = \{ x \in \mathbb{R}^d \, | \, \forall \,\epsilon > 0 \, B_{\epsilon}(x) \cap A \neq \emptyset \};$$
\item[-] the \emph{boundary} of $A$ as the set $\partial A = \bar{A} \setminus int(A).$
\end{itemize}

Given a measure $m$ on $(\mathbb{R}^d, \mathcal{B}(\mathbb{R}^d))$ and $A \in \mathcal{B}(\mathbb{R}^d)$ we say that $A$ is an $m$-continuity set if $m(\partial A)=0$.

\section{Reparametrizations}

See Figure \ref{fig:reparam}.

\begin{figure}
  \centering
  \begin{align*}
  x = Bernoulli(y) \quad &\rightarrow \begin{cases} z = Uniform(0,1) \\ if \, z < y \, \{x=1\} \, else \, \{x=0\} \end{cases} \\
  x = Normal(y, z) \quad &\rightarrow \quad x = y + z\cdot Normal(0,1) \\
  x = Uniform(y, z) \quad &\rightarrow \quad x = y + (z-y) \cdot Uniform(0,1) \\
  x = Laplace(y, c) \quad &\rightarrow \quad x = y + Laplace(0,c) \\
  x = Exponential(c/y) \quad &\rightarrow \quad x = y \cdot Exponential(c)
  \end{align*}
  \caption{Reparametrizations for transforming random assignments involving distributions depending on variable parameters $y,z$ into assignments only involving distributions with constant parameters.}
  \label{fig:reparam}
\end{figure}

\section{Auxiliary Proofs}

\begin{proposition}
\label{prop:appexist}
For any $r \in \mathbb{N}_0$, there exists an operator $\textbf{match}_r$ satisfying R1 and R2.
\end{proposition}

\proof
We need to show that for any distribution $D$ we are able to find a GM matching the first $r$-th order moments of $D$. To do this, consider a $d$-dimensional random variable $x \sim D$ and let $r \in \mathbb{N}_0$ be fixed. Let us define the set
$$ \mathcal{N}(r) = \left\{ \alpha = (\alpha_1, \ldots, \alpha_d) : \sum_{i=1}^d \alpha_i \le r \right\}$$
and the associated \emph{truncated moment sequence} $(s_\alpha)_{\alpha \in \mathcal{N}(r)}$, with $s_\alpha = \mathbb{E}[X^\alpha]$. 
By Theorem 17.2 in~\cite{schmudgen2017moment}, there exists a $C$-atomic positive measure (i.e. a discrete measure placing positive probability mass on $C$ points), with $C \le |\mathcal{N}(r)|$, such that 
\begin{equation} \label{eq:apprepmes}
    \int_{\mathbb{R}^d} x^\alpha dm = s_\alpha \, \forall \, \alpha \in \mathcal{N}.
\end{equation}
Since $\mathcal{N}(r)$ is finite for any $r$, for any truncated moment sequence $(s_\alpha)_{\alpha \in \mathcal{N}(r)}$ there exist a $C$-atomic measure $m_r$ satisfying \eqref{eq:apprepmes} for every $\alpha \in \mathcal{N}(r)$. Moreover, since $s_0 = 1$, $m_r$ is a probability measure, and therefore it is induced by a finite mixture of Dirac deltas. Since any finite mixture of Dirac deltas is a GM, the proof concludes.  
\endproof

\begin{proposition} \label{prop:appwelldef}
For any $D$ and $r$ the Criterion of Choice uniquely identifies $\textbf{match}_r(D)$.
\end{proposition}

\proof
By Proposition \ref{prop:appexist} $GM_r(D)$ is non-empty, moreover by Theorem 17.2 in \cite{schmudgen2017moment} there is at least one moment-matching mixture such that $C < |\mathcal{N}(r)|$, therefore $c^*$ is well-defined. Once $c^*$ is fixed, the set of parameters $\mathcal{P}$ satisfying the moments conditions is the set of solutions of a system of polynomial equations equating the moments of the mixture of $c^*$ components, expressed as functions of $p_i, \mu_i$ and $\Sigma_i$, to the moments of $D$ \cite{wang2015estimating}. Being the set of solutions of a polynomial system, $\mathcal{P}$ is closed. Moreover, since by Example 12.2.8 in \cite{cover1999elements} for fixed moments the entropy is bounded from above, $-H$ is bounded from below, and we can always choose $M > 0$ so that $\mathcal{P}^*$ is contained in $\mathcal{P}\cap \left([0,1]^{c^*}\times[0,M]^{dc^* + \frac{1}{2}d(d+1)c^*}\right)$. It follows that $\mathcal{P}^*$ is compact. Finally, the maximum with respect to the lexicographic ordering can be seen as maximising projections of the vector of parameters $P$ on different coordinates, in a given order. Since $\mathcal{P}^*$ is compact, the set of maximals with respect to the lexicographic ordering is non-empty, but since the lexicographic ordering is a total order the set of maximals can have only one element which is uniquely defined.
\endproof

\begin{lemma}
\label{lem:apporder2}
The following two properties hold:
\begin{itemize}
    \item[i)] when $r=2$, $\textbf{match}_2(D)$ is a single Gaussian variable with mean and covariance matrix equal to those of $D$;
    \item[ii)] if $D$ is Gaussian, for any $r \ge 2$ $\textbf{match}_r(D) = D.$
\end{itemize}
\end{lemma}

\proof The first point follows observing that, since we want to match the first two order moments, a single Gaussian variable can be used, so $c^*=1$. Moreover, since we have a single component with mean and covariance matrix fixed, the set $\mathcal{P}$ has a single set of parameters and our criterion of choice reduces to approximating $D$ with a Gaussian having the same mean and covariance matrix. 

On the other hand, if $D$ is Gaussian then for any $r \ge 2$ we always have $c^*=1$ and the set $\mathcal{P}$ has a single set of parameters, so $\textbf{match}_r(D) = D$. 
\endproof

\section{Proof of the Universal Approximation Theorem}

\begin{theorem} \label{thm:appconv}
Assume that $P=(V,E)$ is a program such that for each $v \in V$ and each path $\pi \in \Pi^P$ the output distribution $D$ of $\llbracket v \rrbracket_\pi$ satisfies the following:
\begin{itemize}
    \item[H1)] $D$ is determined by its moments;
    \item[H2)] if $D$ is the input distribution for a test or observe node $v'$, then the set defined by the LBC labelling $v'$ is an $m_{D_z}$-continuity set.
\end{itemize}

Then there exists a sequence of maps $(R^k: V \to \mathbb{N}_0)_{k \in \mathbb{N}}$ such that:
\begin{equation} \label{eq:appconv}
\llbracket P \rrbracket^{R^k} \xrightarrow{k \to \infty}  \llbracket P \rrbracket. \end{equation}
where the convergence is intended in the weak topology, or equivalently, in the Levy-Prokhorov metric.
\end{theorem}

\subsection{Preliminary Results}

The proof of the main theorem relies on two auxiliary results: first, we show that the exact semantics preserves weak convergence (Lemma~\ref{lem:app1}); second we prove that, given a weakly converging sequence of distributions $D_n$, it is always possible to choose a sequence of integers $r_n$ such that $T^{GM}_{r_n}(D_n)$ converges to the same limit (Lemma~\ref{lem:app2}). 

\begin{lemma} \label{lem:app1}
Let $P$ be a program, $\pi = (v_0, \ldots, v_n) \in \Pi^P$ and $v_i \in \pi$ be fixed.
Suppose $(p_n, D_n)$ is a sequence of pairs such that the following conditions are satisfied:
\begin{itemize}
    \item $0 \le p_n \le 1 \, \forall \, n$ and $p_n \xrightarrow{n \to \infty} p$ in $\mathbb{R}$;
    \item $D_n \xrightarrow{n \to \infty} D$;
    \item $D$ is determined by its moments;
    \item if $v_i \colon test$ or $v_i \colon observe$ and $v_i$ is labelled by an LBC defining the set $\mathtt{\llbracket B(z) \rrbracket}$, $\mathtt{\llbracket B(z) \rrbracket}$ is an $m_{D_z}$-continuity set;
    \item $\llbracket v_i \rrbracket_\pi(p, D) = (p', D')$ such that $p' \neq 0$.
\end{itemize}
Then $\llbracket v_i \rrbracket_\pi (p_n, D_n) \to \llbracket v_i \rrbracket_\pi(p, D).$
\end{lemma}

\proof

Let us consider separately the possible types of $v_i$.

If $v_i \colon entry, exit$ there is nothing to prove.

Suppose $v_i \colon state$. If $v_i$ is indexed by $\mathtt{skip}$ there is nothing to prove. If it is indexed by an assignment instruction $\mathtt{x_j=E(z)}$ the conclusion follows from the Mapping Theorem \cite[Theorem 29.2]{billingsley2013convergence} with $h:\mathbb{R}^n \to \mathbb{R}^d$ such that
$$[h(z)]_k = \begin{cases} x_k &\text{ if } k \neq j \\ E(z) &\text{ if } k=j. \end{cases}$$ 

Suppose $v_i \colon test$ and $cond(v_i) = true$ and set $\llbracket v_i \rrbracket_\pi (p_n, D_n) = (p_n', D_n')$ and $\llbracket v_i \rrbracket_\pi (p, D) = (p', D')$. By hypothesis $p' \neq 0$.
First observe that $p_n' = p_n \cdot P_{D_{n,z}}(\mathtt{\llbracket B(z) \rrbracket}) \to p \cdot P_{D_z}(\mathtt{\llbracket B(z) \rrbracket})$ because of Theorem 29.1 from~\cite{billingsley2013convergence} and the fact that by hypothesis the set $\mathtt{\llbracket B(z) \rrbracket}$ must be an $m_{D_z}$-continuity set. From this it follows that, starting from some $n_0 > 0$, $p_n' > 0, \, \forall \, n>n_0$. 

Let $z^* \in \mathbb{R}^n$ be such that $m_{D_{n,z}}(z^*)=0 \, \forall \, n$. %In particular, observe that if for each $n$ $D_{n,z}$ is a finite mixture of Gaussians, it is always possible to find such a point, since there is at most a countable number of Dirac deltas involved in the sequence. 
We can then define the map
$$ h(z) = \begin{cases} z & \text{ if } z \in \mathtt{\llbracket B(z) \rrbracket} \\
z^* & \text{ else. } \end{cases}$$

$h$ is $m_{D_z}$-measurable and its set of discontinuity points is given by $\partial\mathtt{\llbracket B(z) \rrbracket}$, so that $m_{D_z}(\partial\mathtt{\llbracket B(z) \rrbracket}) = 0$ because we are assuming that the sets $\mathtt{\llbracket B(z) \rrbracket}$ are $m_{D_z}$-continuity sets. So applying again the Mapping Theorem we have that
$$ \frac{1}{p_{n}'} m_{D_{n,z}} \circ h^{-1} \to \frac{1}{p'} m_{D_z} \circ h^{-1}.$$
Moreover, $\mathbb{R}^{n-d}$ is a continuity set for any measure (since it has no border), so when applying the operator $\mathit{Marg}_x$ the weak convergence is preserved. The conclusion follows observing that $\mathit{Marg}_x(\frac{1}{p_n'} m_{D_n} \circ h^{-1}) = m_{D_n'}$ and $\mathit{Marg}_x(\frac{1}{p'} m_{D} \circ h^{-1}) = m_{D'}$. Convergence for $cond(v_i) = false$ follows from the same argument. 

If $v_i \colon observe$, we can apply the same argument used for $v \colon test$.
\endproof 

\begin{lemma} \label{lem:app2}
Given a weakly converging sequence of distributions $D_n \to D$ for each $n \in \mathbb{N}$ it is possible to find  an integer $r_n \in \mathbb{N}_0$ such that 
$$ T^{GM}_{r_n}(D_n) \xrightarrow{n \to \infty} D.$$
\end{lemma}

\proof
Consider the space of distribution over $\mathbb{R}^d$ with the Levy-Prokhorov metric $d_{LP}$. Consider the family of sequences $T^{GM}_{r}(D_n)$ where $n, r \in \mathbb{N}$. We want to show that for each $n$ it is possible to fix $r_n$ such that 
$$ \forall \, \epsilon > 0 \, \exists \, n_0 \text{ s.t. } \forall \, n \ge n_0 \quad d_{LP}\left(T^{GM}_{r_n}(D_n), D\right) < \epsilon. $$
Let $\epsilon > 0$ be fixed and $\epsilon_n$ be a real sequence such that $\epsilon_n \to 0$ and $|\epsilon_n|< \frac{\epsilon}{2} \, \forall \, n.$
By Theorem 30.2 from \cite{billingsley2013convergence} $T^{GM}_r(D^n) \xrightarrow{r \to \infty} D_n \, \forall \, n$, so there exists $\bar{r}_n$ such that $\forall \, r > \bar{r}_n$ $d_{LP}\left(T^{GM}_r(D_n), D_n\right) < \epsilon_n.$
Moreover let $n_0$ be such that $\forall \, n > n_0$ $d_{LP}(D^n, D) < \frac{\epsilon}{2}$.
Then, for each $n$ we can choose $r_n > \bar{r}_n$ and we have that $\forall \, n > n_0$:
$$ d_{LP}\left(T^{GM}_{r_n}(D_n), D\right), < d_{LP}\left(T^{GM}_{r_n}(D_n), D_n\right) + d_{LP}(D_n, D) < \epsilon_n + \frac{\epsilon}{2} < \epsilon.$$
\endproof

\subsection{Proof of Theorem~\ref{thm:appconv}} 

We first prove that the theorem is true for programs $P$ such that $\forall \, \pi \in \Pi^P$ if $\llbracket \pi \rrbracket = (p', D')$ it holds $p' \neq 0$. Then we prove that, given this, the conclusion generalizes to any $P$ in the hypotheses of the theorem. 

Suppose that $P$ is such that $\forall \, \pi \in \Pi^P$ such that $\llbracket \pi \rrbracket = (p', D')$ it holds $p' \neq 0$. Let $\pi = v_0 \cdots v_n \in \Pi^P$ be fixed such that $\llbracket \pi \rrbracket = (p', D')$ with $p' \neq 0$. We want to prove that:
\begin{equation} \label{eq:apppconv} \llbracket P \rrbracket^{R^k} = \sum_{\substack{(p_k,D_k) = \llbracket \pi \rrbracket^{R^k} \\ \pi \in \Pi^P}} \frac{p_k}{\displaystyle{\sum_{\substack{(p'_k,D'_k) = \llbracket \pi \rrbracket^{R^k} \\ \pi \in \Pi^P}} p'_k}}D_k \to \sum_{\substack{(p,D) = \llbracket \pi \rrbracket :\\ \pi \in \Pi^P}} \frac{p}{\displaystyle{\sum_{\substack{(p',D') = \llbracket \pi \rrbracket \\ \pi \in \Pi^P}} p'}}D = \llbracket P \rrbracket. \end{equation}

Observe that since $| \Pi^P | < \infty$ this is implied by:
\begin{equation} \label{eq:apppiconv} \llbracket \pi \rrbracket^{R^k} \to \llbracket \pi \rrbracket. \end{equation}

By definition of path semantics we can prove \eqref{eq:apppiconv} by showing that
for every $i = 0, \ldots, n$ it is possible to choose $R^k(v_i) \, \forall \, k \in \mathbb{N}$ such that the output of $\llbracket v_i \rrbracket^{R^k}_\pi$ converges to the output $\llbracket v_i \rrbracket_\pi$.

For $i=0$ we can set $R^k(v_0)$ to any value, as this does not affect the final distribution. In fact $\llbracket v_0 \rrbracket^{R^k}_\pi = \llbracket v_0 \rrbracket_\pi = (1, \delta_0).$

For $i=1$ if $v_1:exit$ there is nothing to prove.
If not we set $R^k(v_1) = k$ and let $(p_0', D_0')$ be $\llbracket v_1 \rrbracket_\pi(1, \delta_0)$. Then:
$$ \llbracket v_1 \rrbracket^{R^k}_\pi(1,\delta_0) = \left(\mathbb{I}, T^{GM}_k\right) \circ \llbracket v_1 \rrbracket_\pi(1,\delta_0) = \left(\mathbb{I}, T^{GM}_k\right)(p_0', D_0') = \left(p_0', T^{GM}_k(D_0')\right)$$
and by Theorem 30.2 of \cite{billingsley2013convergence} $T^{GM}_k(D_0') \to D_0'$.

So we have proved that the statement holds for $i=0$ and $i=1$. Now suppose that it holds for some $i > 1$ and let us prove that it holds for $i+1$.

If $v_{i+1} \colon exit$ there is nothing to prove. If not let $(p_k, D_k)$ be output of $\llbracket v_i \rrbracket^{R^k}_\pi$. By inductive hypothesis $(p_k, D_k) \to (\bar{p}, \bar{D}) = \llbracket v_i \rrbracket_\pi$. Then:
$$ \llbracket v_{i+1} \rrbracket^{R^k}_\pi (p_k, D_k) = \left(\mathbb{I}, T^{GM}_{R^k(v_{i+1})}\right) \circ \llbracket v_{i+1} \rrbracket_\pi (p_k, D_k).$$

Let $(p_k', D_k')$ be $\llbracket v_{i+1} \rrbracket_\pi (p_k, D_k)$. By hypothesis $p_k' \neq 0$ (or $\llbracket \pi \rrbracket(D) = (0, D')$) so we can apply Lemma \ref{lem:app1} to get $(p_k', D_k') \to \llbracket v_{i+1} \rrbracket_\pi(\bar{p}, \bar{D}) = (\bar{p}', \bar{D}')$, so 
$$ \llbracket v_{i+1} \rrbracket^{R^k}_\pi (p_k, D_k) = \left(\mathbb{I}, T^{GM}_{R^k(v_{i+1})}\right)(p_k', D_k') = \left(p_k',T^{GM}_{R^k(v_{i+1})}(D_k')\right).$$
Then, by Lemma \ref{lem:app2} we can choose a sequence of integers $s_k$ such that setting $R^k(v_{i+1}) = s_k$ $T^{GM}_{R^k(v_{i+1})}(D_k') \to \bar{D}'$.
Thus, we have set $R^k(v_{i+1}) = s_k$ so that $\llbracket v_{i+1} \rrbracket^{R^k}_\pi (p_k, D_k) \to \llbracket v_{i+1} \rrbracket_\pi (\bar{p}, \bar{D}).$

Now suppose that for $\llbracket \pi \rrbracket = (0, D')$ for some $\pi$, we want to prove that \eqref{eq:apppconv} still holds. In this case the path $\pi$ does not contribute to the output distribution of computed by $\llbracket P \rrbracket(D)$. Moreover, there exist $i$ such that at $v_i$ the output pair is $(0, D'')$ while for all $j < i$ the output at $v_j$ is $(p^{(j)}, D^{(j)})$ with $p^{(j)} \neq 0$. The statement then holds up to node $v_i$, that takes in input a sequence $(p_k, D_k) \to (p'', D'')$.  Letting $ \llbracket v_{i} \rrbracket^{R^k}_\pi (p_k, D_k) = (p_k', D_k')$ and using the same argument as in the proof of Lemma \ref{lem:app1} we can prove $p_k' \to 0$. So \eqref{eq:apppconv} will hold even if \eqref{eq:apppiconv} does not.
\endproof

\section{Lexicographic Ordering for GMs}
\label{app:lexordering}

Consider a Gaussian mixture $p_1D_1 + \hdots + p_CD_C$ where $D_i \sim \mathcal{N}(\mu_i, \Sigma_i)$, $i = 1, \ldots, C$. Since it is uniquely identified by its parameters we can order them in a vector  $(p_i, \mu_i, \Sigma_i)_{i=1,\ldots,C}$  in the following way:
\begin{itemize}
    \item $p_1 \ge p_2 \ge \ldots \ge p_{C}$;
    \item if $p_i = p_{i+1}$ then either of the following two conditions holds:
    \begin{itemize}
        \item there exists $j \in \{1, \ldots,d\}$ such that $\mu_i(s) = \mu_{i+1}(s) \, \forall \, s < j$ and $\mu_i(j) > \mu_{i+1}(j)$ (means are ordered according to the lexicographic order);
        \item if $\mu_i = \mu_{i+1}$ there exists $j \in \{1, \ldots, d^2\}$ such that
        $ \Sigma_i(s) = \Sigma_{i+1}(s) \, \forall \, s < j$ and $\Sigma_i(j) > \Sigma_{i+1}(j)$, where $\Sigma$ is converted into a vector using lexicographic ordering, i.e. $$\Sigma = (\Sigma(0,0), \ldots, \Sigma(0, c^*), \Sigma(1,0), \dots, \Sigma(1,c^*), \ldots, \Sigma(c^*,c^*))$$
    \end{itemize}
\end{itemize}
This procedure allows us to consider a set of parameters $P$ as a single vector 
\begin{align*}
     P &= (p_1, p_2, \ldots, p_{C}, \mu_1(0), \mu_1(1), \ldots, \mu_{C}(d), \\
     & \hspace{1cm} \Sigma_1(0,0), \ldots , \Sigma_1(d,d), \ldots, \Sigma_{C}(0,0), \ldots, \Sigma_{C}(d,d)).
\end{align*}
For two set of parameters $P_1, P_2$ we say that $P_1 \succ P_2$ if $P_1$ is greater then $P_2$ according to the lexicographic ordering, i.e. if exists $i$ such that
$P_1(j) = P_2(j) \, \forall j < i$ and $P_1(i) > P_2(i)$.
Observe that $\succ$ is a total order, i.e. if $P_1 \neq P_2$ necessarily either $P_1 \succ P_2$ or $P_2 \succ P_1$.

\section{SOGA implementation}

We assume that each node in the control-flow graph has two attribute lists of \textit{children} and \textit{parents}, whose elements point, respectively, to children and parent nodes. Furthermore, each node has two attributes, $p$ and $dist$: $p$ is a non-negative scalar proportional to the probability of reaching that node, while $dist$ stores the output distribution (in the form of a GM) computed by that semantics of the node. Nodes have type-specific attributes: nodes of type \textit{test} and \textit{observe} have an attribute \textit{LBC} storing an LBC expression; nodes of type \textit{state} have an attribute \textit{cond} taking value \textit{true, false} or \textit{none} and an attribute \textit{expr} storing an assignment expression. 

To apply SOGA we create a queue $visit\_queue$ containing the \textit{entry} node. Then we apply iteratively SOGA on pop($visit\_queue$). 
When called on a new node, the algorithm first accesses the attributes $p$ and $dist$ of its parents, and invokes \textit{merge\_dist} on the list of pairs $(p,D)$. Then, computes the semantics corresponding to the node type as follows:
\begin{itemize}
    \item if $v \colon entry$, it initializes $node.p$ to 1, $node.dist$ to $\delta_0$ and $node.trunc$ to $none$;

    \item if $v \colon observe$, it saves the LBC in $node.trunc$, and calls the function \textit{approx\_trunc};
    
    \item if $v \colon test$, it does nothing;
    
    \item if $v \colon state$, it checks if $node.cond = true$ or $node.cond = false$ and in that case retrieves the LBC condition from the parent $test$ node. Then it calls the function \textit{approx\_trunc}. This results in a new pair $(p, dist)$ on which the function \textit{apply\_rule} is applied. Finally, the output is stored in $node.p$, $node.dist$;

    \item if $v \colon exit$, after merging the resulting distribution is returned as the approximated output distribution of the whole program.
\end{itemize}

After executing the semantics of the node the queue is updated, by pushing the children nodes of the current node. This is detailed in Algorithms~\ref{alg:sogaentry}-\ref{alg:sogaexit}.

\begin{algorithm}[]
\caption{\textsl{SOGA}($node\colon entry$):} 
\label{alg:sogaentry}
\begin{algorithmic}
    \STATE $node.p = 1$;
    \STATE $node.dist = [d \cdot gm([1], [0], [0])]$;
    \STATE $node.trunc = none$
    \FOR{$child$ in $node.children$}
        \STATE push($visit\_queue$, $child$)
    \ENDFOR
\end{algorithmic}
\end{algorithm}

\begin{algorithm}[]
\caption{\textsl{SOGA}($node\colon observe$):} 
\label{alg:sogaobserve}
\begin{algorithmic}
    \STATE $input\_list = []$
    \FOR{$par$ \textbf{in} $node.parents$}
    \STATE $input\_list.append((par.p,par.dist))$
    \ENDFOR
    \STATE $node.p, node.dist =$ merge\_dist($input\_list)$
    \STATE $node.trunc = node.LBC$
    \STATE $I, node.dist =$ approx\_trunc($node.dist, node.trunc$)
    \STATE $node.p = node.p \cdot I$
    \FOR{$child$ in $node.children$}
        \STATE push($visit\_queue$, $child$)
    \ENDFOR
\end{algorithmic}
\end{algorithm}

\begin{algorithm}[]
\caption{\textsl{SOGA}($node \colon test$):} 
\label{alg:sogatest}
\begin{algorithmic}
    \FOR{$par$ \textbf{in} $node.parents$}
    \STATE $input\_list.append((par.p,par.dist))$
    \ENDFOR
    \STATE $node.p, node.dist =$ merge\_dist($input\_list)$
    \FOR{$child$ in $node.children$}
        \STATE push($visit\_queue$, $child$)
    \ENDFOR
\end{algorithmic}
\end{algorithm}

\begin{algorithm}[H]
\caption{\textsl{SOGA}($node\colon state$):} 
\label{alg:sogastate}
\begin{algorithmic}
    \FOR{$par$ \textbf{in} $node.parents$}
    \STATE $input\_list.append((par.p,par.dist))$
    \ENDFOR
    \STATE $node.p, node.dist =$ merge\_dist($input\_list)$
    \NoThen \IF{$node.cond == true$}
        \STATE $ trunc = node.parent.LBC$
    \ELSIF{$node.cond == false$}
    \STATE $ trunc = \textbf{not } node.parent.LBC$
    \ENDIF
    \NoThen \IF{$node.cond != none$}
    \STATE $p', node.dist = \text{approx\_trunc}(node.dist, trunc)$
    \STATE $node.p = node.p \cdot p'$
    \ENDIF
    \STATE $node.dist =$ apply\_rule($node.dist, node.expr$)
    \FOR{$child$ in $node.children$}
        \STATE push($visit\_queue$, $child$)
    \ENDFOR
\end{algorithmic}
\end{algorithm}

\begin{algorithm}[]
\caption{\textsl{SOGA}($node\colon exit, p, dist, trunc$):} 
\label{alg:sogaexit}
\begin{algorithmic}
    \FOR{$par$ \textbf{in} $node.parents$}
    \STATE $input\_list.append((par.p,par.dist))$
    \ENDFOR
    \STATE $node.p, node.dist =$ merge\_dist($input\_list)$
    \STATE \textbf{return} $node.dist$
\end{algorithmic}
\end{algorithm}

\section{Computational Cost for Moments of Truncated Gaussians}
\label{app:cost}

We derive the computational cost of computing the first two order central moments of a $d$-dimensional Gaussian distribution truncated to an hyper-rectangle $[\underline{a}, \underline{b}] = [a_1, b_1] \times \hdots \times [a_d, b_d]$ in the special case in which $a_1 > -\infty$, $a_i = -\infty \, \forall \, i=2, \ldots, d$ and $b = \infty \, \forall \, i = 1, \ldots, d$. Observe that this case and the symmetric one with $b_1 < \infty$ are the only ones arising in the execution of SOGA, due to the fact that we restrict conditional branches to have the form in (4). To carry out the computation we use the recursive formulas from \cite{kan2017moments} reported below.

Let $\underline{r} = (r_1, \hdots, r_d) \in \mathbb{N}^d_0$. We define:
$$ F^d_{\underline{r}}(\underline{a},\underline{b},\mu,\Sigma) = \int_{[\underline{a},\underline{b}]} x^{\underline{r}}f_{\mathcal{N}(\mu, \Sigma)}(x) dx.$$
If $X$ is a Gaussian with mean $\mu$ and covariance matrix $\Sigma$ truncated to $[\underline{a}, \underline{b}]$ we have that $$\mathbb{E}[X^{\underline{r}}] = \frac{F^d_{\underline{r}}(\underline{a},\underline{b},\mu,\Sigma)}{F^d_{\underline{0}}(\underline{a},\underline{b},\mu,\Sigma)}$$
so, if we compute $F^d_{\underline{r}}$ for all $\underline{r}$ such that $\sum_{i=1}^d r_i \le 2$ we can retrieve the first two order moments of the truncated Gaussian in $O(d^2)$ operations. 

Observe that due to the particular form of our hyper-rectangles $F^d_{\underline{0}}$ can be computed in costant time, as if $d=1$.

To compute $F^d_{\underline{r}}$ for other value of $\underline{r}$ we use the recursive formula:
\begin{equation} \label{eq:appkan} F^d_{\underline{r}+e_i}(\underline{a},\underline{b},\mu,\Sigma) = \mu_iF^d_{\underline{r}}(\underline{a},\underline{b},\mu,\Sigma) + e_i^T\Sigma c_{\underline{r}}
\end{equation}
where 
\begin{align}
    \label{eq:appkanc}
    c_{\underline{r},j} & = k_j F^d_{\underline{r}-e_j}(\underline{a},\underline{b},\mu,\Sigma) + a_j^{k_j}f_{\mathcal{N}(\mu_j,\Sigma_{j,j})}(a_j)F^{d-1}_{\underline{r}_{(j)}}(\underline{a}_{(j)}, \underline{b}_{(j)}, \tilde{\mu}^a_j, \tilde{\Sigma}_j) \\
    \label{eq:appkanmu}
    \tilde{\mu}^a_j & = \mu_{(j)} + \Sigma_{(j),j}\frac{a_j-\mu_j}{\Sigma_{j,j}} \\
    \label{eq:appkansigma}
    \tilde{\Sigma}_j & = \Sigma_{(j),(j)} - \frac{1}{\Sigma_{j,j}}\Sigma_{(j),j}\Sigma_{j,(j)}
\end{align}
and for a vector $\underline{v}$ the notation $\underline{v}_{(j)}$ denotes the vector obtained from $\underline{v}$ suppressing the index $j$. Moreover, it is understood that when $a_j = -\infty$ the second term at the right hand side of \eqref{eq:appkanc} is 0.

To compute moments of order 1, i.e. $F^d_{e_i}(\underline{a},\underline{b}, \mu, \Sigma)$ for $i=1, \ldots, d$, we set $\underline{r}=\underline{0}$ in \eqref{eq:appkan}. We first compute $c_{\underline{0}}$ for which we have
$$c_{\underline{0},1} = f_{\mathcal{N}(\mu_i,\Sigma_{1,1})}(a_1)F^{d-1}_{\underline{r}_{(1)}}(\underline{a}_{(1)},\underline{b}_{(1)},\tilde{\mu}^a_1, \tilde{\Sigma}_1) = f_{\mathcal{N}(\mu_i,\Sigma_{1,1})}(a_1)$$
since $[\underline{a}_{(1)}, \underline{b}_{(1)}] = \mathbb{R}^{d-1}$ and
$$c_{\underline{0},j} = 0.$$
Therefore $c_{\underline{0}}$ is computed in constant time and the only computational cost in computing the first order moments is due to the $(d \times d)\cdot(d \times 1)$ matrix multiplication $e_i^T\Sigma c_{\underline{0}}$, which is $O(d^2)$. Since we need to compute $d$ first order moments, the total cost is $O(d^3)$.

To compute moment of order 2, we set $\underline{r} = e_s$ and compute $F^d_{e_s + e_i}(\underline{a},\underline{b},\mu,\Sigma)$ as $s,i = 1, \ldots, d$. We first need to compute $c_{e_s}$ for which we have:
\begin{equation} \label{eq:appkanc2}
c_{e_s,1} = \begin{cases} F^d_{\underline{0}}(\underline{a}, \underline{b}, \mu, \Sigma) + a_1 f_{\mathcal{N}(\mu_1, \Sigma_{1,1})}(a_1) & \text{ if } s=0 \\
f_{\mathcal{N}(\mu_1, \Sigma_{1,1})}(a_1)F^{d-1}_{e_{s_{(1)}}}(\underline{a}_{(1)}, \underline{b}_{(1)}, \tilde{\mu}^a_1, \tilde{\Sigma}_1) & \text{ if } s \neq 0. \end{cases}
\end{equation}
\begin{equation*} 
c_{e_s,j} = \begin{cases} F^d_{\underline{0}}(\underline{a}, \underline{b}, \mu, \Sigma) & \text{ if } s=j \\
0 & \text{ if } s \neq j. \end{cases}
\end{equation*}
As $s = 2, \hdots, d$ we need to compute $F^{d-1}_{e_{s_{(1)}}}(\underline{a}_{(1)}, \underline{b}_{(1)}, \tilde{\mu}^a_1, \tilde{\Sigma}_1)$. However since this are the first order moments of a gaussian with mean $\mu^a_1$ in  $[\underline{a}_{(1)}, \underline{b}_{(1)}] = \mathbb{R}^{d-1}$, computing the previous quantity amounts to computing $\tilde{\mu}^a_1$ which can be done in $O(d)$ operations. Once this is done, $c_{e_s,j}$ can be computed for every $s$ and $j$ in $O(d^2)$ operations. Finally, we need to perform again the matrix multiplication $e_1^T \Sigma c_{e_s}$, this time for $\frac{d(d-1)}{2}$ times, for a total computational cost of $O(d^4)$.

\section{Additional experimental results}

\subsection{Variational Inference}

See Table \ref{tab:appvi}.

\subsection{MAP Estimation}

See Table \ref{tab:appmap}.

\subsection{Variational Inference for Collaborative Filtering}

See Table \ref{tab:appcf}.

\begin{table}[H]
    \centering
    \resizebox{\textwidth}{!}{\begin{tabular}{cccccccc}
    \toprule 
    & \multicolumn{4}{c}{Pyro (VI)} & \multicolumn{2}{c}{SOGA} & \multirow{2}{*}{True value} \\
    \cmidrule(l){2-5} \cmidrule(l){6-7}  
    Model & l.r. & $\#\_$steps & value & time & value & time &  \\
    \midrule
    Bernoulli & 0.01 & 2000 & 0.247 & 4.51 & \multirow{3}{*}{0.252} & \multirow{3}{*}{1.28} & \multirow{3}{*}{0.25} \\
    & 0.005 & 500 & 0.282 & 1.438 & & & \\
    & 0.001 & 1700 & 0.306 & 4.44 & & & \\
    \midrule
    BayesPointMachine & 0.01 & 5800 & 0.046 & 60.49 & \multirow{3}{*}{0.011} & \multirow{3}{*}{2.20} & \multirow{3}{*}{0.056*} \\
    & 0.005 & \multicolumn{3}{c}{not converged} \\
    & 0.001 & \multicolumn{3}{c}{not converged} \\
    \midrule
    ClickGraph & 0.01 & 200 & 0.566 & 3.13 & \multirow{3}{*}{0.63} & \multirow{3}{*}{208} & \multirow{3}{*}{0.614} \\
    & 0.005 & 200 & 0.504 & 3.03 \\
    & 0.001 & 400 & 0.490 & 6.36 \\
    \midrule
    CoinBias & 0.01 & 200 & 0.419 & 0.91 & \multirow{3}{*}{0.41} & \multirow{3}{*}{0.61} & \multirow{3}{*}{0.41} \\
    & 0.005 & 1100 & 0.425 & 5.233 \\
    & 0.001 & 900 & 0.403 & 4.36 \\
    \midrule
    SurveyUnbias & 0.01 & 500 & 0.770 & 2.89 & \multirow{3}{*}{0.80} & \multirow{3}{*}{1.56} & \multirow{3}{*}{0.80} \\
    & 0.005 & 800 & 0.743 & 4.42 \\
    & 0.001 & 900 & 0.701 & 5.02 \\
    \midrule
    TrueSkills & 0.01 & 200 & 101.4 & 1.30 & \multirow{3}{*}{104.7} & \multirow{3}{*}{0.05} & \multirow{3}{*}{104*} \\
    & 0.005 & 200 & 100.79 & 1.31 \\
    & 0.001 & 200 & 100.16 & 1.48 \\
    \midrule
    Altermu & 0.01 & \multicolumn{3}{c}{not converged} & \multirow{3}{*}{0.000} & \multirow{3}{*}{0.16} & \multirow{3}{*}{0*} \\
    & 0.005 & 1400 & 0.030 & 33.1 \\
    & 0.001 & \multicolumn{3}{c}{not converged} \\
    \midrule
    Altermu2 & 0.01 & 1300 & 8.713 & 29.52  & \multirow{3}{*}{0.156} & \multirow{3}{*}{0.36} & \multirow{3}{*}{0.155} \\
    & 0.005 & 5700 & -9.624 & 150.39 \\
    & 0.001 & 200 & 0.098 & 5.50 \\
    \midrule
    NormalMixtures & 0.01 & 400 & 0.344 & 28.80 & \multirow{3}{*}{0.298} & \multirow{3}{*}{50.4} & \multirow{3}{*}{0.286*} \\
    & 0.005 & 1700 & 0.295 & 104.89 \\
    & 0.001 & 300 & 0.500 & 17.77 \\
    \midrule
    TimeSeries & 0.01 & 2800 & -1.832 & 59.128 & \multirow{3}{*}{-1.590} & \multirow{3}{*}{3.79} & \multirow{3}{*}{-1.600*} \\
    & 0.005 & 900 & -2.257 & 19.87 \\
    & 0.001 & 1100 & -1.701 & 26.15 \\
    \bottomrule    
    \end{tabular}}
    \caption{Comparison between Pyro's Variational Inference, SOGA and true values of the models of Table 3 with continuous posterior. For Pyro's VI we report the values and the runtimes for 3 different learning rates (l.r.), together with the number of steps needed to meet our stopping criterion ($\#\_$steps). By 'not converged', we mean that the stopping criterion was not met after 10k steps of gradient descent. True values, are obtained using PSI or, then not available, using STAN or AQUA, denoted by $*$.}
    \label{tab:appvi}
\end{table}

\begin{table}[ht]
    \centering
    \resizebox{\textwidth}{!}{\begin{tabular}{cccccccc}
    \toprule
    & \multicolumn{4}{c}{\emph{Pyro}} & \multicolumn{2}{c}{\emph{SOGA}} & \multirow{2}{*}{\emph{True value}} \\
    \cmidrule(l){2-5} \cmidrule(l){6-7}  
    \emph{Model} & \emph{l.r.} & \emph{$\#\_$steps} &\emph{value} & \emph{time} & \emph{value} & \emph{time} \\
    \midrule
    \textsl{Bernoulli} & 0.01 & 200 & 0.200 & 0.20 & \multirow{3}{*}{0.220} & \multirow{3}{*}{11.97} & \multirow{3}{*}{0.200} \\
    & 0.005 & 200 & 0.200 & 0.20 \\
    & 0.001 & 800 & 0.200 & 0.83 \\
    \midrule
    \textsl{Bernoulli (P)} & 0.01 & 200 & 0.200 & 0.20 & \multirow{3}{*}{0.290} & \multirow{3}{*}{1.28} & \multirow{3}{*}{0.200} \\
    & 0.005 & 200 & 0.200 & 0.20 \\
    & 0.001 & 800 & 0.200 & 0.83 \\
    \midrule
    \textsl{BayesPointMachine} & 0.01 & 900 & 0.000 & 7.83 & \multirow{3}{*}{0.011} & \multirow{3}{*}{2.20} & \multirow{3}{*}{0.032 $\pm$ 0.002*}\\
    & 0.005 & \multicolumn{3}{c}{not converged} \\
    & 0.001 & \multicolumn{3}{c}{not converged} \\
    \midrule
    \textsl{ClickGraph} & 0.01 & 700 & 0.417 & 9.51 & \multirow{3}{*}{0.861} & \multirow{3}{*}{208} & \multirow{3}{*}{1.000} \\
    & 0.005 & 300 & 0.490 & 4.23 \\
    & 0.001 & 200 & 0.501 & 2.98 \\
    \midrule
    \textsl{CoinBias} & 0.01 & 200 & 0.400 & 0.63 & \multirow{3}{*}{0.493} & \multirow{3}{*}{0.61} & \multirow{3}{*}{0.400} \\
    & 0.005 & 200 & 0.400 & 0.65 \\
    & 0.001 & 700 & 0.398 & 2.32 \\
    \midrule
    \textsl{SurveyUnbias} & 0.01 & 700 & 0.964 & 3.46 & \multirow{3}{*}{0.755} & \multirow{3}{*}{1.56} & \multirow{3}{*}{1.000}\\
    & 0.005 & 1000 & 0.943 & 4.88 \\
    & 0.001 & 2200 & 0.847 & 11.53 \\
    \midrule
    \textsl{TrueSkills} & 0.01 & 200 & 101.6 & 0.99 & \multirow{3}{*}{104.7} & \multirow{3}{*}{0.05} & \multirow{3}{*}{104.8 $\pm$ 0.681*} \\
    & 0.005 & 200 & 100.8 & 0.97 \\
    & 0.001 & 200 & 100.2 & 1.03 \\
    \midrule
    \textsl{Altermu} & 0.01 & \multicolumn{3}{c}{not converged} &\multirow{3}{*}{ 0.000} & \multirow{3}{*}{0.16} & \multirow{3}{*}{0.114 $\pm$ 0.092*}  \\
    & 0.005 & \multicolumn{3}{c}{not converged} \\
    & 0.001 & \multicolumn{3}{c}{not converged} \\
    \midrule
    \textsl{NormalMixtures (P)} & 0.01 &  1100 & 0.236 & 49.48 & \multirow{3}{*}{0.276} & \multirow{3}{*}{50.4} & \multirow{3}{*}{0.275 $\pm$ 0.005*}\\
    & 0.005 & 300 & 0.477 & 14.98 \\
    & 0.001 & 200 & 0.500 & 9.72 \\
    \midrule
    \textsl{TimeSeries} & 0.01 & 3100 & -1.564 & 55.37 & \multirow{3}{*}{-1.494} & \multirow{3}{*}{3.79} & \multirow{3}{*}{-1.604 $\pm$ 0.021*} \\
    & 0.005 & 1100 & -1.497 & 26.64 \\
    & 0.001 & 2800 & -1.347 & 2800\\
    \bottomrule
    \end{tabular}}
    \caption{Comparison between Pyro and SOGA for MAP estimation. Models with `(P)' were pruned when SOGA was applied. True values are derived optimizing the exact posterior, or from samples (denoted with `*').}
    \label{tab:appmap}
\end{table}

\begin{table}[ht]
    \centering
    \begin{tabular}{ccccccc}
    \toprule
    & & \multicolumn{2}{c}{\emph{SOGA}} & \multicolumn{3}{c}{\emph{Pyro}} \\
    \cmidrule(l){3-4} \cmidrule(l){5-7}  
    \emph{k} & \emph{Ground truth} & \emph{time} &\emph{value} & \emph{l.r.} & \emph{time} & \emph{value} \\
    \midrule
    \multirow{3}{*}{1} & \multirow{3}{*}{2} & \multirow{3}{*}{0.16} & \multirow{3}{*}{1.86} & 0.01 & 12.86 & 1.79 \\
    & & & & 0.005 & 18.90 & 1.84 \\
    & & & & 0.001 & 20.21 & 0.51 \\
    \midrule
    \multirow{3}{*}{2} & \multirow{3}{*}{25} & \multirow{3}{*}{0.18} & \multirow{3}{*}{24.28} & 0.01 & 14.07 & 23.49 \\
    & & & & 0.005 & 25.10 & 23.93\\
    & & & & 0.001 & 30.49 & 6.75 \\
    \midrule
    \multirow{3}{*}{3} & \multirow{3}{*}{-5} & \multirow{3}{*}{0.19} & \multirow{3}{*}{-5.79} & 0.01 & 15.72 & -5.51 \\
    & & & & 0.005 & 26.40 & -5.82 \\
    & & & & 0.001 & 30.61 & -1.86 \\
    \midrule
    \multirow{3}{*}{5} & \multirow{3}{*}{-30} & \multirow{3}{*}{0.22} & \multirow{3}{*}{-31.98} & 0.01 & 18.49 & -32.21 \\
    & & & & 0.005 & 23.19 & -31.47 \\
    & & & & 0.001 & 31.49 & -6.21 \\
    \midrule
    \multirow{3}{*}{10} & \multirow{3}{*}{151} & \multirow{3}{*}{0.30} & \multirow{3}{*}{149.75} & 0.01 & 11.16 & 148.92 \\
    & & & & 0.005 & 20.04 & 146.39 \\
    & & & & 0.001 & 30.71 & 9.83 \\
    \midrule
    \multirow{3}{*}{20} & \multirow{3}{*}{70} & \multirow{3}{*}{0.60} & \multirow{3}{*}{73.76} & 0.01 & 10.94 & 68.51 \\
    & & & & 0.005 & 23.18 & 69.92 \\
    & & & & 0.001 & 30.73 & 6.87 \\
    \bottomrule
    \end{tabular}
    \caption{Comparison between Pyro and SOGA for Variational Inference on Collaborative Filtering models.}
    \label{tab:appcf}
\end{table}

%% file: main.bbl

\begin{thebibliography}{70}

%%% the Web address.

\ifx \showCODEN    \undefined \def \showCODEN     #1{\unskip}     \fi
\ifx \showDOI      \undefined \def \showDOI       #1{#1}\fi
\ifx \showISBNx    \undefined \def \showISBNx     #1{\unskip}     \fi
\ifx \showISBNxiii \undefined \def \showISBNxiii  #1{\unskip}     \fi
\ifx \showISSN     \undefined \def \showISSN      #1{\unskip}     \fi
\ifx \showLCCN     \undefined \def \showLCCN      #1{\unskip}     \fi
\ifx \shownote     \undefined \def \shownote      #1{#1}          \fi
\ifx \showarticletitle \undefined \def \showarticletitle #1{#1}   \fi
\ifx \showURL      \undefined \def \showURL       {\relax}        \fi
% invisible to TeX
\providecommand\bibfield[2]{#2}
\providecommand\bibinfo[2]{#2}
\providecommand\natexlab[1]{#1}
\providecommand\showeprint[2][]{arXiv:#2}

\bibitem[Albarghouthi et~al\mbox{.}(2017)]%
        {albarghouthi2017fairsquare}
\bibfield{author}{\bibinfo{person}{Aws Albarghouthi}, \bibinfo{person}{Loris D'Antoni}, \bibinfo{person}{Samuel Drews}, {and} \bibinfo{person}{Aditya~V Nori}.} \bibinfo{year}{2017}\natexlab{}.
\newblock \showarticletitle{Fairsquare: probabilistic verification of program fairness}.
\newblock \bibinfo{journal}{\emph{Proceedings of the ACM on Programming Languages}} \bibinfo{volume}{1}, \bibinfo{number}{OOPSLA} (\bibinfo{year}{2017}), \bibinfo{pages}{1--30}.
\newblock


\bibitem[Barthe et~al\mbox{.}(2016)]%
        {barthe2016synthesizing}
\bibfield{author}{\bibinfo{person}{Gilles Barthe}, \bibinfo{person}{Thomas Espitau}, \bibinfo{person}{Luis~Mar{\'\i}a Ferrer~Fioriti}, {and} \bibinfo{person}{Justin Hsu}.} \bibinfo{year}{2016}\natexlab{}.
\newblock \showarticletitle{Synthesizing probabilistic invariants via Doob’s decomposition}. In \bibinfo{booktitle}{\emph{International Conference on Computer Aided Verification}}. Springer, \bibinfo{pages}{43--61}.
\newblock


\bibitem[Bartocci et~al\mbox{.}(2020)]%
        {bartocci2020mora}
\bibfield{author}{\bibinfo{person}{Ezio Bartocci}, \bibinfo{person}{Laura Kov{\'a}cs}, {and} \bibinfo{person}{Miroslav Stankovi{\v{c}}}.} \bibinfo{year}{2020}\natexlab{}.
\newblock \showarticletitle{Mora-automatic generation of moment-based invariants}. In \bibinfo{booktitle}{\emph{International Conference on Tools and Algorithms for the Construction and Analysis of Systems}}. Springer, \bibinfo{pages}{492--498}.
\newblock


\bibitem[Billingsley(2008)]%
        {billingsley2008probability}
\bibfield{author}{\bibinfo{person}{Patrick Billingsley}.} \bibinfo{year}{2008}\natexlab{}.
\newblock \bibinfo{booktitle}{\emph{Probability and measure}}.
\newblock \bibinfo{publisher}{John Wiley \& Sons}.
\newblock


\bibitem[Billingsley(2013)]%
        {billingsley2013convergence}
\bibfield{author}{\bibinfo{person}{Patrick Billingsley}.} \bibinfo{year}{2013}\natexlab{}.
\newblock \bibinfo{booktitle}{\emph{Convergence of probability measures}}.
\newblock \bibinfo{publisher}{John Wiley \& Sons}.
\newblock


\bibitem[Bingham et~al\mbox{.}(2019)]%
        {bingham2019pyro}
\bibfield{author}{\bibinfo{person}{Eli Bingham}, \bibinfo{person}{Jonathan~P Chen}, \bibinfo{person}{Martin Jankowiak}, \bibinfo{person}{Fritz Obermeyer}, \bibinfo{person}{Neeraj Pradhan}, \bibinfo{person}{Theofanis Karaletsos}, \bibinfo{person}{Rohit Singh}, \bibinfo{person}{Paul Szerlip}, \bibinfo{person}{Paul Horsfall}, {and} \bibinfo{person}{Noah~D Goodman}.} \bibinfo{year}{2019}\natexlab{}.
\newblock \showarticletitle{Pyro: Deep universal probabilistic programming}.
\newblock \bibinfo{journal}{\emph{The Journal of Machine Learning Research}} \bibinfo{volume}{20}, \bibinfo{number}{1} (\bibinfo{year}{2019}), \bibinfo{pages}{973--978}.
\newblock


\bibitem[Bishop and Nasrabadi(2006)]%
        {bishop2006pattern}
\bibfield{author}{\bibinfo{person}{Christopher~M Bishop} {and} \bibinfo{person}{Nasser~M Nasrabadi}.} \bibinfo{year}{2006}\natexlab{}.
\newblock \bibinfo{booktitle}{\emph{Pattern Recognition and Machine Learning}}. Vol.~\bibinfo{volume}{4}.
\newblock \bibinfo{publisher}{Springer}.
\newblock


\bibitem[Boyen and Koller(1998)]%
        {boyen1998tractable}
\bibfield{author}{\bibinfo{person}{Xavier Boyen} {and} \bibinfo{person}{Daphne Koller}.} \bibinfo{year}{1998}\natexlab{}.
\newblock \showarticletitle{Tractable inference for complex stochastic processes}. In \bibinfo{booktitle}{\emph{Proceedings of the Fourteenth conference on Uncertainty in artificial intelligence}}. \bibinfo{pages}{33--42}.
\newblock


\bibitem[Carpenter et~al\mbox{.}(2017)]%
        {carpenter2017stan}
\bibfield{author}{\bibinfo{person}{Bob Carpenter}, \bibinfo{person}{Andrew Gelman}, \bibinfo{person}{Matthew~D Hoffman}, \bibinfo{person}{Daniel Lee}, \bibinfo{person}{Ben Goodrich}, \bibinfo{person}{Michael Betancourt}, \bibinfo{person}{Marcus Brubaker}, \bibinfo{person}{Jiqiang Guo}, \bibinfo{person}{Peter Li}, {and} \bibinfo{person}{Allen Riddell}.} \bibinfo{year}{2017}\natexlab{}.
\newblock \showarticletitle{Stan: A probabilistic programming language}.
\newblock \bibinfo{journal}{\emph{Journal of Statistical Software}} \bibinfo{volume}{76}, \bibinfo{number}{1} (\bibinfo{year}{2017}).
\newblock


\bibitem[Chaganty et~al\mbox{.}(2013)]%
        {chaganty2013efficiently}
\bibfield{author}{\bibinfo{person}{Arun Chaganty}, \bibinfo{person}{Aditya Nori}, {and} \bibinfo{person}{Sriram Rajamani}.} \bibinfo{year}{2013}\natexlab{}.
\newblock \showarticletitle{Efficiently sampling probabilistic programs via program analysis}. In \bibinfo{booktitle}{\emph{Artificial Intelligence and Statistics}}. PMLR, \bibinfo{pages}{153--160}.
\newblock


\bibitem[Chakarov and Sankaranarayanan(2014)]%
        {chakarov2014expectation}
\bibfield{author}{\bibinfo{person}{Aleksandar Chakarov} {and} \bibinfo{person}{Sriram Sankaranarayanan}.} \bibinfo{year}{2014}\natexlab{}.
\newblock \showarticletitle{Expectation invariants for probabilistic program loops as fixed points}. In \bibinfo{booktitle}{\emph{International Static Analysis Symposium}}. Springer, \bibinfo{pages}{85--100}.
\newblock


\bibitem[Chaudhuri and Solar-Lezama(2010)]%
        {chaudhuri2010smooth}
\bibfield{author}{\bibinfo{person}{Swarat Chaudhuri} {and} \bibinfo{person}{Armando Solar-Lezama}.} \bibinfo{year}{2010}\natexlab{}.
\newblock \showarticletitle{Smooth interpretation}.
\newblock \bibinfo{journal}{\emph{ACM Sigplan Notices}} \bibinfo{volume}{45}, \bibinfo{number}{6} (\bibinfo{year}{2010}), \bibinfo{pages}{279--291}.
\newblock


\bibitem[Chaudhuri and Solar-Lezama(2011)]%
        {chaudhuri2011smoothing}
\bibfield{author}{\bibinfo{person}{Swarat Chaudhuri} {and} \bibinfo{person}{Armando Solar-Lezama}.} \bibinfo{year}{2011}\natexlab{}.
\newblock \showarticletitle{Smoothing a program soundly and robustly}. In \bibinfo{booktitle}{\emph{International Conference on Computer Aided Verification}}. Springer, \bibinfo{pages}{277--292}.
\newblock


\bibitem[Chen et~al\mbox{.}(2022)]%
        {chen2022does}
\bibfield{author}{\bibinfo{person}{Mingshuai Chen}, \bibinfo{person}{Joost-Pieter Katoen}, \bibinfo{person}{Lutz Klinkenberg}, {and} \bibinfo{person}{Tobias Winkler}.} \bibinfo{year}{2022}\natexlab{}.
\newblock \showarticletitle{Does a program yield the right distribution? Verifying probabilistic programs via generating functions}. In \bibinfo{booktitle}{\emph{International Conference on Computer Aided Verification}}. Springer, \bibinfo{pages}{79--101}.
\newblock


\bibitem[Cousot and Cousot(1977)]%
        {cousot1977abstract}
\bibfield{author}{\bibinfo{person}{Patrick Cousot} {and} \bibinfo{person}{Radhia Cousot}.} \bibinfo{year}{1977}\natexlab{}.
\newblock \showarticletitle{Abstract interpretation: a unified lattice model for static analysis of programs by construction or approximation of fixpoints}. In \bibinfo{booktitle}{\emph{Proceedings of the 4th ACM SIGACT-SIGPLAN Symposium on Principles of Programming Languages}}. \bibinfo{pages}{238--252}.
\newblock


\bibitem[Cover(1999)]%
        {cover1999elements}
\bibfield{author}{\bibinfo{person}{Thomas~M Cover}.} \bibinfo{year}{1999}\natexlab{}.
\newblock \bibinfo{booktitle}{\emph{Elements of information theory}}.
\newblock \bibinfo{publisher}{John Wiley \& Sons}.
\newblock


\bibitem[Ethier and Kurtz(2009)]%
        {ethier2009markov}
\bibfield{author}{\bibinfo{person}{Stewart~N Ethier} {and} \bibinfo{person}{Thomas~G Kurtz}.} \bibinfo{year}{2009}\natexlab{}.
\newblock \bibinfo{booktitle}{\emph{Markov processes: characterization and convergence}}.
\newblock \bibinfo{publisher}{John Wiley \& Sons}.
\newblock


\bibitem[Filieri et~al\mbox{.}(2013)]%
        {filieri2013reliability}
\bibfield{author}{\bibinfo{person}{Antonio Filieri}, \bibinfo{person}{Corina~S P{\u{a}}s{\u{a}}reanu}, {and} \bibinfo{person}{Willem Visser}.} \bibinfo{year}{2013}\natexlab{}.
\newblock \showarticletitle{Reliability analysis in symbolic pathfinder}. In \bibinfo{booktitle}{\emph{2013 35th International Conference on Software Engineering (ICSE)}}. IEEE, \bibinfo{pages}{622--631}.
\newblock


\bibitem[Florescu(2014)]%
        {florescu2014probability}
\bibfield{author}{\bibinfo{person}{Ionut Florescu}.} \bibinfo{year}{2014}\natexlab{}.
\newblock \bibinfo{booktitle}{\emph{Probability and stochastic processes}}.
\newblock \bibinfo{publisher}{John Wiley \& Sons}.
\newblock


\bibitem[Gao et~al\mbox{.}(2017)]%
        {gao2017estimating}
\bibfield{author}{\bibinfo{person}{Weihao Gao}, \bibinfo{person}{Sreeram Kannan}, \bibinfo{person}{Sewoong Oh}, {and} \bibinfo{person}{Pramod Viswanath}.} \bibinfo{year}{2017}\natexlab{}.
\newblock \showarticletitle{Estimating mutual information for discrete-continuous mixtures}.
\newblock \bibinfo{journal}{\emph{Advances in Neural Information Processing Systems}}  \bibinfo{volume}{30} (\bibinfo{year}{2017}).
\newblock


\bibitem[Gehr et~al\mbox{.}(2016)]%
        {gehr2016psi}
\bibfield{author}{\bibinfo{person}{Timon Gehr}, \bibinfo{person}{Sasa Misailovic}, {and} \bibinfo{person}{Martin Vechev}.} \bibinfo{year}{2016}\natexlab{}.
\newblock \showarticletitle{PSI: Exact symbolic inference for probabilistic programs}. In \bibinfo{booktitle}{\emph{International Conference on Computer Aided Verification}}. Springer, \bibinfo{pages}{62--83}.
\newblock


\bibitem[Gelman et~al\mbox{.}(2013)]%
        {gelman2013bayesian}
\bibfield{author}{\bibinfo{person}{Andrew Gelman}, \bibinfo{person}{John~B Carlin}, \bibinfo{person}{Hal~S Stern}, \bibinfo{person}{David~B Dunson}, \bibinfo{person}{Aki Vehtari}, {and} \bibinfo{person}{Donald~B Rubin}.} \bibinfo{year}{2013}\natexlab{}.
\newblock \bibinfo{booktitle}{\emph{Bayesian data analysis}}.
\newblock \bibinfo{publisher}{CRC press}.
\newblock


\bibitem[Goodman et~al\mbox{.}(2008)]%
        {goodman2008church}
\bibfield{author}{\bibinfo{person}{Noah~D Goodman}, \bibinfo{person}{Vikash~K Mansinghka}, \bibinfo{person}{Daniel Roy}, \bibinfo{person}{Keith Bonawitz}, {and} \bibinfo{person}{Joshua~B Tenenbaum}.} \bibinfo{year}{2008}\natexlab{}.
\newblock \showarticletitle{Church: a language for generative models}. In \bibinfo{booktitle}{\emph{Proceedings of the Twenty-Fourth Conference on Uncertainty in Artificial Intelligence}}. \bibinfo{pages}{220--229}.
\newblock


\bibitem[Gordon et~al\mbox{.}(2014)]%
        {gordon2014probabilistic}
\bibfield{author}{\bibinfo{person}{Andrew~D Gordon}, \bibinfo{person}{Thomas~A Henzinger}, \bibinfo{person}{Aditya~V Nori}, {and} \bibinfo{person}{Sriram~K Rajamani}.} \bibinfo{year}{2014}\natexlab{}.
\newblock \showarticletitle{Probabilistic programming}.
\newblock In \bibinfo{booktitle}{\emph{Future of Software Engineering Proceedings}}. \bibinfo{pages}{167--181}.
\newblock


\bibitem[Gu and Eisenstat(1995)]%
        {gu1995divide}
\bibfield{author}{\bibinfo{person}{Ming Gu} {and} \bibinfo{person}{Stanley~C Eisenstat}.} \bibinfo{year}{1995}\natexlab{}.
\newblock \showarticletitle{A divide-and-conquer algorithm for the symmetric tridiagonal eigenproblem}.
\newblock \bibinfo{journal}{\emph{SIAM J. Matrix Anal. Appl.}} \bibinfo{volume}{16}, \bibinfo{number}{1} (\bibinfo{year}{1995}), \bibinfo{pages}{172--191}.
\newblock


\bibitem[Hansen(2010)]%
        {hansen2010generalized}
\bibfield{author}{\bibinfo{person}{Lars~Peter Hansen}.} \bibinfo{year}{2010}\natexlab{}.
\newblock \showarticletitle{Generalized method of moments estimation}.
\newblock In \bibinfo{booktitle}{\emph{Macroeconometrics and Time series Analysis}}. \bibinfo{publisher}{Springer}, \bibinfo{pages}{105--118}.
\newblock


\bibitem[Hastings(1970)]%
        {hastings1970monte}
\bibfield{author}{\bibinfo{person}{W~Keith Hastings}.} \bibinfo{year}{1970}\natexlab{}.
\newblock \showarticletitle{Monte Carlo sampling methods using Markov chains and their applications}.
\newblock  (\bibinfo{year}{1970}).
\newblock


\bibitem[Hoffman et~al\mbox{.}(2013)]%
        {hoffman2013stochastic}
\bibfield{author}{\bibinfo{person}{Matthew~D Hoffman}, \bibinfo{person}{David~M Blei}, \bibinfo{person}{Chong Wang}, {and} \bibinfo{person}{John Paisley}.} \bibinfo{year}{2013}\natexlab{}.
\newblock \showarticletitle{Stochastic variational inference}.
\newblock \bibinfo{journal}{\emph{Journal of Machine Learning Research}} (\bibinfo{year}{2013}).
\newblock


\bibitem[Hofmann and Puzicha(1999)]%
        {hofmann1999latent}
\bibfield{author}{\bibinfo{person}{Thomas Hofmann} {and} \bibinfo{person}{Jan Puzicha}.} \bibinfo{year}{1999}\natexlab{}.
\newblock \showarticletitle{Latent class models for collaborative filtering}. In \bibinfo{booktitle}{\emph{IJCAI}}, Vol.~\bibinfo{volume}{99}.
\newblock


\bibitem[Holtzen et~al\mbox{.}(2020)]%
        {holtzen2020scaling}
\bibfield{author}{\bibinfo{person}{Steven Holtzen}, \bibinfo{person}{Guy Van~den Broeck}, {and} \bibinfo{person}{Todd Millstein}.} \bibinfo{year}{2020}\natexlab{}.
\newblock \showarticletitle{Scaling exact inference for discrete probabilistic programs}.
\newblock \bibinfo{journal}{\emph{Proceedings of the ACM on Programming Languages}} \bibinfo{volume}{4}, \bibinfo{number}{OOPSLA} (\bibinfo{year}{2020}), \bibinfo{pages}{1--31}.
\newblock


\bibitem[Hornik et~al\mbox{.}(1989)]%
        {hornik1989multilayer}
\bibfield{author}{\bibinfo{person}{Kurt Hornik}, \bibinfo{person}{Maxwell Stinchcombe}, {and} \bibinfo{person}{Halbert White}.} \bibinfo{year}{1989}\natexlab{}.
\newblock \showarticletitle{Multilayer feedforward networks are universal approximators}.
\newblock \bibinfo{journal}{\emph{Neural Networks}} \bibinfo{volume}{2}, \bibinfo{number}{5} (\bibinfo{year}{1989}), \bibinfo{pages}{359--366}.
\newblock


\bibitem[Huang et~al\mbox{.}(2021)]%
        {huang2021aqua}
\bibfield{author}{\bibinfo{person}{Zixin Huang}, \bibinfo{person}{Saikat Dutta}, {and} \bibinfo{person}{Sasa Misailovic}.} \bibinfo{year}{2021}\natexlab{}.
\newblock \showarticletitle{Aqua: Automated quantized inference for probabilistic programs}. In \bibinfo{booktitle}{\emph{International Symposium on Automated Technology for Verification and Analysis}}. Springer, \bibinfo{pages}{229--246}.
\newblock


\bibitem[Jordan et~al\mbox{.}(1999)]%
        {jordan1999introduction}
\bibfield{author}{\bibinfo{person}{Michael~I Jordan}, \bibinfo{person}{Zoubin Ghahramani}, \bibinfo{person}{Tommi~S Jaakkola}, {and} \bibinfo{person}{Lawrence~K Saul}.} \bibinfo{year}{1999}\natexlab{}.
\newblock \showarticletitle{An introduction to variational methods for graphical models}.
\newblock \bibinfo{journal}{\emph{Machine Learning}} \bibinfo{volume}{37}, \bibinfo{number}{2} (\bibinfo{year}{1999}), \bibinfo{pages}{183--233}.
\newblock


\bibitem[Kan and Robotti(2017)]%
        {kan2017moments}
\bibfield{author}{\bibinfo{person}{Raymond Kan} {and} \bibinfo{person}{Cesare Robotti}.} \bibinfo{year}{2017}\natexlab{}.
\newblock \showarticletitle{On moments of folded and truncated multivariate normal distributions}.
\newblock \bibinfo{journal}{\emph{Journal of Computational and Graphical Statistics}} \bibinfo{volume}{26}, \bibinfo{number}{4} (\bibinfo{year}{2017}), \bibinfo{pages}{930--934}.
\newblock


\bibitem[Katoen et~al\mbox{.}(2010)]%
        {katoen2010linear}
\bibfield{author}{\bibinfo{person}{Joost-Pieter Katoen}, \bibinfo{person}{Annabelle~K McIver}, \bibinfo{person}{Larissa~A Meinicke}, {and} \bibinfo{person}{Carroll~C Morgan}.} \bibinfo{year}{2010}\natexlab{}.
\newblock \showarticletitle{Linear-invariant generation for probabilistic programs}. In \bibinfo{booktitle}{\emph{International Static Analysis Symposium}}. Springer, \bibinfo{pages}{390--406}.
\newblock


\bibitem[Kharchenko et~al\mbox{.}(2014)]%
        {kharchenko2014bayesian}
\bibfield{author}{\bibinfo{person}{Peter~V Kharchenko}, \bibinfo{person}{Lev Silberstein}, {and} \bibinfo{person}{David~T Scadden}.} \bibinfo{year}{2014}\natexlab{}.
\newblock \showarticletitle{Bayesian approach to single-cell differential expression analysis}.
\newblock \bibinfo{journal}{\emph{Nature Methods}} \bibinfo{volume}{11}, \bibinfo{number}{7} (\bibinfo{year}{2014}), \bibinfo{pages}{740--742}.
\newblock


\bibitem[Koren et~al\mbox{.}(2021)]%
        {koren2021advances}
\bibfield{author}{\bibinfo{person}{Yehuda Koren}, \bibinfo{person}{Steffen Rendle}, {and} \bibinfo{person}{Robert Bell}.} \bibinfo{year}{2021}\natexlab{}.
\newblock \showarticletitle{Advances in collaborative filtering}.
\newblock \bibinfo{journal}{\emph{Recommender systems handbook}} (\bibinfo{year}{2021}), \bibinfo{pages}{91--142}.
\newblock


\bibitem[Kozen(1979)]%
        {kozen1979semantics}
\bibfield{author}{\bibinfo{person}{Dexter Kozen}.} \bibinfo{year}{1979}\natexlab{}.
\newblock \showarticletitle{Semantics of probabilistic programs}. In \bibinfo{booktitle}{\emph{20th Annual Symposium on Foundations of Computer Science (FOCS 1979)}}. IEEE, \bibinfo{pages}{101--114}.
\newblock


\bibitem[Kozen(1983)]%
        {kozen1983probabilistic}
\bibfield{author}{\bibinfo{person}{Dexter Kozen}.} \bibinfo{year}{1983}\natexlab{}.
\newblock \showarticletitle{A probabilistic {PDL}}. In \bibinfo{booktitle}{\emph{Proceedings of the fifteenth annual ACM Symposium on Theory of computing}}. \bibinfo{pages}{291--297}.
\newblock


\bibitem[Kucukelbir et~al\mbox{.}(2015)]%
        {kucukelbir2015automatic}
\bibfield{author}{\bibinfo{person}{Alp Kucukelbir}, \bibinfo{person}{Rajesh Ranganath}, \bibinfo{person}{Andrew Gelman}, {and} \bibinfo{person}{David Blei}.} \bibinfo{year}{2015}\natexlab{}.
\newblock \showarticletitle{Automatic variational inference in Stan}.
\newblock \bibinfo{journal}{\emph{Advances in Neural Information Processing Systems}}  \bibinfo{volume}{28} (\bibinfo{year}{2015}).
\newblock


\bibitem[Kullback and Leibler(1951)]%
        {kullback1951information}
\bibfield{author}{\bibinfo{person}{Solomon Kullback} {and} \bibinfo{person}{Richard~A Leibler}.} \bibinfo{year}{1951}\natexlab{}.
\newblock \showarticletitle{On information and sufficiency}.
\newblock \bibinfo{journal}{\emph{The Annals of Mathematical Statistics}} \bibinfo{volume}{22}, \bibinfo{number}{1} (\bibinfo{year}{1951}), \bibinfo{pages}{79--86}.
\newblock


\bibitem[Lasserre(2009)]%
        {lasserre2009moments}
\bibfield{author}{\bibinfo{person}{Jean~Bernard Lasserre}.} \bibinfo{year}{2009}\natexlab{}.
\newblock \bibinfo{booktitle}{\emph{Moments, positive polynomials and their applications}}. Vol.~\bibinfo{volume}{1}.
\newblock \bibinfo{publisher}{World Scientific}.
\newblock


\bibitem[Laurel and Misailovic(2020)]%
        {laurel2020continualization}
\bibfield{author}{\bibinfo{person}{Jacob Laurel} {and} \bibinfo{person}{Sasa Misailovic}.} \bibinfo{year}{2020}\natexlab{}.
\newblock \showarticletitle{Continualization of probabilistic programs with correction}. In \bibinfo{booktitle}{\emph{European Symposium on Programming}}. Springer, Cham, \bibinfo{pages}{366--393}.
\newblock


\bibitem[Lo(1972)]%
        {lo1972finite}
\bibfield{author}{\bibinfo{person}{J Lo}.} \bibinfo{year}{1972}\natexlab{}.
\newblock \showarticletitle{Finite-dimensional sensor orbits and optimal nonlinear filtering}.
\newblock \bibinfo{journal}{\emph{IEEE Transactions on information theory}} \bibinfo{volume}{18}, \bibinfo{number}{5} (\bibinfo{year}{1972}), \bibinfo{pages}{583--588}.
\newblock


\bibitem[Mansinghka et~al\mbox{.}(2014)]%
        {mansinghka2014venture}
\bibfield{author}{\bibinfo{person}{Vikash Mansinghka}, \bibinfo{person}{Daniel Selsam}, {and} \bibinfo{person}{Yura Perov}.} \bibinfo{year}{2014}\natexlab{}.
\newblock \showarticletitle{Venture: a higher-order probabilistic programming platform with programmable inference}.
\newblock \bibinfo{journal}{\emph{arXiv preprint arXiv:1404.0099}} (\bibinfo{year}{2014}).
\newblock


\bibitem[Milch et~al\mbox{.}(2004)]%
        {milch2004blog}
\bibfield{author}{\bibinfo{person}{Brian Milch}, \bibinfo{person}{Bhaskara Marthi}, {and} \bibinfo{person}{Stuart Russell}.} \bibinfo{year}{2004}\natexlab{}.
\newblock \showarticletitle{BLOG: Relational modeling with unknown objects}. In \bibinfo{booktitle}{\emph{ICML 2004 workshop on statistical relational learning and its connections to other fields}}. \bibinfo{pages}{67--73}.
\newblock


\bibitem[Moosbrugger et~al\mbox{.}(2022)]%
        {moosbrugger2022moment}
\bibfield{author}{\bibinfo{person}{Marcel Moosbrugger}, \bibinfo{person}{Miroslav Stankovi{\v{c}}}, \bibinfo{person}{Ezio Bartocci}, {and} \bibinfo{person}{Laura Kov{\'a}cs}.} \bibinfo{year}{2022}\natexlab{}.
\newblock \showarticletitle{This is the moment for probabilistic loops}.
\newblock \bibinfo{journal}{\emph{Proceedings of the ACM on Programming Languages}} \bibinfo{volume}{6}, \bibinfo{number}{OOPSLA2} (\bibinfo{year}{2022}), \bibinfo{pages}{1497--1525}.
\newblock


\bibitem[Narayanan et~al\mbox{.}(2016)]%
        {narayanan2016probabilistic}
\bibfield{author}{\bibinfo{person}{Praveen Narayanan}, \bibinfo{person}{Jacques Carette}, \bibinfo{person}{Wren Romano}, \bibinfo{person}{Chung-chieh Shan}, {and} \bibinfo{person}{Robert Zinkov}.} \bibinfo{year}{2016}\natexlab{}.
\newblock \showarticletitle{Probabilistic inference by program transformation in {H}akaru (system description)}. In \bibinfo{booktitle}{\emph{International Symposium on Functional and Logic Programming}}. Springer, \bibinfo{pages}{62--79}.
\newblock


\bibitem[Neal(1999)]%
        {neal1999erroneous}
\bibfield{author}{\bibinfo{person}{Radford~M Neal}.} \bibinfo{year}{1999}\natexlab{}.
\newblock \showarticletitle{Erroneous results in “Marginal likelihood from the Gibbs output”}.
\newblock \bibinfo{journal}{\emph{minmeo, University of Toronto}} (\bibinfo{year}{1999}).
\newblock


\bibitem[Nishihara et~al\mbox{.}(2013)]%
        {nishihara2013detecting}
\bibfield{author}{\bibinfo{person}{Robert Nishihara}, \bibinfo{person}{Thomas Minka}, {and} \bibinfo{person}{Daniel Tarlow}.} \bibinfo{year}{2013}\natexlab{}.
\newblock \showarticletitle{Detecting parameter symmetries in probabilistic models}.
\newblock \bibinfo{journal}{\emph{arXiv preprint arXiv:1312.5386}} (\bibinfo{year}{2013}).
\newblock


\bibitem[Nitti et~al\mbox{.}(2016)]%
        {nitti2016probabilistic}
\bibfield{author}{\bibinfo{person}{Davide Nitti}, \bibinfo{person}{Tinne De~Laet}, {and} \bibinfo{person}{Luc De~Raedt}.} \bibinfo{year}{2016}\natexlab{}.
\newblock \showarticletitle{Probabilistic logic programming for hybrid relational domains}.
\newblock \bibinfo{journal}{\emph{Machine Learning}} \bibinfo{volume}{103}, \bibinfo{number}{3} (\bibinfo{year}{2016}), \bibinfo{pages}{407--449}.
\newblock


\bibitem[Nori et~al\mbox{.}(2014)]%
        {nori2014r2}
\bibfield{author}{\bibinfo{person}{Aditya Nori}, \bibinfo{person}{Chung-Kil Hur}, \bibinfo{person}{Sriram Rajamani}, {and} \bibinfo{person}{Selva Samuel}.} \bibinfo{year}{2014}\natexlab{}.
\newblock \showarticletitle{R2: An efficient MCMC sampler for probabilistic programs}. In \bibinfo{booktitle}{\emph{Proceedings of the AAAI Conference on Artificial Intelligence}}, Vol.~\bibinfo{volume}{28}.
\newblock


\bibitem[Obermeyer et~al\mbox{.}(2019)]%
        {obermeyer2019tensor}
\bibfield{author}{\bibinfo{person}{Fritz Obermeyer}, \bibinfo{person}{Eli Bingham}, \bibinfo{person}{Martin Jankowiak}, \bibinfo{person}{Neeraj Pradhan}, \bibinfo{person}{Justin Chiu}, \bibinfo{person}{Alexander Rush}, {and} \bibinfo{person}{Noah Goodman}.} \bibinfo{year}{2019}\natexlab{}.
\newblock \showarticletitle{Tensor variable elimination for plated factor graphs}. In \bibinfo{booktitle}{\emph{International Conference on Machine Learning}}. PMLR, \bibinfo{pages}{4871--4880}.
\newblock


\bibitem[P{\'e}rez and Quintana(2006)]%
        {perez2006survey}
\bibfield{author}{\bibinfo{person}{Dilcia P{\'e}rez} {and} \bibinfo{person}{Yamilet Quintana}.} \bibinfo{year}{2006}\natexlab{}.
\newblock \showarticletitle{A survey on the Weierstrass approximation theorem}.
\newblock \bibinfo{journal}{\emph{arXiv preprint math/0611038}} (\bibinfo{year}{2006}).
\newblock


\bibitem[Pfeffer(2001)]%
        {pfeffer2001ibal}
\bibfield{author}{\bibinfo{person}{Avi Pfeffer}.} \bibinfo{year}{2001}\natexlab{}.
\newblock \showarticletitle{IBAL: A probabilistic rational programming language}. In \bibinfo{booktitle}{\emph{IJCAI}}. Citeseer, \bibinfo{pages}{733--740}.
\newblock


\bibitem[Pierson and Yau(2015)]%
        {pierson2015zifa}
\bibfield{author}{\bibinfo{person}{Emma Pierson} {and} \bibinfo{person}{Christopher Yau}.} \bibinfo{year}{2015}\natexlab{}.
\newblock \showarticletitle{ZIFA: Dimensionality reduction for zero-inflated single-cell gene expression analysis}.
\newblock \bibinfo{journal}{\emph{Genome Biology}} \bibinfo{volume}{16}, \bibinfo{number}{1} (\bibinfo{year}{2015}), \bibinfo{pages}{1--10}.
\newblock


\bibitem[Saad et~al\mbox{.}(2021)]%
        {saad2021sppl}
\bibfield{author}{\bibinfo{person}{Feras~A Saad}, \bibinfo{person}{Martin~C Rinard}, {and} \bibinfo{person}{Vikash~K Mansinghka}.} \bibinfo{year}{2021}\natexlab{}.
\newblock \showarticletitle{SPPL: probabilistic programming with fast exact symbolic inference}. In \bibinfo{booktitle}{\emph{Proceedings of the 42nd ACM SIGPLAN International Conference on Programming Language Design and Implementation}}. \bibinfo{pages}{804--819}.
\newblock


\bibitem[Schm{\"u}dgen(2017)]%
        {schmudgen2017moment}
\bibfield{author}{\bibinfo{person}{Konrad Schm{\"u}dgen}.} \bibinfo{year}{2017}\natexlab{}.
\newblock \bibinfo{booktitle}{\emph{The moment problem}}. Vol.~\bibinfo{volume}{9}.
\newblock \bibinfo{publisher}{Springer}.
\newblock


\bibitem[Skiena(2008)]%
        {skiena2008algorithm}
\bibfield{author}{\bibinfo{person}{SS Skiena}.} \bibinfo{year}{2008}\natexlab{}.
\newblock \bibinfo{title}{The Algorithm Design Manual. Springer Publishing Company}.
\newblock
\newblock


\bibitem[Tierney and Kadane(1986)]%
        {tierney1986accurate}
\bibfield{author}{\bibinfo{person}{Luke Tierney} {and} \bibinfo{person}{Joseph~B Kadane}.} \bibinfo{year}{1986}\natexlab{}.
\newblock \showarticletitle{Accurate approximations for posterior moments and marginal densities}.
\newblock \bibinfo{journal}{\emph{J. Amer. Statist. Assoc.}} \bibinfo{volume}{81}, \bibinfo{number}{393} (\bibinfo{year}{1986}), \bibinfo{pages}{82--86}.
\newblock


\bibitem[Tolpin et~al\mbox{.}(2016)]%
        {tolpin2016design}
\bibfield{author}{\bibinfo{person}{David Tolpin}, \bibinfo{person}{Jan-Willem van~de Meent}, \bibinfo{person}{Hongseok Yang}, {and} \bibinfo{person}{Frank Wood}.} \bibinfo{year}{2016}\natexlab{}.
\newblock \showarticletitle{Design and implementation of probabilistic programming language anglican}. In \bibinfo{booktitle}{\emph{Proceedings of the 28th Symposium on the Implementation and Application of Functional programming Languages}}. \bibinfo{pages}{1--12}.
\newblock


\bibitem[Tsiatis(1975)]%
        {tsiatis1975nonidentifiability}
\bibfield{author}{\bibinfo{person}{Anastasios Tsiatis}.} \bibinfo{year}{1975}\natexlab{}.
\newblock \showarticletitle{A nonidentifiability aspect of the problem of competing risks.}
\newblock \bibinfo{journal}{\emph{Proceedings of the National Academy of Sciences}} \bibinfo{volume}{72}, \bibinfo{number}{1} (\bibinfo{year}{1975}), \bibinfo{pages}{20--22}.
\newblock


\bibitem[Virtanen et~al\mbox{.}(2020)]%
        {2020SciPy-NMeth}
\bibfield{author}{\bibinfo{person}{Pauli Virtanen}, \bibinfo{person}{Ralf Gommers}, \bibinfo{person}{Travis~E. Oliphant}, \bibinfo{person}{Matt Haberland}, \bibinfo{person}{Tyler Reddy}, \bibinfo{person}{David Cournapeau}, \bibinfo{person}{Evgeni Burovski}, \bibinfo{person}{Pearu Peterson}, \bibinfo{person}{Warren Weckesser}, \bibinfo{person}{Jonathan Bright}, \bibinfo{person}{St{\'e}fan~J. {van der Walt}}, \bibinfo{person}{Matthew Brett}, \bibinfo{person}{Joshua Wilson}, \bibinfo{person}{K.~Jarrod Millman}, \bibinfo{person}{Nikolay Mayorov}, \bibinfo{person}{Andrew R.~J. Nelson}, \bibinfo{person}{Eric Jones}, \bibinfo{person}{Robert Kern}, \bibinfo{person}{Eric Larson}, \bibinfo{person}{C~J Carey}, \bibinfo{person}{{\.I}lhan Polat}, \bibinfo{person}{Yu Feng}, \bibinfo{person}{Eric~W. Moore}, \bibinfo{person}{Jake {VanderPlas}}, \bibinfo{person}{Denis Laxalde}, \bibinfo{person}{Josef Perktold}, \bibinfo{person}{Robert Cimrman}, \bibinfo{person}{Ian Henriksen}, \bibinfo{person}{E.~A. Quintero},
  \bibinfo{person}{Charles~R. Harris}, \bibinfo{person}{Anne~M. Archibald}, \bibinfo{person}{Ant{\^o}nio~H. Ribeiro}, \bibinfo{person}{Fabian Pedregosa}, \bibinfo{person}{Paul {van Mulbregt}}, {and} \bibinfo{person}{{SciPy 1.0 Contributors}}.} \bibinfo{year}{2020}\natexlab{}.
\newblock \showarticletitle{{{SciPy} 1.0: Fundamental Algorithms for Scientific Computing in Python}}.
\newblock \bibinfo{journal}{\emph{Nature Methods}}  \bibinfo{volume}{17} (\bibinfo{year}{2020}), \bibinfo{pages}{261--272}.
\newblock
\urldef\tempurl%
\url{https://doi.org/10.1038/s41592-019-0686-2}
\showDOI{\tempurl}


\bibitem[Wang et~al\mbox{.}(2015)]%
        {wang2015estimating}
\bibfield{author}{\bibinfo{person}{Sida Wang}, \bibinfo{person}{Arun~Tejasvi Chaganty}, {and} \bibinfo{person}{Percy~S Liang}.} \bibinfo{year}{2015}\natexlab{}.
\newblock \showarticletitle{Estimating mixture models via mixtures of polynomials}.
\newblock \bibinfo{journal}{\emph{Advances in Neural Information Processing Systems}}  \bibinfo{volume}{28} (\bibinfo{year}{2015}).
\newblock


\bibitem[Wick(1950)]%
        {wick1950evaluation}
\bibfield{author}{\bibinfo{person}{Gian-Carlo Wick}.} \bibinfo{year}{1950}\natexlab{}.
\newblock \showarticletitle{The evaluation of the collision matrix}.
\newblock \bibinfo{journal}{\emph{Physical Review}} \bibinfo{volume}{80}, \bibinfo{number}{2} (\bibinfo{year}{1950}), \bibinfo{pages}{268}.
\newblock


\bibitem[{Wolfram Research, Inc.}({[n.\,d.]})]%
        {Mathematica}
\bibfield{author}{\bibinfo{person}{{Wolfram Research, Inc.}}} \bibinfo{year}{[n.\,d.]}\natexlab{}.
\newblock \bibinfo{booktitle}{\emph{Mathematica}}.
\newblock
\urldef\tempurl%
\url{https://www.wolfram.com/mathematica}
\showURL{%
\tempurl}


\bibitem[Wu et~al\mbox{.}(2018)]%
        {wu2018discrete}
\bibfield{author}{\bibinfo{person}{Yi Wu}, \bibinfo{person}{Siddharth Srivastava}, \bibinfo{person}{Nicholas Hay}, \bibinfo{person}{Simon Du}, {and} \bibinfo{person}{Stuart Russell}.} \bibinfo{year}{2018}\natexlab{}.
\newblock \showarticletitle{Discrete-continuous mixtures in probabilistic programming: Generalized semantics and inference algorithms}. In \bibinfo{booktitle}{\emph{International Conference on Machine Learning}}. PMLR, \bibinfo{pages}{5343--5352}.
\newblock


\bibitem[Zhao et~al\mbox{.}(2013)]%
        {zhao2013interactive}
\bibfield{author}{\bibinfo{person}{Xiaoxue Zhao}, \bibinfo{person}{Weinan Zhang}, {and} \bibinfo{person}{Jun Wang}.} \bibinfo{year}{2013}\natexlab{}.
\newblock \showarticletitle{Interactive collaborative filtering}. In \bibinfo{booktitle}{\emph{Proceedings of the 22nd ACM International Conference on Information \& Knowledge Management}}. \bibinfo{pages}{1411--1420}.
\newblock


\bibitem[Zhou(2020)]%
        {zhou2020universality}
\bibfield{author}{\bibinfo{person}{Ding-Xuan Zhou}.} \bibinfo{year}{2020}\natexlab{}.
\newblock \showarticletitle{Universality of deep convolutional neural networks}.
\newblock \bibinfo{journal}{\emph{Applied and Computational Harmonic Analysis}} \bibinfo{volume}{48}, \bibinfo{number}{2} (\bibinfo{year}{2020}), \bibinfo{pages}{787--794}.
\newblock


\bibitem[Zhou et~al\mbox{.}(2020)]%
        {zhou2020divide}
\bibfield{author}{\bibinfo{person}{Yuan Zhou}, \bibinfo{person}{Hongseok Yang}, \bibinfo{person}{Yee~Whye Teh}, {and} \bibinfo{person}{Tom Rainforth}.} \bibinfo{year}{2020}\natexlab{}.
\newblock \showarticletitle{Divide, conquer, and combine: a new inference strategy for probabilistic programs with stochastic support}. In \bibinfo{booktitle}{\emph{International Conference on Machine Learning}}. PMLR, \bibinfo{pages}{11534--11545}.
\newblock


\end{thebibliography}
